\documentclass[prd,aps,nofootinbib,floatfix,11pt]{revtex4}

\usepackage{amsmath,graphicx,epsfig,amssymb,dsfont,mathtools,}
\usepackage[usenames]{color}
\usepackage{ulem} 
\usepackage{bigstrut}
\usepackage{slashed}
\usepackage{multirow}
\usepackage{makecell}

\allowdisplaybreaks

\begin{document}
\title{Weak Decays of Triply Heavy Baryons }
\author{Wei Wang~\footnote{Email:wei.wang@sjtu.edu.cn} and  Ji Xu~\footnote{Email:xuji1991@sjtu.edu.cn}}
\affiliation{
INPAC, Shanghai Key Laboratory for Particle Physics and Cosmology, MOE Key Laboratory for Particle Physics, Astrophysics and Cosmology,  \\ School of Physics and Astronomy, Shanghai Jiao Tong University, Shanghai  200240,   China }

\begin{abstract}
After the experimental establishment  of doubly heavy baryons,   baryons with three quarks  are the last  missing pieces  of the lowest-lying baryon multiplets in   quark model. In this work  we study semileptonic  and nonleptonic weak decays of triply heavy baryons, $\Omega_{ccc}^{++}, \Omega_{ccb}^{+}, \Omega_{cbb}^{0}, \Omega_{bbb}^{-}$.   Decay amplitudes for various channels are parametrized in terms of a few  SU(3) irreducible amplitudes.  We point out  that branching fractions for Cabibbo allowed processes,  $\Omega_{ccc}\to (\Xi_{cc}^{++} \overline K^0, \Xi_{cc}^{++}K^-\pi^+, \Omega_{cc}^{+}\pi^+, \Xi_{c}^+ D^+, \Xi_{c}^{\prime} D^+, \Lambda_c D^+\overline K^0, \Xi_{c}^+ D^0 \pi^+, \Xi_{c}^0 D^+\pi^+)$ may reach a few percents.  We suggest our experimental colleagues to perform a search at hadron colliders and  the electron and positron collisions in future, which will presumably   lead to   discoveries of triply heavy baryons and complete the baryon multiplets. Using the expanded   amplitudes, we derive a number of relations for the partial  widths which can be examined in future.
\end{abstract}

\maketitle

\section{Introduction}

In the past decades,   hadron spectroscopy has experienced a rapid renaissance, predominantly   propelled by   discoveries of a number of hadron resonances  that defy the standard quark model interpretations.  These resonant states are generically  classified  as hadron exotics, and for  reviews on recent progresses, please see Refs.~\cite{Chen:2016qju,Chen:2016spr,Ali:2017jda,Guo:2017jvc}.   Among all  exotic hadrons, the $X(3872)$ plays a most important  role due to its special properties. Aside from these unexpected discoveries, there are also gradual progresses in the traditional sector of the charmonium and bottomonium spectroscopy.
One of the highlights in recent years  is the discovery of $\Xi_{cc}^{++}$ by the LHCb collaboration~\cite{Aaij:2017ueg}:
\begin{eqnarray}
m_{\Xi_{cc}^{++}} =  (3621.40\pm0.72\pm0.27\pm0.14) {\rm MeV}.
\end{eqnarray}
This observed state  fills well in the quark model as the lowest-lying  $ccu$ baryon~\cite{Patrignani:2016xqp}.
After the tentative establishment of $\Xi_{cc}^{++}$,  it is plausible to fill the baryon family with the last missing members, namely, baryons made of three heavy quarks. Charm and bottom quarks are much heavier than the $u,d,s$, thus  baryons with three heavy quarks will refrain from light quark contaminations and  the study of  triply heavy baryons  can   help us to better understand  the dynamics of strong interactions.

Previous studies  of  triply heavy baryons  concentrated  on three    facets:   spectroscopy, production and decays. Most theoretical investigations  in the literature, especially in recent years, have  focused on the masses and magnetic moments~\cite{Jia:2006gw,Martynenko:2007je,Patel:2008mv,Zhang:2009re,Meinel:2010pw,LlanesEstrada:2011kc,Wang:2011ae,Meinel:2012qz,Aliev:2012tt,Padmanath:2013zfa,Brown:2014ena,Wei:2015gsa,Wei:2016jyk,Thakkar:2016sog}, while less attentions have  been paid to the production and decay properties.  The only available  estimate of the production is  conducted in Refs.~\cite{Chen:2011mb,GomshiNobary:2004mq}, where the cross sections at the LHC with $\sqrt s=7$ TeV are found to reach the 0.1 nb level depending on different kinematics cuts.  In the $b\to c$ transitions among triply heavy baryons,   Ref.~\cite{Flynn:2011gf}  has discussed the implications of heavy quark spin symmetry.  Some   decay modes of the $\Omega_{ccc}$ are analyzed recently in Ref.~\cite{Geng:2017mxn}.

The main focus of this paper is to provide a systematic analysis of   weak decays of the lowest-lying triply heavy baryons, $\Omega_{ccc, ccb, cbb, bbb}$.  The $\Omega_{ccc}$ and $\Omega_{bbb}$ have spin $3/2$, while the $\Omega_{ccb}$ and $\Omega_{cbb}$ can be  the  $J^P=1/2^+$ or $J^P=3/2^+$ state.  As we will show,  various types of weak decays of   triply heavy baryons occur, but  unfortunately, a universal dynamical (factorization) approach can not be established yet.  This gives a barrier for us to predict their decay widths.  Instead we will use an optional  theoretical tools to analyze   heavy quark decays,  the flavor SU(3) symmetry~\cite{Zeppenfeld:1980ex,Savage:1989ub,Gronau:1994rj,Grinstein:1996us,He:1998rq,Deshpande:2000jp,He:2000dg,Deshpande:1994ii,He:2000ys,Fu:2002nr,Chiang:2003pm,Chiang:2004nm,Chiang:2006ih,Li:2007bh,Wang:2008rk,Chiang:2008zb,Cheng:2014rfa,He:2014xha,He:2015fwa,Hsiao:2015iiu,He:2016xvd,He:2017fln,Lu:2016ogy,Cheng:2016ejf,Cheng:2012xb,Cheng:2011qh,Wang:2017azm,Geng:2017mxn,Shi:2017dto,Geng:2017esc,Geng:2018plk}. One advantage of the SU(3) analysis is that it is independent of the factorization details.  In this work,  we consider    semileptonic decay channels with one or two hadrons in the final state,  while for nonleptonic decays,  the two-body and three-body   modes will be analyzed.

The rest of this paper is organized as follows.
In Sec.~\ref{sec:particle_multiplet}, we will collect the representation matrices for various particle multiplets in the SU(3) group. In Sec.~\ref{sec:semileptonic},   semileptonic   decay modes with one or two hadrons in the final state are analyzed. In Sec.~\ref{sec:ccc_nonleptonic}, Sec.~\ref{sec:bbb_nonleptonic} and Sec.~\ref{sec:ccb_cbb_nonleptonic}, nonleptonic decays of $\Omega_{ccc}$, $\Omega_{bbb}$, $\Omega_{ccb}$ and $\Omega_{cbb}$ will be studied  in order.  In Sec.~\ref{eq:goldenmodes}, we shall present a collection of golden modes which are most likely to discover the triply heavy baryons.  A short summary is given in the last section.

\section{Particle Multiplets}
\label{sec:particle_multiplet}

In this section, we will collect the representations for   hadron multiplets under  the flavor SU(3) group.
We start with the baryon sector.
Light baryons made of three light quarks can group into   an SU(3) octet and a decuplet. The light baryon decuplet  is  symmetric in  SU(3) flavor space and is characterized  by the following matrix
\begin{eqnarray}
(T_{10})^{111} &=&  \Delta^{++},\;\;\; (T_{10})^{112}= (T_{10})^{121}=(T_{10})^{211}= \frac{1}{\sqrt3} \Delta^+,\nonumber\\
(T_{10})^{222} &=&  \Delta^{-},\;\;\; (T_{10})^{122}= (T_{10})^{212}=(T_{10})^{221}= \frac{1}{\sqrt3} \Delta^0, \nonumber\\
(T_{10})^{113} &=& (T_{10})^{131}=(T_{10})^{311}= \frac{1}{\sqrt3} \Sigma^{\prime+},\;\;(T_{10})^{223} = (T_{10})^{232}=(T_{10})^{322}= \frac{1}{\sqrt3} \Sigma^{\prime-},\nonumber\\
(T_{10})^{123} &=& (T_{10})^{132}=(T_{10})^{213}=(T_{10})^{231}=(T_{10})^{312}=(T_{10})^{321}= \frac{1}{\sqrt6} \Sigma^{\prime0},\nonumber\\
(T_{10})^{133} &=& (T_{10})^{313}=(T_{10})^{331}= \frac{1}{\sqrt3} \Xi^{\prime0},\;\;(T_{10})^{233} = (T_{10})^{323}=(T_{10})^{332}= \frac{1}{\sqrt3}  \Xi^{\prime-}, \nonumber\\
(T_{10})^{333}&=& \Omega^-.
\end{eqnarray}
The octet has the expression:
\begin{eqnarray}
T_8= \left(\begin{array}{ccc} \frac{1}{\sqrt{2}}\Sigma^0+\frac{1}{\sqrt{6}}\Lambda^0 & \Sigma^+  &  p  \\ \Sigma^-  &  -\frac{1}{\sqrt{2}}\Sigma^0+\frac{1}{\sqrt{6}}\Lambda^0 & n \\ \Xi^-   & \Xi^0  & -\sqrt{\frac{2}{3}}\Lambda^0
  \end{array} \right).
\end{eqnarray}

The  triply heavy baryons form an SU(3) singlet, while
doubly heavy baryons  are an SU(3) triplet:
\begin{eqnarray}
 T_{cc}  = \left(\begin{array}{c}  \Xi^{++}_{cc}(ccu)  \\  \Xi^+_{cc}(ccd)  \\  \Omega^+_{cc}(ccs)
\end{array}\right)\,,\;\;
  T_{bc} = \left(\begin{array}{c}  \Xi^+_{bc}(bcu)  \\  \Xi^0_{bc}(bcd)  \\  \Omega^0_{bc}(bcs)
\end{array}\right)\,,\;\;
  T_{bb} = \left(\begin{array}{c}  \Xi^0_{bb}(bbu)  \\  \Xi^-_{bb}(bbd)  \\  \Omega^-_{bb}(bbs)
\end{array}\right).
\end{eqnarray}
Singly charmed and bottom baryons with two light quarks  can form an anti-triplet or sextet. For the anti-triplet and sextet, we have the matrix expression:
\begin{eqnarray}
 T_{\bf{c\bar 3}}= \left(\begin{array}{ccc} 0 & \Lambda_c^+  &  \Xi_c^+  \\ -\Lambda_c^+ & 0 & \Xi_c^0 \\ -\Xi_c^+   &  -\Xi_c^0  & 0
  \end{array} \right),\;\;
 T_{\bf{c6}} = \left(\begin{array}{ccc} \Sigma_c^{++} &  \frac{1}{\sqrt{2}}\Sigma_c^+   & \frac{1}{\sqrt{2}} \Xi_c^{\prime+}\\
  \frac{1}{\sqrt{2}}\Sigma_c^+& \Sigma_c^{0} & \frac{1}{\sqrt{2}} \Xi_c^{\prime0} \\
  \frac{1}{\sqrt{2}} \Xi_c^{\prime+}   &  \frac{1}{\sqrt{2}} \Xi_c^{\prime0}  & \Omega_c^0
  \end{array} \right)\,.
\end{eqnarray}

In the meson sector, a light pseudo-scalar meson is formed by a light quark and one light antiquark. It forms an octet:
\begin{eqnarray}
 M_{8}=\begin{pmatrix}
 \frac{\pi^0}{\sqrt{2}}+\frac{\eta}{\sqrt{6}}   &\pi^+ & K^+\\
 \pi^-&-\frac{\pi^0}{\sqrt{2}}+\frac{\eta}{\sqrt{6}}&{K^0}\\
 K^-&\overline K^0 &-\frac{2\eta}{\sqrt{6}}
 \end{pmatrix},
\end{eqnarray}
while the singlet $\eta_1$ is not considered.
Our   analysis is also applicable to  light vector mesons and other light mesons.
The  charmed meson forms an SU(3) anti-triplet:
\begin{eqnarray}
D_i=\left(\begin{array}{ccc} D^0, & D^+, & D^+_s  \end{array} \right),
\end{eqnarray}
and the anti-charmed meson forms an SU(3) triplet:
\begin{eqnarray}
\overline D^i=\left(\begin{array}{ccc}\overline D^0, & D^-, & D^-_s  \end{array} \right).
\end{eqnarray}
The above two SU(3) triplets are also applicable to the bottom mesons.

In the following we will construct the hadron-level effective Hamiltonian for various decay modes. It is necessary to point out that a hadron in the final state must be created by its anti-particle field. For instance, in order to produce a $\Xi_{ccu}^{++}$, one must use the $\overline \Xi_{ccu}^{--}$ in the Hamiltonian, and the   doubly heavy baryon anti-triplet is abbreviated  as $\overline T_{cc}$.

\section{Semi-Leptonic decay channels}
\label{sec:semileptonic}

\subsection{Semileptonic $\Omega_{ccc}$ decays}

\begin{table}
\caption{Amplitudes for semileptonic $ \Omega_{ccc}$  decays derived  from Eq.~\eqref{eq:ccc_semi}.  The light meson in the final state can   be replaced by its vector counterpart.  For instance, the $K^0$ can be replaced by a $K^{*0}$ decaying into $K^-\pi^+$.   }\label{tab:ccc_semi}
\begin{tabular}{cc|cc ccc|c}\hline\hline
channel & amplitude &channel & amplitude  \\\hline
$\Omega_{ccc}^{++}\to \Xi_{cc}^{+}\ell^+\nu_{\ell} $ & $ a_1 V_{cd}^*$ & $\Omega_{ccc}^{++}\to \Lambda_c^+  D^0 \ell^+\nu_{\ell} $ & $ a_3 V_{cd}^*$\\\hline
$\Omega_{ccc}^{++}\to \Omega_{cc}^{+}\ell^+\nu_{\ell} $ & $ a_1 V_{cs}^*$ & $\Omega_{ccc}^{++}\to \Xi_c^+  D^0 \ell^+\nu_{\ell} $ & $ a_3 V_{cs}^*$\\\hline
\Xcline{1-2}{1.2pt}
$\Omega_{ccc}^{++}\to \Xi_{cc}^{++}  \pi^-  \ell^+\nu_{\ell} $ & $ a_2 V_{cd}^*$ & $\Omega_{ccc}^{++}\to \Xi_c^0  D^+ \ell^+\nu_{\ell} $ & $ a_3 V_{cs}^*$\\\hline
$\Omega_{ccc}^{++}\to \Xi_{cc}^{++}  K^-  \ell^+\nu_{\ell} $ & $ a_2 V_{cs}^*$ & $\Omega_{ccc}^{++}\to \Xi_c^0  D^+_s \ell^+\nu_{\ell} $ & $ -a_3 V_{cd}^*$\\\hline
\Xcline{3-4}{1.2pt}
 $\Omega_{ccc}^{++}\to \Xi_{cc}^{+}  \pi^0  \ell^+\nu_{\ell} $ & $ -\frac{a_2 V_{cd}^*}{\sqrt{2}}$   & $\Omega_{ccc}^{++}\to \Sigma_{c}^{+}  D^0 \ell^+\nu_{\ell} $ & $ \frac{a_4 V_{cd}^*}{\sqrt{2}}$\\\hline
 $\Omega_{ccc}^{++}\to \Xi_{cc}^{+}  \overline K^0  \ell^+\nu_{\ell} $ & $ a_2 V_{cs}^*$   & $\Omega_{ccc}^{++}\to \Sigma_{c}^{0}  D^+ \ell^+\nu_{\ell} $ & $ a_4 V_{cd}^*$\\\hline
 $\Omega_{ccc}^{++}\to \Xi_{cc}^{+}  \eta  \ell^+\nu_{\ell} $ & $ \frac{a_2 V_{cd}^*}{\sqrt{6}}$    & $\Omega_{ccc}^{++}\to \Xi_{c}^{\prime+}  D^0 \ell^+\nu_{\ell} $ & $ \frac{a_4 V_{cs}^*}{\sqrt{2}}$\\\hline
 $\Omega_{ccc}^{++}\to \Omega_{cc}^{+}  K^0  \ell^+\nu_{\ell} $ & $ a_2 V_{cd}^*$   & $\Omega_{ccc}^{++}\to \Xi_{c}^{\prime0}  D^+ \ell^+\nu_{\ell} $ & $ \frac{a_4 V_{cs}^*}{\sqrt{2}}$\\\hline
 $\Omega_{ccc}^{++}\to \Omega_{cc}^{+}  \eta  \ell^+\nu_{\ell} $ & $ -\sqrt{\frac{2}{3}} a_2 V_{cs}^*$   & $\Omega_{ccc}^{++}\to \Xi_{c}^{\prime0}  D^+_s \ell^+\nu_{\ell} $ & $ \frac{a_4 V_{cd}^*}{\sqrt{2}}$\\\hline
   & & $\Omega_{ccc}^{++}\to \Omega_{c}^{0}  D^+_s \ell^+\nu_{\ell} $ & $ a_4 V_{cs}^*$\\\hline
\hline
\end{tabular}
\end{table}

The $c\to q\bar\ell\nu$ transition is induced by the effective electro-weak Hamiltonian:
\begin{eqnarray}
{\cal H}_{e.w.}&=&\frac{G_F}{\sqrt2} \left[V_{cq}^* \bar q  \gamma^\mu(1-\gamma_5)c \bar \nu_{\ell}\gamma_\mu(1-\gamma_5) \ell\right] +h.c.,
\end{eqnarray}
where $q=d,s$ and  $\ell=e,\mu$.  The $V_{cd}$ and  $V_{cs}$    are   CKM matrix elements. Heavy-to-light quark operators
are  an SU(3) triplet,  denoted as $H_{  3}$ with the nonzero components $(H_{  3})^2=V_{cd}^*,~(H_{  3})^3=V_{cs}^*$.
At   hadron level, the effective Hamiltonian  for  three-body and four-body semileptonic $\Omega_{ccc}$   decays   can be constructed as:
\begin{eqnarray}
\label{eq:ccc_semi}
  {\cal H}_{\rm{eff}}&=& a_1 \Omega_{ccc} (\overline T_{cc})_i (H_{  3})^i ~\bar\nu_\ell \ell + a_2 \Omega_{ccc} (\overline T_{cc})_i (M_8)^i_j (H_{3})^j ~\bar\nu_\ell \ell \nonumber\\
 && +a_3\Omega_{ccc} (\overline T_{c\bar 3})_{[ij]} \overline D^i (H_{3})^j ~\bar\nu_\ell \ell +a_4 \Omega_{ccc}  (\overline T_{c6})_{\{ij\}}\overline D^i (H_{  3})^j ~\bar\nu_\ell \ell.
\end{eqnarray}
where the $a_i$s  are   SU(3) irreducible amplitudes. Decay amplitudes for different channels can be deduced from the Hamiltonian in Eq.~\eqref{eq:ccc_semi}, and collected in Tab.~\ref{tab:ccc_semi}. A few remarks are given in order.
\begin{itemize}
\item
In this table and following ones, the light pseudoscalar mesons can be replaced by their light counterparts.  For instance  the $K^0$ can   be replaced by a $K^{*0}$, which is reconstructed by the $K^-\pi^+$ final state.  The $\eta$ meson is difficult to reconstruct at hadron colliders, while the vector $\phi$ meson can be reconstructed in the $K^+K^-$ final state with a high efficiency.

\item The $c\to s$ transition is proportional to the $V_{cs}\sim 1$, and  the $c\to d$ transition  has a smaller CKM matrix element $V_{cd}\sim 0.2$.  Inspired by the experimental  data on $D$ meson decays~\cite{Patrignani:2016xqp}, we can infer that  branching fractions  for the $c\to s$ channels are about  a few percents, and the ones for the $c\to d$ transitions are at the order $10^{-3}$.

\item A number of relations for decay widths can be easily read off from Tab.~\ref{tab:ccc_semi}.  For instance for  the $c\to s$ decays into a doubly charmed baryons, we have
\begin{eqnarray}
\Gamma( \Omega_{ccc}^{++}\to \Xi_{cc}^{++}  K^-  \ell^+\nu_{\ell}) =
\Gamma( \Omega_{ccc}^{++}\to \Xi_{cc}^{+}  \overline K^0  \ell^+\nu_{\ell}) = \frac{3}{2}\Gamma( \Omega_{ccc}^{++}\to \Omega_{cc}^{+}  \eta \ell^+\nu_{\ell}).
\end{eqnarray}
However it is necessary to point out that the above relations will be modified due to the different masses of the final hadrons. Once the mass of $\Omega_{ccc}$ is experimentally measured in future,   kinematical corrections can be included, and these relations can be refined.

\end{itemize}


\subsection{Semileptonic $\Omega_{bbb}$ decays}
 \begin{table}
\caption{Decay amplitudes for  $\Omega_{bbb}$ semileptonic decays with one or two hadrons in the final state.   }\label{tab:bbb_semi}
\begin{tabular}{ cc|cc|c c c|c}\hline\hline
channel & amplitude &channel & amplitude&channel & amplitude \\\hline
$\Omega_{bbb}^{-}\to \Omega_{cbb}^{0}\ell^-\bar\nu_{\ell} $ & $ b_1 V_{{cb}}$ &$\Omega_{bbb}^{-}\to \Xi_{bb}^{0}\ell^-\bar\nu_{\ell} $ & $ b_4 V_{{ub}}$\\\hline
$\Omega_{bbb}^{-}\to \Xi_{bb}^{0}  D^0 \ell^-\bar\nu_{\ell} $ & $ b_2 V_{{cb}}$ & $\Omega_{bbb}^{-}\to \Xi_{bb}^{-}  D^+ \ell^-\bar\nu_{\ell} $ & $ b_2 V_{{cb}}$&$\Omega_{bbb}^{-}\to \Omega_{bb}^{-}  D^+_s \ell^-\bar\nu_{\ell} $ & $ b_2 V_{{cb}}$ \\\hline
$\Omega_{bbb}^{-}\to \Xi_{bc}^{0}  \overline B^0 \ell^-\bar\nu_{\ell} $ & $ b_3 V_{{cb}}$ &
$\Omega_{bbb}^{-}\to \Omega_{bc}^{0}  \overline B^0_s \ell^-\bar\nu_{\ell} $ & $ b_3 V_{{cb}}$ & $\Omega_{bbb}^{-}\to \Xi_{bc}^{+}  B^- \ell^-\bar\nu_{\ell} $ & $ b_3 V_{{cb}}$  \\\hline
 $\Omega_{bbb}^{-}\to \Xi_{bb}^{0}  \pi^0  \ell^-\bar\nu_{\ell} $ & $ \frac{b_5 V_{{ub}}}{\sqrt{2}}$ & $\Omega_{bbb}^{-}\to \Xi_{bb}^{0}  \eta  \ell^-\bar\nu_{\ell} $ & $ \frac{b_5 V_{{ub}}}{\sqrt{6}}$ &
  $\Omega_{bbb}^{-}\to \Xi_{bb}^{-}  \pi^+  \ell^-\bar\nu_{\ell} $ & $ b_5 V_{{ub}}$\\\hline
  $\Omega_{bbb}^{-}\to \Omega_{bb}^{-}  K^+  \ell^-\bar\nu_{\ell} $ & $ b_5 V_{{ub}}$  &   $\Omega_{bbb}^{-}\to \Lambda_b^0  \overline B^0 \ell^-\bar\nu_{\ell} $ & $ -b_6 V_{{ub}}$ &
 $\Omega_{bbb}^{-}\to \Xi_b^0  \overline B^0_s \ell^-\bar\nu_{\ell} $ & $ -b_6 V_{{ub}}$\\\hline
$\Omega_{bbb}^{-}\to \Sigma_{b}^{+}  B^- \ell^-\bar\nu_{\ell} $ & $ b_7 V_{{ub}}$ & $\Omega_{bbb}^{-}\to \Sigma_{b}^{0}  \overline B^0 \ell^-\bar\nu_{\ell} $ & $ \frac{b_7 V_{{ub}}}{\sqrt{2}}$& $\Omega_{bbb}^{-}\to \Xi_{b}^{\prime0}  \overline B^0_s \ell^-\bar\nu_{\ell} $ & $ \frac{b_7 V_{{ub}}}{\sqrt{2}}$\\\hline
\hline
\end{tabular}
\end{table}

The $b$ quark decay is controlled by the electro-weak  Hamiltonian
\begin{eqnarray}
 {\cal H}_{e.e.} &=& \frac{G_F}{\sqrt2} \left[V_{q'b} \bar q' \gamma^\mu(1-\gamma_5)b \bar  \ell\gamma_\mu(1-\gamma_5) \nu_\ell\right] +h.c.,
\end{eqnarray}
with $q'=u,c$, and here $\ell=e,\mu,\tau$.  The $b\to c$ transition is an SU(3) singlet and thus the transition is simply a singlet. The $b\to u$ transition forms an SU(3) triplet  $H_{3}'$ with $(H_3')^1=1$ and $(H_3')^{2,3}=0$.   The hadron level  Hamiltonian is given as
\begin{eqnarray}
 {\cal H}_{\rm{eff}}&=&b_1\Omega_{bbb}\overline\Omega_{cbb}~\bar \ell \nu_\ell+b_2\Omega_{bbb}(\overline T_{bb})_i \overline D^i~\bar \ell \nu_\ell+b_3\Omega_{bbb}(\overline T_{bc})_i\overline B^i~\bar \ell \nu_\ell \nonumber\\
   &&+b_4 \Omega_{bbb}(\overline T_{bb})_i (H_{  3}')^i ~\bar \ell \nu_\ell +b_5\Omega_{bbb} (\overline T_{bb})_i  (M_8)^i_j (H_{3}')^j ~\bar  \ell \nu_l \nonumber\\
  &&+ b_6 \Omega_{bbb}   (\overline T_{b\bar 3})_{[ij]}\overline B^i(H_{3}')^j ~\bar  \ell \nu_\ell+ b_7 \Omega_{bbb}  (\overline T_{b6})_{\{ij\}} \overline B^i(H_{3}')^j ~\bar  \ell \nu_\ell.
\end{eqnarray}
The $b_i$s are the SU(3) independent amplitudes like the  $a_i$s  in Eq.~\eqref{eq:ccc_semi}.
The decay amplitudes can be deduced from this Hamiltonian, and the results are given in Tab.~\ref{tab:bbb_semi}.

A few remarks are given in order.
\begin{itemize}
\item The $\Omega_{cbb}$ in the final state can be $1/2^+$ or $3/2^+$.  It is similar for the $T_{bb}$, $T_{bc}$ and others.
\item The $b\to c$ transition has a larger  CKM matrix element $V_{cb}\sim 0.04$, and the typical branching fractions might reach  the order  $10^{-3}$ to $10^{-2}$.  However such decay modes still contain a triply heavy baryon which  must be detected through its   subsequent weak decays.

\item The $b\to u$ transition is suppressed due to $V_{ub}\sim 10^{-3}$. Typical branching fractions are at the order $10^{-4}$.

\end{itemize}


\subsection{Semileptonic $\Omega_{ccb}$ decays }

 \begin{table}
\caption{Decay amplitudes for $ \Omega_{ccb}^+$ semileptonic decays. }\label{tab:ccb_semi}
\begin{tabular}{|cc|cc|c|c|c|c}\hline\hline
channel & amplitude &channel & amplitude \\\hline
$\Omega_{ccb}^{+}\to \Omega_{ccc}^{++}\ell^-\bar\nu_{\ell} $ & $ c_7 V_{{cb}}$ & $\Omega_{ccb}^{+}\to \Sigma_{b}^{0}  D^0 \ell^+\nu_{\ell} $ & $ \frac{c_5 V_{cd}^*}{\sqrt{2}}$\\\hline
\Xcline{1-2}{1.2pt}
$\Omega_{ccb}^{+}\to \Xi_{cc}^{++}\ell^-\bar\nu_{\ell} $ & $ c_9 V_{{ub}}$ & $\Omega_{ccb}^{+}\to \Sigma_{b}^{-}  D^+ \ell^+\nu_{\ell} $ & $ c_5 V_{cd}^*$\\\hline
\Xcline{1-2}{1.2pt}
$\Omega_{ccb}^{+}\to \Xi_{bc}^{0}\ell^+\nu_{\ell} $ & $ c_1 V_{cd}^*$ & $\Omega_{ccb}^{+}\to \Xi_{b}^{\prime0}  D^0 \ell^+\nu_{\ell} $ & $ \frac{c_5 V_{cs}^*}{\sqrt{2}}$\\\hline
$\Omega_{ccb}^{+}\to \Omega_{bc}^{0}\ell^+\nu_{\ell} $ & $ c_1 V_{cs}^*$ & $\Omega_{ccb}^{+}\to \Xi_{b}^{\prime-}  D^+ \ell^+\nu_{\ell} $ & $ \frac{c_5 V_{cs}^*}{\sqrt{2}}$\\\hline
\Xcline{1-2}{1.2pt}
$\Omega_{ccb}^{+}\to \Xi_{bc}^{+}  \pi^-  \ell^+\nu_{\ell} $ & $ c_2 V_{cd}^*$ & $\Omega_{ccb}^{+}\to \Xi_{b}^{\prime-}  D^+_s \ell^+\nu_{\ell} $ & $ \frac{c_5 V_{cd}^*}{\sqrt{2}}$\\\hline
$\Omega_{ccb}^{+}\to \Xi_{bc}^{+}  K^-  \ell^+\nu_{\ell} $ & $ c_2 V_{cs}^*$ & $\Omega_{ccb}^{+}\to \Omega_{b}^{-}  D^+_s \ell^+\nu_{\ell} $ & $ c_5 V_{cs}^*$\\\hline
\Xcline{3-4}{1.2pt}
$\Omega_{ccb}^{+}\to \Xi_{bc}^{0}  \pi^0  \ell^+\nu_{\ell} $ & $ -\frac{c_2 V_{cd}^*}{\sqrt{2}}$ & $\Omega_{ccb}^{+}\to \Sigma_{c}^{+}  B^- \ell^+\nu_{\ell} $ & $ \frac{c_6 V_{cd}^*}{\sqrt{2}}$\\\hline
$\Omega_{ccb}^{+}\to \Xi_{bc}^{0}  \overline K^0  \ell^+\nu_{\ell} $ & $ c_2 V_{cs}^*$ & $\Omega_{ccb}^{+}\to \Sigma_{c}^{0}  \overline B^0 \ell^+\nu_{\ell} $ & $ c_6 V_{cd}^*$\\\hline
$\Omega_{ccb}^{+}\to \Xi_{bc}^{0}  \eta  \ell^+\nu_{\ell} $ & $ \frac{c_2 V_{cd}^*}{\sqrt{6}}$ & $\Omega_{ccb}^{+}\to \Xi_{c}^{\prime+}  B^- \ell^+\nu_{\ell} $ & $ \frac{c_6 V_{cs}^*}{\sqrt{2}}$\\\hline
$\Omega_{ccb}^{+}\to \Omega_{bc}^{0}  K^0  \ell^+\nu_{\ell} $ & $ c_2 V_{cd}^*$ & $\Omega_{ccb}^{+}\to \Xi_{c}^{\prime0}  \overline B^0 \ell^+\nu_{\ell} $ & $ \frac{c_6 V_{cs}^*}{\sqrt{2}}$\\\hline
$\Omega_{ccb}^{+}\to \Omega_{bc}^{0}  \eta  \ell^+\nu_{\ell} $ & $ -\sqrt{\frac{2}{3}} c_2 V_{cs}^*$ & $\Omega_{ccb}^{+}\to \Xi_{c}^{\prime0}  \overline B^0_s \ell^+\nu_{\ell} $ & $ \frac{c_6 V_{cd}^*}{\sqrt{2}}$\\\hline
\Xcline{1-2}{1.2pt}
$\Omega_{ccb}^{+}\to \Lambda_b^0  D^0 \ell^+\nu_{\ell} $ & $ c_3 V_{cd}^*$ & $\Omega_{ccb}^{+}\to \Omega_{c}^{0}  \overline B^0_s \ell^+\nu_{\ell} $ & $ c_6 V_{cs}^*$\\\hline
\Xcline{3-4}{1.2pt}
$\Omega_{ccb}^{+}\to \Xi_b^0  D^0 \ell^+\nu_{\ell} $ & $ c_3 V_{cs}^*$ & $\Omega_{ccb}^{+}\to \Xi_{cc}^{++}  D^0 \ell^-\bar\nu_{\ell} $ & $ c_8 V_{{cb}}$\\\hline
$\Omega_{ccb}^{+}\to \Xi_b^-  D^+ \ell^+\nu_{\ell} $ & $ c_3 V_{cs}^*$ & $\Omega_{ccb}^{+}\to \Xi_{cc}^{+}  D^+ \ell^-\bar\nu_{\ell} $ & $ c_8 V_{{cb}}$\\\hline
$\Omega_{ccb}^{+}\to \Xi_b^-  D^+_s \ell^+\nu_{\ell} $ & $ -c_3 V_{cd}^*$ & $\Omega_{ccb}^{+}\to \Omega_{cc}^{+}  D^+_s \ell^-\bar\nu_{\ell} $ & $ c_8 V_{{cb}}$\\\hline
\Xcline{1-4}{1.2pt}
$\Omega_{ccb}^{+}\to \Lambda_c^+  B^- \ell^+\nu_{\ell} $ & $ c_4 V_{cd}^*$ & $\Omega_{ccb}^{+}\to \Xi_{cc}^{++}  \pi^0  \ell^-\bar\nu_{\ell} $ & $ \frac{c_{10} V_{{ub}}}{\sqrt{2}}$\\\hline
$\Omega_{ccb}^{+}\to \Xi_c^+  B^- \ell^+\nu_{\ell} $ & $ c_4 V_{cs}^*$ & $\Omega_{ccb}^{+}\to \Xi_{cc}^{++}  \eta  \ell^-\bar\nu_{\ell} $ & $ \frac{c_{10} V_{{ub}}}{\sqrt{6}}$\\\hline
$\Omega_{ccb}^{+}\to \Xi_c^0  \overline B^0 \ell^+\nu_{\ell} $ & $ c_4 V_{cs}^*$ & $\Omega_{ccb}^{+}\to \Xi_{cc}^{+}  \pi^+  \ell^-\bar\nu_{\ell} $ & $ c_{10} V_{{ub}}$\\\hline
$\Omega_{ccb}^{+}\to \Xi_c^0  \overline B^0_s \ell^+\nu_{\ell} $ & $ -c_4 V_{cd}^*$ & $\Omega_{ccb}^{+}\to \Omega_{cc}^{+}  K^+  \ell^-\bar\nu_{\ell} $ & $ c_{10} V_{{ub}}$\\\hline
\Xcline{1-4}{1.2pt}
$\Omega_{ccb}^{+}\to \Sigma_{c}^{++}  D^0 \ell^-\bar\nu_{\ell} $ & $ c_{12} V_{{ub}}$ & $\Omega_{ccb}^{+}\to \Lambda_c^+  D^+ \ell^-\bar\nu_{\ell} $ & $ -c_{11} V_{{ub}}$\\\hline
$\Omega_{ccb}^{+}\to \Sigma_{c}^{+}  D^+ \ell^-\bar\nu_{\ell} $ & $ \frac{c_{12} V_{{ub}}}{\sqrt{2}}$ & $\Omega_{ccb}^{+}\to \Xi_c^+  D^+_s \ell^-\bar\nu_{\ell} $ & $ -c_{11} V_{{ub}}$\\\hline
$\Omega_{ccb}^{+}\to \Xi_{c}^{\prime+}  D^+_s \ell^-\bar\nu_{\ell} $ & $ \frac{c_{12} V_{{ub}}}{\sqrt{2}}$ && \\\hline
\hline
\end{tabular}
\end{table}

Both charm quark and bottom quark in $\Omega_{ccb}$ can weakly decay. Thus
the hadron level  Hamiltonian for semileptonic $\Omega_{ccb}$ decays is given as
\begin{eqnarray}
  {\cal H}_{\rm{eff}}&=& c_1 \Omega_{ccb} (\overline T_{bc})_i (H_{3})^i ~\bar\nu_\ell \ell + c_2 \Omega_{ccb} (\overline T_{bc})_i (M_8)^i_j (H_{3})^j ~\bar\nu_\ell \ell \nonumber\\
 && +  c_3\Omega_{ccb} (\overline T_{b\bar 3})_{[ij]} \overline D^i (H_{3})^j ~\bar\nu_\ell \ell+ c_4\Omega_{ccb} (\overline T_{c\bar 3})_{[ij]} \overline B^i (H_{3})^j ~\bar\nu_\ell \ell \nonumber\\
 &&+ c_5\Omega_{ccb} (\overline T_{b6})_{\{ij\}} \overline D^i (H_{3})^j ~\bar\nu_\ell \ell+ c_6\Omega_{ccb} (\overline T_{c6})_{\{ij\}} \overline B^i (H_{3})^j ~\bar\nu_\ell \ell\nonumber\\
 && +c_7 \Omega_{ccb}\overline  \Omega_{ccc}~\bar \ell \nu_\ell +c_8\Omega_{ccb}(\overline T_{cc})_i\overline D^i~\bar \ell \nu_\ell  \nonumber\\
  &&+c_{9} \Omega_{ccb}(\overline T_{cc})_i (H_{3}')^i ~\bar \ell \nu_\ell +c_{10}\Omega_{ccb} (\overline T_{cc})_i  (M_8)^i_j (H_{3}')^j ~\bar  \ell \nu_l \nonumber\\
  &&+ c_{11} \Omega_{ccb}   (\overline T_{c\bar 3})_{[ij]} \overline D^i (H_{3}')^j ~\bar  \ell \nu_\ell+ c_{12} \Omega_{ccb}   (\overline T_{c6})_{\{ij\}} \overline D^i (H_{3}')^j ~\bar  \ell \nu_\ell.
\end{eqnarray}
Decay amplitudes for different channels are obtained by expanding the above Hamiltonian and  are collected in Tab.~\ref{tab:ccb_semi}.


\subsection{Semileptonic $\Omega_{cbb}$ decays }

 \begin{table}
\caption{Amplitudes for $\Omega_{cbb}$ semileptonic decay}\label{tab:cbb_semi}
\begin{tabular}{|cc|cc|c|c|c|c}\hline\hline
channel & amplitude &channel & amplitude \\\hline
$\Omega_{cbb}^{0}\to \Omega_{ccb}^{+}\ell^-\bar\nu_{\ell} $ & $ d_1 V_{{cb}}$ & $\Omega_{cbb}^{0}\to \Xi_{cc}^{++}  B^- \ell^-\bar\nu_{\ell} $ & $ d_2 V_{{cb}}$\\\hline
\Xcline{1-2}{1.2pt}
$\Omega_{cbb}^{0}\to \Xi_{bc}^{+}\ell^-\bar\nu_{\ell} $ & $ d_4 V_{{ub}}$ & $\Omega_{cbb}^{0}\to \Xi_{cc}^{+}  \overline B^0 \ell^-\bar\nu_{\ell} $ & $ d_2 V_{{cb}}$\\\hline
\Xcline{1-2}{1.2pt}
$\Omega_{cbb}^{0}\to \Xi_{bb}^{-}\ell^+\nu_{\ell} $ & $ d_{10} V_{cd}^*$ & $\Omega_{cbb}^{0}\to \Omega_{cc}^{+}  \overline B^0_s \ell^-\bar\nu_{\ell} $ & $ d_2 V_{{cb}}$\\\hline
\Xcline{3-4}{1.2pt}
$\Omega_{cbb}^{0}\to \Omega_{bb}^{-}\ell^+\nu_{\ell} $ & $ d_{10} V_{cs}^*$ & $\Omega_{cbb}^{0}\to \Xi_{bc}^{+}  D^0 \ell^-\bar\nu_{\ell} $ & $ d_3 V_{{cb}}$\\\hline
\Xcline{1-2}{1.2pt}
$\Omega_{cbb}^{0}\to \Xi_{bb}^{0}  \pi^-  \ell^+\nu_{\ell} $ & $ d_{11} V_{cd}^*$ & $\Omega_{cbb}^{0}\to \Xi_{bc}^{0}  D^+ \ell^-\bar\nu_{\ell} $ & $ d_3 V_{{cb}}$\\\hline
$\Omega_{cbb}^{0}\to \Xi_{bb}^{0}  K^-  \ell^+\nu_{\ell} $ & $ d_{11} V_{cs}^*$ & $\Omega_{cbb}^{0}\to \Omega_{bc}^{0}  D^+_s \ell^-\bar\nu_{\ell} $ & $ d_3 V_{{cb}}$\\\hline
\Xcline{3-4}{1.2pt}
$\Omega_{cbb}^{0}\to \Xi_{bb}^{-}  \pi^0  \ell^+\nu_{\ell} $ & $ -\frac{d_{11} V_{cd}^*}{\sqrt{2}}$ & $\Omega_{cbb}^{0}\to \Xi_{bc}^{+}  \pi^0  \ell^-\bar\nu_{\ell} $ & $ \frac{d_5 V_{{ub}}}{\sqrt{2}}$\\\hline
$\Omega_{cbb}^{0}\to \Xi_{bb}^{-}  \overline K^0  \ell^+\nu_{\ell} $ & $ d_{11} V_{cs}^*$ & $\Omega_{cbb}^{0}\to \Xi_{bc}^{+}  \eta  \ell^-\bar\nu_{\ell} $ & $ \frac{d_5 V_{{ub}}}{\sqrt{6}}$\\\hline
$\Omega_{cbb}^{0}\to \Xi_{bb}^{-}  \eta  \ell^+\nu_{\ell} $ & $ \frac{d_{11} V_{cd}^*}{\sqrt{6}}$ & $\Omega_{cbb}^{0}\to \Xi_{bc}^{0}  \pi^+  \ell^-\bar\nu_{\ell} $ & $ d_5 V_{{ub}}$\\\hline
$\Omega_{cbb}^{0}\to \Omega_{bb}^{-}  K^0  \ell^+\nu_{\ell} $ & $ d_{11} V_{cd}^*$ & $\Omega_{cbb}^{0}\to \Omega_{bc}^{0}  K^+  \ell^-\bar\nu_{\ell} $ & $ d_5 V_{{ub}}$\\\hline
\Xcline{3-4}{1.2pt}
$\Omega_{cbb}^{0}\to \Omega_{bb}^{-}  \eta  \ell^+\nu_{\ell} $ & $ -\sqrt{\frac{2}{3}} d_{11} V_{cs}^*$ & $\Omega_{cbb}^{0}\to \Lambda_b^0  D^+ \ell^-\bar\nu_{\ell} $ & $ -d_6 V_{{ub}}$\\\hline
\Xcline{1-2}{1.2pt}
$\Omega_{cbb}^{0}\to \Lambda_b^0  B^- \ell^+\nu_{\ell} $ & $ d_{12} V_{cd}^*$ & $\Omega_{cbb}^{0}\to \Xi_b^0  D^+_s \ell^-\bar\nu_{\ell} $ & $ -d_6 V_{{ub}}$\\\hline
\Xcline{3-4}{1.2pt}
$\Omega_{cbb}^{0}\to \Xi_b^0  B^- \ell^+\nu_{\ell} $ & $ d_{12} V_{cs}^*$ & $\Omega_{cbb}^{0}\to \Lambda_c^+  \overline B^0 \ell^-\bar\nu_{\ell} $ & $ -d_7 V_{{ub}}$\\\hline
$\Omega_{cbb}^{0}\to \Xi_b^-  \overline B^0 \ell^+\nu_{\ell} $ & $ d_{12} V_{cs}^*$ & $\Omega_{cbb}^{0}\to \Xi_c^+  \overline B^0_s \ell^-\bar\nu_{\ell} $ & $ -d_7 V_{{ub}}$\\\hline
\Xcline{3-4}{1.2pt}
$\Omega_{cbb}^{0}\to \Xi_b^-  \overline B^0_s \ell^+\nu_{\ell} $ & $ -d_{12} V_{cd}^*$ & $\Omega_{cbb}^{0}\to \Sigma_{b}^{+}  D^0 \ell^-\bar\nu_{\ell} $ & $ d_8 V_{{ub}}$\\\hline
\Xcline{1-2}{1.2pt}
$\Omega_{cbb}^{0}\to \Sigma_{b}^{0}  B^- \ell^+\nu_{\ell} $ & $ \frac{d_{13} V_{cd}^*}{\sqrt{2}}$ & $\Omega_{cbb}^{0}\to \Sigma_{b}^{0}  D^+ \ell^-\bar\nu_{\ell} $ & $ \frac{d_8 V_{{ub}}}{\sqrt{2}}$\\\hline
$\Omega_{cbb}^{0}\to \Sigma_{b}^{-}  \overline B^0 \ell^+\nu_{\ell} $ & $ d_{13} V_{cd}^*$ & $\Omega_{cbb}^{0}\to \Xi_{b}^{\prime0}  D^+_s \ell^-\bar\nu_{\ell} $ & $ \frac{d_8 V_{{ub}}}{\sqrt{2}}$\\\hline
\Xcline{3-4}{1.2pt}
$\Omega_{cbb}^{0}\to \Xi_{b}^{\prime0}  B^- \ell^+\nu_{\ell} $ & $ \frac{d_{13} V_{cs}^*}{\sqrt{2}}$ & $\Omega_{cbb}^{0}\to \Sigma_{c}^{++}  B^- \ell^-\bar\nu_{\ell} $ & $ d_9 V_{{ub}}$\\\hline
$\Omega_{cbb}^{0}\to \Xi_{b}^{\prime-}  \overline B^0 \ell^+\nu_{\ell} $ & $ \frac{d_{13} V_{cs}^*}{\sqrt{2}}$ & $\Omega_{cbb}^{0}\to \Sigma_{c}^{+}  \overline B^0 \ell^-\bar\nu_{\ell} $ & $ \frac{d_9 V_{{ub}}}{\sqrt{2}}$\\\hline
$\Omega_{cbb}^{0}\to \Xi_{b}^{\prime-}  \overline B^0_s \ell^+\nu_{\ell} $ & $ \frac{d_{13} V_{cd}^*}{\sqrt{2}}$ & $\Omega_{cbb}^{0}\to \Xi_{c}^{\prime+}  \overline B^0_s \ell^-\bar\nu_{\ell} $ & $ \frac{d_9 V_{{ub}}}{\sqrt{2}}$\\\hline
$\Omega_{cbb}^{0}\to \Omega_{b}^{-}  \overline B^0_s \ell^+\nu_{\ell} $ & $ d_{13} V_{cs}^*$ && \\\hline
\hline
\end{tabular}
\end{table}

Similarly the hadron level  Hamiltonian for  semileptonic $\Omega_{cbb}$ decay is given as
\begin{eqnarray}
 {\cal H}_{\rm{eff}}&=&d_1\Omega_{cbb}\Omega_{ccb}~\bar \ell \nu_\ell+d_2\Omega_{cbb}(\overline T_{cc})_i \overline B^i~\bar \ell \nu_\ell+d_3\Omega_{cbb}(\overline T_{bc})_i\overline D^i~\bar \ell \nu_\ell \nonumber\\
   &&+d_4 \Omega_{cbb}(T_{bc})_i (H_{  3}')^i ~\bar \ell \nu_\ell +d_5\Omega_{cbb} (T_{bc})_i  M^i_j (H_{  3}')^j ~\bar  \ell \nu_l \nonumber\\
  &&+ d_6 \Omega_{cbb}  (\overline T_{b\bar 3})_{[ij]}  \overline D^i (H_{3}')^j ~\bar  \ell \nu_\ell+ d_7 \Omega_{cbb}   (\overline T_{c\bar 3})_{[ij]} \overline B^i (H_{3}')^j ~\bar  \ell \nu_\ell \nonumber\\
  && +d_8\Omega_{cbb}(\overline T_{b6})_{\{ij\}} \overline D^i (H_{3}')^j ~\bar  \ell \nu_\ell +d_{9}\Omega_{cbb} (\overline T_{c6})_{\{ij\}} \overline B^i (H_{3}')^j ~\bar  \ell \nu_\ell  \nonumber\\
   &&+d_{10} \Omega_{cbb} (T_{bb})_i (H_{  3})^i ~\bar\nu_\ell \ell + d_{11} \Omega_{cbb} (T_{bb})_i M^i_j (H_{3})^j ~\bar\nu_\ell \ell \nonumber\\
 && +d_{12}\Omega_{cbb} (\overline T_{b\bar 3})_{[ij]} \overline B^i (H_{3})^j ~\bar\nu_\ell \ell+  d_{13} \Omega_{cbb}  (\overline T_{b6})_{\{ij\}}\overline B^i (H_{3})^j ~\bar\nu_\ell \ell.
\end{eqnarray}
Expanding the above equations, we will obtain the decay amplitudes given in Tab.~\ref{tab:cbb_semi}.

\section{Non-Leptonic  $\Omega_{ccc}$ decays}
\label{sec:ccc_nonleptonic}

\begin{figure}
\includegraphics[width=0.8\columnwidth]{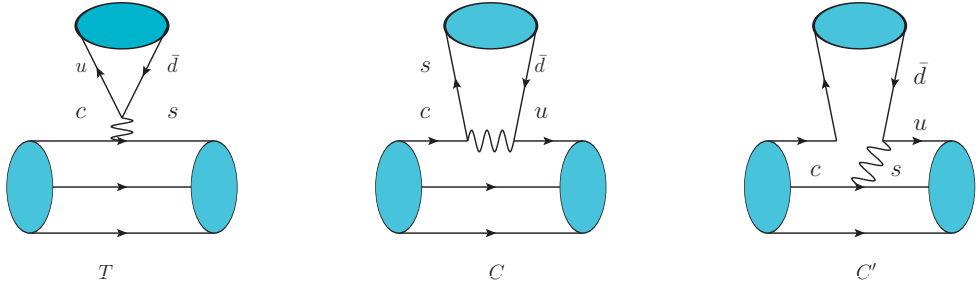}
\caption{ Feynman diagrams for  two-body decay modes induced Cabibbo-allowed transitions. In the first two panels, the final state contains  a doubly charmed baryon and a light meson, described by Eq.~\eqref{eq:ccc2ccqM8}.  The last panel corresponds to decays into  a charmed baryon and a charmed meson.   The first panel is color-allowed, while the last two panels are suppressed by $1/N_c$.  }
\label{fig:topology_2body}
\end{figure}

Nonleptonic charm quark decays into light quarks are classified   into three groups:
\begin{eqnarray}
 c\to s \bar d u,  \;\;\; c\to u \bar dd/\bar ss, \;\;\; c\to  d \bar s u.
\end{eqnarray}
Feynman diagrams for two-body decays induced by the $c\to s \bar d u$    are given in Fig.~\ref{fig:topology_2body}.   In the first two panels, the final state contains  a doubly charmed baryon and a light meson, while the last panel corresponds to decays into  a charmed baryon and a charmed meson.  The first panel is color-allowed, while the last two panels are suppressed by $1/N_c$.

Penguin contributions in charm quark decays are highly suppressed, and thus are neglected in our analysis.
Tree operators  transform
under the flavor SU(3) symmetry as ${\bf  3}\otimes {\bf\bar 3}\otimes {\bf
3}={\bf  3}\oplus {\bf  3}\oplus {\bf\bar 6}\oplus {\bf {15}}$. For charm quark decays, the vector representation $H_3$ will vanishes as an approximation.
For the $c\to s  u \bar d$ transition, we have
\begin{eqnarray}
(H_{\overline 6})^{31}_2=-(H_{\overline 6})^{13}_2=1,\;\;\;
 (H_{15})^{31}_2= (H_{15})^{13}_2=1,\label{eq:H3615_c_allowed}
\end{eqnarray}
while for the  doubly Cabibbo suppressed $c\to d  u \bar s$ transition, we have
\begin{eqnarray}
(H_{\overline 6})^{21}_3=-(H_{\overline 6})^{12}_3=\sin^2\theta_C,\;\;
 (H_{15})^{21}_3= (H_{15})^{12}_3=\sin^2\theta_C. \label{eq:H3615_c_doubly_suprressed}
\end{eqnarray}
CKM matrix elements for  $c\to u \bar dd$ and $c\to u \bar ss$ transitions are approximately equal in magnitude but different in sign. With both contributions,   one has the nonzero components:
\begin{eqnarray}
(H_{\overline 6})^{31}_3 =-(H_{\overline 6})^{13}_3 =(H_{\overline 6})^{12}_2 =-(H_{\overline 6})^{21}_2 =\sin(\theta_C),\nonumber\\
 (H_{15})^{31}_3= (H_{15})^{13}_3=-(H_{15})^{12}_2=-(H_{15})^{21}_2= \sin(\theta_C).\label{eq:H3615_cc_singly_suppressed}
\end{eqnarray}
In the following we will use $s_C$ to abbreviate the sine of Cabibbo angle $\theta$.

\subsection{Decays into a doubly-charmed baryon and one (two)   light meson(s)}

 \begin{table}
\caption{Amplitudes for $\Omega_{ccc}$ decays into doubly-charmed baryon and a light meson.  Cabibbo-allowed, singly Cabibbo-suppressed and doubly Cabibbo-suppressed decay channels are included in this table and the following ones. The  amplitudes $a_i$s are different with the ones in Tab.~\ref{tab:ccc_semi}.   }\label{tab:ccc_Atwobody}
\begin{tabular}{|cc|cc|c|c|c|c}\hline\hline
channel & amplitude &channel & amplitude \\\hline
$\Omega_{ccc}^{++}\to \Xi_{cc}^{++}  \overline K^0  $ & $ a_2-a_1$ & $\Omega_{ccc}^{++}\to \Xi_{cc}^{++}  \pi^0  $ & $ \frac{\left(a_2-a_1\right) \text{sC}}{\sqrt{2}}$\\\hline
$\Omega_{ccc}^{++}\to \Omega_{cc}^{+}  \pi^+  $ & $ a_1+a_2$ & $\Omega_{ccc}^{++}\to \Xi_{cc}^{++}  \eta  $ & $ \sqrt{\frac{3}{2}} \left(a_1-a_2\right) \text{sC}$\\\hline
\Xcline{1-2}{1.2pt}
$\Omega_{ccc}^{++}\to \Xi_{cc}^{++}  K^0  $ & $ \left(a_1-a_2\right) \left(-\text{sC}^2\right)$ & $\Omega_{ccc}^{++}\to \Xi_{cc}^{+}  \pi^+  $ & $ \left(a_1+a_2\right) (-\text{sC})$\\\hline
$\Omega_{ccc}^{++}\to \Xi_{cc}^{+}  K^+  $ & $ \left(a_1+a_2\right) \text{sC}^2$ & $\Omega_{ccc}^{++}\to \Omega_{cc}^{+}  K^+  $ & $ \left(a_1+a_2\right) \text{sC}$\\\hline
\hline
\end{tabular}
\end{table}

\begin{table}
\caption{Amplitudes for $\Omega_{ccc}$ decays into doubly-charmed baryon and two light mesons.  The amplitude $b_2$ defined in Eq.~\eqref{eq:ccc2ccq2M8} is not shown since it always accompanies  with $b_1$ in the form $b_1-b_2$.   }\label{tab:ccc_Athreebody}
\begin{tabular}{|cc|cc|c|c|c|c}\hline\hline
channel & amplitude &channel & amplitude \\\hline
$\Omega_{ccc}^{++}\to \Xi_{cc}^{++}  \overline K^0   \pi^0  $ & $ \frac{b_1 -b_3+b_4}{\sqrt{2}}$ & $\Omega_{ccc}^{++}\to \Xi_{cc}^{++}  \pi^0   \pi^0  $ & $ \left(b_1 -b_3+b_4\right) \text{sC}$\\\hline
$\Omega_{ccc}^{++}\to \Xi_{cc}^{++}  K^-   \pi^+  $ & $ -b_1 +b_3+b_4$ & $\Omega_{ccc}^{++}\to \Xi_{cc}^{++}  \pi^-   \pi^+  $ & $ \left(b_1 -b_3-b_4\right) \text{sC}$\\\hline
$\Omega_{ccc}^{++}\to \Xi_{cc}^{++}  \eta   \overline K^0  $ & $ \frac{b_1 -b_3+b_4}{\sqrt{6}}$ & $\Omega_{ccc}^{++}\to \Xi_{cc}^{++}  K^-   K^+  $ & $ \left(-b_1 +b_3+b_4\right) \text{sC}$\\\hline
$\Omega_{ccc}^{++}\to \Xi_{cc}^{+}  \pi^+   \overline K^0  $ & $ 2 b_4$ & $\Omega_{ccc}^{++}\to \Xi_{cc}^{++}  \eta   \pi^0  $ & $ -\frac{\left(b_1 -b_3+b_4\right) \text{sC}}{\sqrt{3}}$\\\hline
$\Omega_{ccc}^{++}\to \Omega_{cc}^{+}  \pi^+   \eta  $ & $ \sqrt{\frac{2}{3}} \left(b_1 +b_3-b_4\right)$ & $\Omega_{ccc}^{++}\to \Xi_{cc}^{++}  \eta   \eta  $ & $ -\left(b_1 -b_3+b_4\right) \text{sC}$\\\hline
$\Omega_{ccc}^{++}\to \Omega_{cc}^{+}  K^+   \overline K^0  $ & $ b_1 +b_3+b_4$ & $\Omega_{ccc}^{++}\to \Xi_{cc}^{+}  \pi^0   \pi^+  $ & $ \sqrt{2} b_4 \text{sC}$\\\hline
\Xcline{1-2}{1.2pt}
$\Omega_{ccc}^{++}\to \Xi_{cc}^{++}  K^+   \pi^-  $ & $ \left(b_1 -b_3-b_4\right) \left(-\text{sC}^2\right)$ & $\Omega_{ccc}^{++}\to \Xi_{cc}^{+}  \overline K^0   K^+  $ & $ \left(b_1 +b_3-b_4\right) (-\text{sC})$\\\hline
$\Omega_{ccc}^{++}\to \Xi_{cc}^{++}  K^0   \pi^0  $ & $ \frac{\left(b_1 -b_3+b_4\right) \text{sC}^2}{\sqrt{2}}$ & $\Omega_{ccc}^{++}\to \Xi_{cc}^{+}  \eta   \pi^+  $ & $ -\sqrt{\frac{2}{3}} \left(b_1 +b_3+2 b_4\right) \text{sC}$\\\hline
$\Omega_{ccc}^{++}\to \Xi_{cc}^{++}  \eta   K^0  $ & $ \frac{\left(b_1 -b_3+b_4\right) \text{sC}^2}{\sqrt{6}}$ & $\Omega_{ccc}^{++}\to \Omega_{cc}^{+}  \pi^+   K^0  $ & $ \left(b_1 +b_3-b_4\right) \text{sC}$\\\hline
$\Omega_{ccc}^{++}\to \Xi_{cc}^{+}  K^+   \pi^0  $ & $ \frac{\left(b_1 +b_3-b_4\right) \text{sC}^2}{\sqrt{2}}$ & $\Omega_{ccc}^{++}\to \Omega_{cc}^{+}  K^+   \pi^0  $ & $ \frac{\left(b_1 +b_3+b_4\right) \text{sC}}{\sqrt{2}}$\\\hline
$\Omega_{ccc}^{++}\to \Xi_{cc}^{+}  K^+   \eta  $ & $ -\frac{\left(b_1 +b_3-b_4\right) \text{sC}^2}{\sqrt{6}}$ & $\Omega_{ccc}^{++}\to \Omega_{cc}^{+}  K^+   \eta  $ & $ -\frac{\left(b_1 +b_3+5 b_4\right) \text{sC}}{\sqrt{6}}$\\\hline
$\Omega_{ccc}^{++}\to \Xi_{cc}^{+}  K^0   \pi^+  $ & $ \left(b_1 +b_3+b_4\right) \text{sC}^2$ & $\Omega_{ccc}^{++}\to \Omega_{cc}^{+}  K^0   \pi^+  $ & $ \left(b_1 +b_3-b_4\right) \text{sC}$\\\hline
$\Omega_{ccc}^{++}\to \Omega_{cc}^{+}  K^+   K^0  $ & $ 2 b_4 \text{sC}^2$ && \\\hline
\end{tabular}
\end{table}

For decays into a doubly-charmed baryon and a   light meson, one may derive the   effective Hamiltonian:
\begin{eqnarray}
 {\cal H}_{eff}&=&  a_1 \Omega_{ccc} (\overline T_{cc})_i     (M_8)^{k}_{j}  (H_{\overline6})^{ij}_{k}+ a_2\Omega_{ccc}(\overline T_{cc})_i    (M_8)^{k}_{j}  (H_{15})^{ij}_{k}, \label{eq:ccc2ccqM8}
\end{eqnarray}
whose Feynman diagrams are given in Fig.~\ref{fig:topology_2body}.  It is necessary to stress that the above SU(3) independent amplitudes $a_i$s are different with the ones in Eq.~\eqref{eq:ccc_semi}.

Expanding the above equations, we will obtain the decay amplitudes given in Tab.~\ref{tab:ccc_Atwobody}.  From Eq.~\eqref{eq:ccc2ccqM8} and Tab.~\ref{tab:ccc_Atwobody}, one can see that there are two SU(3) independent amplitudes, and thus there exists a few relations for decay widths. These relations can be directly  read off from Tab.~\ref{tab:ccc_Atwobody}, and can  be examined by future experiments.

For the reactions  with one additional light meson in the final state, one has the Hamiltonian:
\begin{eqnarray}
 {\cal H}_{eff}&=&  b_1 \Omega_{ccc} (\overline T_{cc})_i   (M_8)^{k}_{j} (M_8)^{j}_l  (H_{\overline6})^{il}_{k} + b_2 \Omega_{ccc} (\overline  T_{cc})_i (M_8)^{i}_{j} (M_8)^k_l (H_{\overline6})^{jl}_{k}  \nonumber\\
 && +  b_3 \Omega_{ccc} (\overline T_{cc})_i (M_8)^{k}_{j} (M_8)^{j}_l  (H_{15})^{il}_{k} + b_4 \Omega_{ccc} (\overline T_{cc})_i (M_8)^{i}_{j} (M_8)^k_l (H_{15})^{jl}_{k}. \label{eq:ccc2ccq2M8}
\end{eqnarray}
Expanding the above equation, we will obtain the decay amplitudes given in Tab.~\ref{tab:ccc_Athreebody}. The following  remarks are in order.
\begin{itemize}
\item From the expanded Hamiltonian, one can find  that the amplitudes $b_1$ and $b_2$ always appear in the combination $b_1-b_2$.  Thus we have removed the amplitude  $b_2$  in Tab.~\ref{tab:ccc_Athreebody}

\item For   channels with two identical particles, there is a factor $1/2$ in the   decay width.
\end{itemize}


\subsection{Decays into a charmed baryon and a charmed meson}


 \begin{table}
\caption{Amplitudes for $\Omega_{ccc}$ decays into a singly-charmed baryon and a charmed meson. $s_C=\sin\theta$. }\label{tab:ccc_Btwobody}
\begin{tabular}{|cc|cc|c|c|c|c}\hline\hline
channel & amplitude &channel & amplitude \\\hline
$\Omega_{ccc}^{++}\to \Xi_c^+  D^+ $ & $ -2 c_1$ & $\Omega_{ccc}^{++}\to \Lambda_c^+  D^+ $ & $ 2 c_1 \text{sC}$\\\hline
$\Omega_{ccc}^{++}\to \Xi_{c}^{\prime+}  D^+ $ & $ \sqrt{2} c_2$ & $\Omega_{ccc}^{++}\to \Xi_c^+  D^+_s $ & $ -2 c_1 \text{sC}$\\\hline
\Xcline{1-2}{1.2pt}
$\Omega_{ccc}^{++}\to \Sigma_{c}^{+}  D^+_s $ & $ \sqrt{2} c_2 \text{sC}^2$ & $\Omega_{ccc}^{++}\to \Sigma_{c}^{+}  D^+ $ & $ -\sqrt{2} c_2 \text{sC}$\\\hline
$\Omega_{ccc}^{++}\to \Lambda_c^+  D^+_s $ & $ -2 c_1 \text{sC}^2$ & $\Omega_{ccc}^{++}\to \Xi_{c}^{\prime+}  D^+_s $ & $ \sqrt{2} c_2 \text{sC}$\\\hline
\hline
\end{tabular}
\end{table}

\begin{table}
\caption{Amplitudes for three-body $\Omega_{ccc}$ decays into  a singly-charmed baryon (anti-triplet),  D meson and a light meson. }\label{tab:ccc_Bthreebody1}
\begin{tabular}{|cc|cc|c|c|c|c}\hline\hline
channel & amplitude &channel & amplitude \\\hline
$\Omega_{ccc}^{++}\to \Lambda_c^+  D^+  \overline K^0  $ & $ -d_2+d_3+d_4-d_5$ & $\Omega_{ccc}^{++}\to \Lambda_c^+  D^0  \pi^+  $ & $ \left(2 d_1+d_2+d_4\right) \text{sC}$\\\hline
$\Omega_{ccc}^{++}\to \Xi_c^+  D^0  \pi^+  $ & $ -2 d_1-d_2-d_4$ & $\Omega_{ccc}^{++}\to \Lambda_c^+  D^+  \pi^0  $ & $ -\frac{\left(2 d_1+d_2-d_4+2 d_5\right) \text{sC}}{\sqrt{2}}$\\\hline
$\Omega_{ccc}^{++}\to \Xi_c^+  D^+  \pi^0  $ & $ \frac{2 d_1+d_3+d_5}{\sqrt{2}}$ &$\Omega_{ccc}^{++}\to \Lambda_c^+  D^+  \eta  $ & $ \frac{\left(2 d_1+3 d_2-2 d_3-3 d_4\right) \text{sC}}{\sqrt{6}}$\\\hline
$\Omega_{ccc}^{++}\to \Xi_c^+  D^+  \eta  $ & $ -\frac{2 d_1+d_3-3 d_5}{\sqrt{6}}$¡¡&¡¡$\Omega_{ccc}^{++}\to \Lambda_c^+  D^+_s  \overline K^0  $ & $ \left(2 d_1+d_3-d_5\right) \text{sC}$\\\hline
$\Omega_{ccc}^{++}\to \Xi_c^+  D^+_s  \overline K^0  $ & $ -2 d_1-d_2+d_4$ & $\Omega_{ccc}^{++}\to \Xi_c^+  D^0  K^+  $ & $ \left(2 d_1+d_2+d_4\right) (-\text{sC})$\\\hline
$\Omega_{ccc}^{++}\to \Xi_c^0  D^+  \pi^+  $ & $ -d_2+d_3-d_4+d_5$ & $\Omega_{ccc}^{++}\to \Xi_c^+  D^+  K^0  $ & $ \left(2 d_1+d_3-d_5\right) (-\text{sC})$\\\hline
\Xcline{1-2}{1.2pt}
$\Omega_{ccc}^{++}\to \Lambda_c^+  D^0  K^+  $ & $ \left(2 d_1+d_2+d_4\right) \left(-\text{sC}^2\right)$ & $\Omega_{ccc}^{++}\to \Xi_c^+  D^+_s  \pi^0  $ & $ \frac{\left(-d_2+d_3+d_4+d_5\right) \text{sC}}{\sqrt{2}}$\\\hline
$\Omega_{ccc}^{++}\to \Lambda_c^+  D^+  K^0  $ & $ \left(2 d_1+d_2-d_4\right) \left(-\text{sC}^2\right)$ & $\Omega_{ccc}^{++}\to \Xi_c^+  D^+_s  \eta  $ & $ \frac{\left(4 d_1+3 d_2-d_3-3 d_4+3 d_5\right) \text{sC}}{\sqrt{6}}$\\\hline
$\Omega_{ccc}^{++}\to \Lambda_c^+  D^+_s  \pi^0  $ & $ \sqrt{2} d_5 \text{sC}^2$ & $\Omega_{ccc}^{++}\to \Xi_c^0  D^+  K^+  $ & $ \left(d_2-d_3+d_4-d_5\right) (-\text{sC})$\\\hline
$\Omega_{ccc}^{++}\to \Lambda_c^+  D^+_s  \eta  $ & $ \sqrt{\frac{2}{3}} \left(2 d_1+d_3\right) \text{sC}^2$ & $\Omega_{ccc}^{++}\to \Xi_c^0  D^+_s  \pi^+  $ & $ \left(d_2-d_3+d_4-d_5\right) (-\text{sC})$\\\hline
$\Omega_{ccc}^{++}\to \Xi_c^+  D^+_s  K^0  $ & $ \left(d_2-d_3-d_4+d_5\right) \left(-\text{sC}^2\right)$ && \\\hline
$\Omega_{ccc}^{++}\to \Xi_c^0  D^+_s  K^+  $ & $ \left(d_2-d_3+d_4-d_5\right) \text{sC}^2$ && \\\hline
\hline
\end{tabular}
\end{table}

  \begin{table}
\caption{Amplitudes for three-body $\Omega_{ccc}$ decays into  a singly-charmed baryon (sextet),  D meson and a light meson.}\label{tab:ccc_Bthreebody2}
\begin{tabular}{|cc|cc|c|c|c|c}\hline\hline
channel & amplitude &channel & amplitude \\\hline
$\Omega_{ccc}^{++}\to \Sigma_{c}^{++}  D^0  \overline K^0  $ & $ e_2-e_4$ & $\Omega_{ccc}^{++}\to \Sigma_{c}^{++}  D^0  \pi^0  $ & $ \frac{\left(e_2-e_4\right) \text{sC}}{\sqrt{2}}$\\\hline
$\Omega_{ccc}^{++}\to \Sigma_{c}^{++}  D^+  K^-  $ & $ e_3-e_5$ & $\Omega_{ccc}^{++}\to \Sigma_{c}^{++}  D^0  \eta  $ & $ -\sqrt{\frac{3}{2}} \left(e_2-e_4\right) \text{sC}$\\\hline
$\Omega_{ccc}^{++}\to \Sigma_{c}^{+}  D^+  \overline K^0  $ & $ \frac{e_2+e_3-e_4-e_5}{\sqrt{2}}$ & $\Omega_{ccc}^{++}\to \Sigma_{c}^{++}  D^+  \pi^-  $ & $ \left(e_5-e_3\right) \text{sC}$\\\hline
$\Omega_{ccc}^{++}\to \Xi_{c}^{\prime+}  D^0  \pi^+  $ & $ \frac{2 e_1+e_2+e_4}{\sqrt{2}}$ & $\Omega_{ccc}^{++}\to \Sigma_{c}^{++}  D^+_s  K^-  $ & $ \left(e_3-e_5\right) \text{sC}$\\\hline
$\Omega_{ccc}^{++}\to \Xi_{c}^{\prime+}  D^+  \pi^0  $ & $ \frac{1}{2} \left(-2 e_1+e_3+e_5\right)$ & $\Omega_{ccc}^{++}\to \Sigma_{c}^{+}  D^0  \pi^+  $ & $ -\frac{\left(2 e_1+e_2+e_4\right) \text{sC}}{\sqrt{2}}$\\\hline
$\Omega_{ccc}^{++}\to \Xi_{c}^{\prime+}  D^+  \eta  $ & $ \frac{2 e_1-e_3+3 e_5}{2 \sqrt{3}}$ & $\Omega_{ccc}^{++}\to \Sigma_{c}^{+}  D^+  \pi^0  $ & $ \frac{1}{2} \left(2 e_1+e_2-e_4-2 e_5\right) \text{sC}$\\\hline
$\Omega_{ccc}^{++}\to \Xi_{c}^{\prime+}  D^+_s  \overline K^0  $ & $ \frac{2 e_1+e_2-e_4}{\sqrt{2}}$ & $\Omega_{ccc}^{++}\to \Sigma_{c}^{+}  D^+  \eta  $ & $ -\frac{\left(2 e_1+3 e_2+2 e_3-3 e_4\right) \text{sC}}{2 \sqrt{3}}$\\\hline
$\Omega_{ccc}^{++}\to \Xi_{c}^{\prime0}  D^+  \pi^+  $ & $ \frac{e_2+e_3+e_4+e_5}{\sqrt{2}}$ & $\Omega_{ccc}^{++}\to \Sigma_{c}^{+}  D^+_s  \overline K^0  $ & $ -\frac{\left(2 e_1-e_3+e_5\right) \text{sC}}{\sqrt{2}}$\\\hline
$\Omega_{ccc}^{++}\to \Omega_{c}^{0}  D^+  K^+  $ & $ e_3+e_5$ & $\Omega_{ccc}^{++}\to \Sigma_{c}^{0}  D^+  \pi^+  $ & $ \left(e_2+e_3+e_4+e_5\right) (-\text{sC})$\\\hline
$\Omega_{ccc}^{++}\to \Omega_{c}^{0}  D^+_s  \pi^+  $ & $ e_2+e_4$ & $\Omega_{ccc}^{++}\to \Xi_{c}^{\prime+}  D^0  K^+  $ & $ \frac{\left(2 e_1+e_2+e_4\right) \text{sC}}{\sqrt{2}}$\\\hline
\Xcline{1-2}{1.2pt}
$\Omega_{ccc}^{++}\to \Sigma_{c}^{++}  D^0  K^0  $ & $ \left(e_2-e_4\right) \text{sC}^2$ & $\Omega_{ccc}^{++}\to \Xi_{c}^{\prime+}  D^+  K^0  $ & $ \frac{\left(2 e_1-e_3+e_5\right) \text{sC}}{\sqrt{2}}$\\\hline
$\Omega_{ccc}^{++}\to \Sigma_{c}^{++}  D^+_s  \pi^-  $ & $ \left(e_3-e_5\right) \text{sC}^2$ & $\Omega_{ccc}^{++}\to \Xi_{c}^{\prime+}  D^+_s  \pi^0  $ & $ \frac{1}{2} \left(e_2+e_3-e_4+e_5\right) \text{sC}$\\\hline
$\Omega_{ccc}^{++}\to \Sigma_{c}^{+}  D^0  K^+  $ & $ \frac{\left(2 e_1+e_2+e_4\right) \text{sC}^2}{\sqrt{2}}$ & $\Omega_{ccc}^{++}\to \Xi_{c}^{\prime+}  D^+_s  \eta  $ & $ -\frac{\left(4 e_1+3 e_2+e_3-3 e_4-3 e_5\right) \text{sC}}{2 \sqrt{3}}$\\\hline
$\Omega_{ccc}^{++}\to \Sigma_{c}^{+}  D^+  K^0  $ & $ \frac{\left(2 e_1+e_2-e_4\right) \text{sC}^2}{\sqrt{2}}$ & $\Omega_{ccc}^{++}\to \Xi_{c}^{\prime0}  D^+  K^+  $ & $ \frac{\left(e_2-e_3+e_4-e_5\right) \text{sC}}{\sqrt{2}}$\\\hline
$\Omega_{ccc}^{++}\to \Sigma_{c}^{+}  D^+_s  \pi^0  $ & $ e_5 \text{sC}^2$ & $\Omega_{ccc}^{++}\to \Xi_{c}^{\prime0}  D^+_s  \pi^+  $ & $ -\frac{\left(e_2-e_3+e_4-e_5\right) \text{sC}}{\sqrt{2}}$\\\hline
$\Omega_{ccc}^{++}\to \Sigma_{c}^{+}  D^+_s  \eta  $ & $ \frac{\left(e_3-2 e_1\right) \text{sC}^2}{\sqrt{3}}$ & $\Omega_{ccc}^{++}\to \Omega_{c}^{0}  D^+_s  K^+  $ & $ \left(e_2+e_3+e_4+e_5\right) \text{sC}$\\\hline
$\Omega_{ccc}^{++}\to \Sigma_{c}^{0}  D^+  K^+  $ & $ \left(e_2+e_4\right) \text{sC}^2$ && \\\hline
$\Omega_{ccc}^{++}\to \Sigma_{c}^{0}  D^+_s  \pi^+  $ & $ \left(e_3+e_5\right) \text{sC}^2$ && \\\hline
$\Omega_{ccc}^{++}\to \Xi_{c}^{\prime+}  D^+_s  K^0  $ & $ \frac{\left(e_2+e_3-e_4-e_5\right) \text{sC}^2}{\sqrt{2}}$ && \\\hline
$\Omega_{ccc}^{++}\to \Xi_{c}^{\prime0}  D^+_s  K^+  $ & $ \frac{\left(e_2+e_3+e_4+e_5\right) \text{sC}^2}{\sqrt{2}}$ && \\\hline
\hline
\end{tabular}
\end{table}

For the two-body decays into a charmed baryon and a charmed meson,
the effective Hamiltonian for the decays of $\Omega_{ccc}$ into a singly charmed baryon and a charmed meson  is given as:
\begin{eqnarray}
{\cal H}_{\rm eff}=c_1 \Omega_{ccc} (\overline  T_{\bf c\bar 3})_{[ij]} \overline D^k (H_{\bar 6})^{ij}_k + c_2\Omega_{ccc} (\overline  T_{\bf c  6})_{[ij]} \overline D^k (H_{15})^{ij}_k.
\end{eqnarray}
The Feynman diagram is shown in the last panel of Fig.~\ref{fig:topology_2body}, and  decay amplitudes are collected  in Tab.~\ref{tab:ccc_Btwobody}.

The three-body decays of $\Xi_{ccc}^{++}$ can  involve an additional  light meson in the final state.
For  the modes with an anti-triplet baryon, we have
\begin{eqnarray}
{\cal H}_{\rm eff}&=&  d_1 \Omega_{ccc} (\overline  T_{\bf c\bar 3})_{[ij]} \overline D^k (M_8)_k^l  (H_{\bar 6})^{ij}_l + d_2 \Omega_{ccc}  (\overline  T_{\bf c\bar 3})_{[ij]} \overline D^j (M_8)_k^l  (H_{\bar 6})^{ik}_l \nonumber\\
&&  +d_3 \Omega_{ccc} (\overline  T_{\bf c\bar 3})_{[ij]} \overline D^l (M_8)_k^i   (H_{\bar 6})^{jk}_l + d_4 \Omega_{ccc}  (\overline  T_{\bf c\bar 3})_{[ij]} \overline D^j (M_8)_k^l  (H_{15})^{ik}_l \nonumber\\
&&+ d_5 \Omega_{ccc} (\overline  T_{\bf c\bar 3})_{[ij]}  \overline D^l (M_8)_k^i  (H_{15})^{jk}_l,
\end{eqnarray}
while the effective Hamiltonian for a sextet baryon is constructed  as:
\begin{eqnarray}
{\cal H}_{\rm eff}&=& e_1 \Omega_{ccc} (\overline  T_{\bf c6})_{\{ij\}} \overline D^k (M_8)_k^l  (H_{15})^{ij}_l +e_2 \Omega_{ccc}  (\overline  T_{\bf c6})_{\{ij\}} \overline D^j (M_8)_k^l  (H_{15})^{ik}_l  \nonumber\\
&& + e_3 \Omega_{ccc} (\overline  T_{\bf c6})_{\{ij\}}  \overline D^l (M_8)_k^i  (H_{15})^{jk}_l + e_4 \Omega_{ccc}  (\overline  T_{\bf c6})_{\{ij\}} \overline D^j (M_8)_k^l  (H_{\bar 6})^{ik}_l \nonumber\\
&&+ e_5 \Omega_{ccc} (\overline  T_{\bf c6})_{\{ij\}}  \overline D^l (M_8)_k^i  (H_{\bar 6})^{jk}_l.
\end{eqnarray}

Expanding the above equations, we will obtain the decay amplitudes given in Tab.~\ref{tab:ccc_Bthreebody1} for anti-triplet baryon and in Tab.~\ref{tab:ccc_Bthreebody2} for sextet baryon.

\section{Non-Leptonic  $\Omega_{bbb}^-$ decays}
\label{sec:bbb_nonleptonic}

For the bottom quark decay, there are generically  4 kinds of quark-level transitions:
\begin{eqnarray}
b\to c\bar c d/s, \;
b\to c \bar u d/s, \;
b\to u \bar c d/s, \;
b\to q\bar qq,
\end{eqnarray}
which will be studied in order.


\subsection{$b\to c\bar c d/s$}

\subsubsection{Decays into  a $J/\psi$ }

\begin{table}
\caption{Amplitudes for $\Omega_{bbb}$ decays into a  $J/\psi$ and a doubly bottom baryon or two bottom hadrons. }
\label{tab:bbb_Jpsi}
\begin{tabular}{|cc|cc|c|c|c|c}\hline\hline
channel & amplitude &channel & amplitude \\\hline
$\Omega_{bbb}^{-}\to \Xi_{bb}^{-}J/\psi $ & $ f_1 V_{cd}^*$ & $\Omega_{bbb}^{-}\to \Lambda_b^0  B^- J/\psi $ & $ f_3 V_{cd}^*$\\\hline
$\Omega_{bbb}^{-}\to \Omega_{bb}^{-}J/\psi $ & $ f_1 V_{cs}^*$ & $\Omega_{bbb}^{-}\to \Xi_b^0  B^- J/\psi $ & $ f_3 V_{cs}^*$\\\hline
\Xcline{1-2}{1.2pt}
$\Omega_{bbb}^{-}\to \Xi_{bb}^{0}  \pi^-  J/\psi $ & $ f_2 V_{cd}^*$ & $\Omega_{bbb}^{-}\to \Xi_b^-  \overline B^0 J/\psi $ & $ f_3 V_{cs}^*$\\\hline
$\Omega_{bbb}^{-}\to \Xi_{bb}^{0}  K^-  J/\psi $ & $ f_2 V_{cs}^*$ & $\Omega_{bbb}^{-}\to \Xi_b^-  \overline B^0_s J/\psi $ & $ -f_3 V_{cd}^*$\\\hline
\Xcline{3-4}{1.2pt}
$\Omega_{bbb}^{-}\to \Xi_{bb}^{-}  \pi^0  J/\psi $ & $ -\frac{f_2 V_{cd}^*}{\sqrt{2}}$ & $\Omega_{bbb}^{-}\to \Sigma_{b}^{0}  B^- J/\psi $ & $ \frac{f_4 V_{cd}^*}{\sqrt{2}}$\\\hline
$\Omega_{bbb}^{-}\to \Xi_{bb}^{-}  \overline K^0  J/\psi $ & $ f_2 V_{cs}^*$ & $\Omega_{bbb}^{-}\to \Sigma_{b}^{-}  \overline B^0 J/\psi $ & $ f_4 V_{cd}^*$\\\hline
$\Omega_{bbb}^{-}\to \Xi_{bb}^{-}  \eta  J/\psi $ & $ \frac{f_2 V_{cd}^*}{\sqrt{6}}$ &$\Omega_{bbb}^{-}\to \Xi_{b}^{\prime0}  B^- J/\psi $ & $ \frac{f_4 V_{cs}^*}{\sqrt{2}}$\\\hline
$\Omega_{bbb}^{-}\to \Omega_{bb}^{-}  K^0  J/\psi $ & $ f_2 V_{cd}^*$¡¡& $\Omega_{bbb}^{-}\to \Xi_{b}^{\prime-}  \overline B^0 J/\psi $ & $ \frac{f_4 V_{cs}^*}{\sqrt{2}}$\\\hline
$\Omega_{bbb}^{-}\to \Omega_{bb}^{-}  \eta  J/\psi $ & $ -\sqrt{\frac{2}{3}} f_2 V_{cs}^*$ & $\Omega_{bbb}^{-}\to \Xi_{b}^{\prime-}  \overline B^0_s J/\psi $ & $ \frac{f_4 V_{cd}^*}{\sqrt{2}}$\\\hline
&& $\Omega_{bbb}^{-}\to \Omega_{b}^{-}  \overline B^0_s J/\psi $ & $ f_4 V_{cs}^*$\\\hline
\hline
\end{tabular}
\end{table}

Such decays will  have the same topology  with  the  $b\to s\ell^+\ell^-$ decays.
The transition operator $b\to c\bar cd/s$  can form an SU(3) triplet:
\begin{eqnarray}
  {\cal H}_{\rm{eff}} &=&   f_1 \Omega_{bbb}(\overline  T_{bb})_i (H_{  3})^i ~ J/\psi+ f_2 \Omega_{bbb}(\overline  T_{bb})_i (M_8)^i_j (H_{3})^j ~ J/\psi\nonumber\\
  &&+ f_3 \Omega_{bbb} (\overline  T_{b\bar 3})_{[ij]} \overline B^i   (H_{3})^j ~ J/\psi+ f_4 \Omega_{bbb} (\overline  T_{b6})_{\{ij\}} \overline B^i   (H_{3})^j ~ J/\psi ,
\end{eqnarray}
with $(H_{  3})_{2}=V_{cd}^*$ and $(H_{  3})_{3}=V_{cs}^*$.
Decay amplitudes for different channels are obtained by expanding the above Hamiltonian and are collected in Tab.~\ref{tab:bbb_Jpsi}.

\subsubsection{Decays into a triply heavy baryon $cbb$  plus a  anti-charmed  meson  }

\begin{table}
\caption{Amplitudes for $\Omega_{bbb}$ decays into $\Omega_{cbb}$ and a anti-charmed meson. }\label{tab:bbb_cbb}
\begin{tabular}{c|c|c|c}\hline\hline
channel & amplitude & channel & amplitude \\\hline
$\Omega_{bbb}^{-}\to \Omega_{cbb}^{0}  D^- $ & $ g_1 V_{cd}^*$&
$\Omega_{bbb}^{-}\to \Omega_{cbb}^{0}  D^-_s $ & $ g_1 V_{cs}^*$\\\hline
\Xcline{1-4}{1.2pt}
$\Omega_{bbb}^{-}\to \Omega_{cbb}^{0}  \overline D^0  \pi^-  $ & $ g_2 V_{cd}^*$&
$\Omega_{bbb}^{-}\to \Omega_{cbb}^{0}  \overline D^0  K^-  $ & $ g_2 V_{cs}^*$\\\hline
$\Omega_{bbb}^{-}\to \Omega_{cbb}^{0}  D^-  \pi^0  $ & $ -\frac{g_2 V_{cd}^*}{\sqrt{2}}$&
$\Omega_{bbb}^{-}\to \Omega_{cbb}^{0}  D^-  \overline K^0  $ & $ g_2 V_{cs}^*$\\\hline
$\Omega_{bbb}^{-}\to \Omega_{cbb}^{0}  D^-  \eta  $ & $ \frac{g_2 V_{cd}^*}{\sqrt{6}}$&
$\Omega_{bbb}^{-}\to \Omega_{cbb}^{0}  D^-_s  K^0  $ & $ g_2 V_{cd}^*$\\\hline
$\Omega_{bbb}^{-}\to \Omega_{cbb}^{0}  D^-_s  \eta  $ & $ -\sqrt{\frac{2}{3}} g_2 V_{cs}^*$\\\hline
\hline
\end{tabular}
\end{table}

The   $b\to c\bar cd/s$   transition can lead to another type of effective Hamiltonian:
\begin{eqnarray}
{\cal H}_{\rm{eff}}=   g_1  \Omega_{bbb}  \overline\Omega_{cbb}D_i (H_{3})^i +  g_2 \Omega_{bbb} \overline \Omega_{cbb}  D_j  (M_8)^j_i (H_{3})^i  .
\end{eqnarray}
which corresponds to the decays into doubly heavy baryon $bcq$  plus a  anti-charmed  meson.
Decay amplitudes for different channels are obtained by expanding the above Hamiltonian and are collected in Tab.~\ref{tab:bbb_cbb}.

\subsection{$b\to c \bar u d/s$ transition}

\subsubsection{Decays into a triply heavy baryon $cbb$ plus light mesons }

\begin{table}
\caption{Amplitudes for $\Omega_{bbb}$ decays into $\Omega_{cbb}$ and light meson(s). }\label{tab:bbb_cbb2}\begin{tabular}{c|c|c|c}\hline\hline
channel & amplitude& channel & amplitude \\\hline
$\Omega_{bbb}^{-}\to \Omega_{bbc}^{0}  \pi^-  $ & $ h_1 V_{cb}V_{ud}^*$ & $\Omega_{bbb}^{-}\to \Omega_{bbc}^{0}  K^-  $ & $ h_1 V_{cb}V_{us}^*$\\\hline
\Xcline{1-4}{1.2pt}
$\Omega_{bbb}^{-}\to \Omega_{bbc}^{0}  \pi^0   K^-  $ & $ \frac{h_2 V_{cb}V_{us}^*}{\sqrt{2}}$ & $\Omega_{bbb}^{-}\to \Omega_{bbc}^{0}  \pi^-   \eta  $ & $ \sqrt{\frac{2}{3}} h_2 V_{cb}V_{ud}^*$\\\hline
$\Omega_{bbb}^{-}\to \Omega_{bbc}^{0}  \pi^-   \overline K^0  $ & $ h_2 V_{cb}V_{us}^*$ & $\Omega_{bbb}^{-}\to \Omega_{bbc}^{0}  K^0   K^-  $ & $ h_2 V_{cb}V_{ud}^*$\\\hline
$\Omega_{bbb}^{-}\to \Omega_{bbc}^{0}  K^-   \eta  $ & $ -\frac{h_2 V_{cb}V_{us}^*}{\sqrt{6}}$ & & \\\hline
\hline
\end{tabular}
\end{table}

The operator to produce a charm quark  from the $b$-quark decay, $\bar c b \bar q u$, is given by
\begin{eqnarray}
{\cal H}_{e.w.} &=& \frac{G_{F}}{\sqrt{2}}
     V_{cb} V_{uq}^{*} \big[
     C_{1}  O^{\bar cu}_{1}
  +  C_{2}  O^{\bar cu}_{2}\Big] +{\rm h.c.} .
\end{eqnarray}
The light quarks in this effective Hamiltonian form an octet with the nonzero entry
\begin{eqnarray}
(H_{{\bf8}})^2_1 =V_{cb}V_{ud}^*,
\end{eqnarray}
for   the $b\to c\bar ud$ transition, and   $(H_{{\bf8}})^3_1 = V_{cb}V_{us}^*$ for  the $b\to c\bar
us$ transition.
\begin{eqnarray}
 {\cal H}_{\rm{eff}}=   h_1  \Omega_{bbb}\overline \Omega_{cbb}  M^j_i  (H_{8})^i_j+  h_2 \Omega_{bbb}\overline \Omega_{cbb} M^j_k M^k_i (H_{8})^i_j .
\end{eqnarray}
Decay amplitudes for different channels are obtained by expanding the above Hamiltonian and are collected in Tab.~\ref{tab:bbb_cbb2}.

\subsubsection{Decays into a charmed meson plus  a doubly bottom baryon or two bottom hadrons.}

 \begin{table}
\caption{Amplitudes for nonleptonic  $\Omega_{bbb}^-$ decays into a charmed meson plus  a doubly bottom baryon or two bottom hadrons. }\label{tab:bbb_B2}
\begin{tabular}{|cc|cc|c|c|c|c}\hline\hline
channel & amplitude &channel & amplitude \\\hline
$\Omega_{bbb}^{-}\to \Xi_{bb}^{-}  D^0 $ & $ j_1V_{cb}V_{ud}^*$ & $\Omega_{bbb}^{-}\to \Lambda_b^0  B^-  D^0 $ & $ -\left(j_5+j_6\right)V_{cb}V_{ud}^*$\\\hline
$\Omega_{bbb}^{-}\to \Omega_{bb}^{-}  D^0 $ & $ j_1V_{cb}V_{us}^*$ & $\Omega_{bbb}^{-}\to \Xi_b^0  B^-  D^0 $ & $ -\left(j_5+j_6\right)V_{cb}V_{us}^*$\\\hline
\Xcline{1-2}{1.2pt}
$\Omega_{bbb}^{-}\to \Xi_{bb}^{0}  D^0  \pi^-  $ & $ \left(j_2+j_4\right)V_{cb}V_{ud}^*$ & $\Omega_{bbb}^{-}\to \Xi_b^-  B^-  D^+ $ & $ -j_5V_{cb}V_{us}^*$\\\hline
$\Omega_{bbb}^{-}\to \Xi_{bb}^{0}  D^0  K^-  $ & $ \left(j_2+j_4\right)V_{cb}V_{us}^*$ & $\Omega_{bbb}^{-}\to \Xi_b^-  B^-  D^+_s $ & $ j_5V_{cb}V_{ud}^*$\\\hline
$\Omega_{bbb}^{-}\to \Xi_{bb}^{-}  D^0  \pi^0  $ & $ \frac{\left(j_3-j_4\right)V_{cb}V_{ud}^*}{\sqrt{2}}$ & $\Omega_{bbb}^{-}\to \Xi_b^-  \overline B^0  D^0 $ & $ -j_6V_{cb}V_{us}^*$\\\hline
$\Omega_{bbb}^{-}\to \Xi_{bb}^{-}  D^0  \overline K^0  $ & $ j_4V_{cb}V_{us}^*$ &$\Omega_{bbb}^{-}\to \Xi_b^-  \overline B^0_s  D^0 $ & $ j_6V_{cb}V_{ud}^*$\\\hline
\Xcline{3-4}{1.2pt}
$\Omega_{bbb}^{-}\to \Xi_{bb}^{-}  D^0  \eta  $ & $ \frac{\left(j_3+j_4\right)V_{cb}V_{ud}^*}{\sqrt{6}}$ & $\Omega_{bbb}^{-}\to \Sigma_{b}^{0}  B^-  D^0 $ & $ \frac{\left(j_7+j_8\right)V_{cb}V_{ud}^*}{\sqrt{2}}$\\\hline
$\Omega_{bbb}^{-}\to \Xi_{bb}^{-}  D^+  \pi^-  $ & $ \left(j_2+j_3\right)V_{cb}V_{ud}^*$ &¡¡$\Omega_{bbb}^{-}\to \Sigma_{b}^{-}  B^-  D^+ $ & $ j_7V_{cb}V_{ud}^*$\\\hline
$\Omega_{bbb}^{-}\to \Xi_{bb}^{-}  D^+  K^-  $ & $ j_2V_{cb}V_{us}^*$¡¡& $\Omega_{bbb}^{-}\to \Sigma_{b}^{-}  \overline B^0  D^0 $ & $ j_8V_{cb}V_{ud}^*$\\\hline
$\Omega_{bbb}^{-}\to \Xi_{bb}^{-}  D^+_s  K^-  $ & $ j_3V_{cb}V_{ud}^*$ & $\Omega_{bbb}^{-}\to \Xi_{b}^{\prime0}  B^-  D^0 $ & $ \frac{\left(j_7+j_8\right)V_{cb}V_{us}^*}{\sqrt{2}}$\\\hline
$\Omega_{bbb}^{-}\to \Omega_{bb}^{-}  D^0  \pi^0  $ & $ \frac{j_3V_{cb}V_{us}^*}{\sqrt{2}}$ & $\Omega_{bbb}^{-}\to \Xi_{b}^{\prime-}  B^-  D^+ $ & $ \frac{j_7V_{cb}V_{us}^*}{\sqrt{2}}$\\\hline
$\Omega_{bbb}^{-}\to \Omega_{bb}^{-}  D^0  K^0  $ & $ j_4V_{cb}V_{ud}^*$ & $\Omega_{bbb}^{-}\to \Xi_{b}^{\prime-}  B^-  D^+_s $ & $ \frac{j_7V_{cb}V_{ud}^*}{\sqrt{2}}$\\\hline
$\Omega_{bbb}^{-}\to \Omega_{bb}^{-}  D^0  \eta  $ & $ \frac{\left(j_3-2 j_4\right)V_{cb}V_{us}^*}{\sqrt{6}}$ & $\Omega_{bbb}^{-}\to \Xi_{b}^{\prime-}  \overline B^0  D^0 $ & $ \frac{j_8V_{cb}V_{us}^*}{\sqrt{2}}$\\\hline
$\Omega_{bbb}^{-}\to \Omega_{bb}^{-}  D^+  \pi^-  $ & $ j_3V_{cb}V_{us}^*$ &¡¡$\Omega_{bbb}^{-}\to \Xi_{b}^{\prime-}  \overline B^0_s  D^0 $ & $ \frac{j_8V_{cb}V_{ud}^*}{\sqrt{2}}$\\\hline
$\Omega_{bbb}^{-}\to \Omega_{bb}^{-}  D^+_s  \pi^-  $ & $ j_2V_{cb}V_{ud}^*$ &¡¡$\Omega_{bbb}^{-}\to \Omega_{b}^{-}  B^-  D^+_s $ & $ j_7V_{cb}V_{us}^*$\\\hline
$\Omega_{bbb}^{-}\to \Omega_{bb}^{-}  D^+_s  K^-  $ & $ \left(j_2+j_3\right)V_{cb}V_{us}^*$ & $\Omega_{bbb}^{-}\to \Omega_{b}^{-}  \overline B^0_s  D^0 $ & $ j_8V_{cb}V_{us}^*$\\\hline
\hline
\end{tabular}
\end{table}

If the $c\bar c$ are separated, then the final state could be a doubly bottom baryon and a charmed meson.  The three-body modes can also   include  decays into a bottom baryon, a bottom meson and a charmed meson. Thus one can have the effective Hamiltonian as:
\begin{eqnarray}
 {\cal  H}_{\rm{eff}} &= &  j_1 \Omega_{bbb}(\overline T_{bb})_i\overline D^j (H_{8})^i_j+ j_2  \Omega_{bbb}  (\overline T_{bb})_k\overline D^k  (M_8)^j_i  (H_{8})^i_j  \nonumber\\
  &&  + j_3 \Omega_{bbb}  (\overline T_{bb})_i\overline D^k   (M_8)^j_k (H_{8})^i_j+ j_4  \Omega_{bbb}  (\overline T_{bb})_i \overline D^k  (M_8)^i_j (H_{8})^j_k    \nonumber\\
  &&+ j_5 \Omega_{bbb}  (\overline T_{b\bar 3})_{[ik]} \overline B^j  \overline D^k (H_{8})^i_j  + j_6 \Omega_{bbb}  (\overline T_{b\bar 3})_{[ik]} \overline B^k \overline D^j (H_{8})^i_j\nonumber\\
  &&+ j_7  \Omega_{bbb} (\overline T_{b6})_{\{ik\}} \overline B^j  \overline D^k(H_{8})^i_j + j_8  \Omega_{bbb} (\overline T_{b6})_{\{ik\}} \overline B^k  \overline D^j (H_{8})^i_j.
\end{eqnarray}
Decay amplitudes for different channels are obtained by expanding the above Hamiltonian and are collected in Tab.~\ref{tab:bbb_B2}.

\subsection{The CKM suppressed $b\to u \bar c d/s$ transition  }

 \begin{table}
\caption{Amplitudes for the  CKM suppressed $\Omega_{bbb}$ decays into anti-charmed meson. }\label{tab:bbb_C}
\begin{tabular}{|cc|cc|c|c|c|c}\hline\hline
channel & amplitude &channel & amplitude \\\hline
$\Omega_{bbb}^{-}\to \Xi_{bb}^{0}  D^- $ & $ \left(k_1+k_2\right) V_{ub}V_{cd}^*$ & $\Omega_{bbb}^{-}\to \Lambda_b^0  B^-  \overline D^0 $ & $ \left(2 k_7+k_8-k_9\right) V_{ub}V_{cd}^*$\\\hline
$\Omega_{bbb}^{-}\to \Xi_{bb}^{0}  D^-_s $ & $ \left(k_1+k_2\right) V_{ub}V_{cs}^*$ & $\Omega_{bbb}^{-}\to \Lambda_b^0  \overline B^0  D^- $ & $ \left(2 k_7+k_8+k_9\right) V_{ub}V_{cd}^*$\\\hline
$\Omega_{bbb}^{-}\to \Xi_{bb}^{-}  \overline D^0 $ & $ \left(k_2-k_1\right) V_{ub}V_{cd}^*$ & $\Omega_{bbb}^{-}\to \Lambda_b^0  \overline B^0  D^-_s $ & $ \left(k_8+k_9\right) V_{ub}V_{cs}^*$\\\hline
$\Omega_{bbb}^{-}\to \Omega_{bb}^{-}  \overline D^0 $ & $ \left(k_2-k_1\right) V_{ub}V_{cs}^*$ & $\Omega_{bbb}^{-}\to \Lambda_b^0  \overline B^0_s  D^-_s $ & $ 2 k_7 V_{ub}V_{cd}^*$\\\hline
\Xcline{1-2}{1.2pt}
$\Omega_{bbb}^{-}\to \Xi_{bb}^{0}  \overline D^0  \pi^-  $ & $ \left(k_3+k_4+k_5+k_6\right) V_{ub}V_{cd}^*$ & $\Omega_{bbb}^{-}\to \Xi_b^0  B^-  \overline D^0 $ & $ \left(2 k_7+k_8-k_9\right) V_{ub}V_{cs}^*$\\\hline
$\Omega_{bbb}^{-}\to \Xi_{bb}^{0}  \overline D^0  K^-  $ & $ \left(k_3+k_4+k_5+k_6\right) V_{ub}V_{cs}^*$ & $\Omega_{bbb}^{-}\to \Xi_b^0  \overline B^0  D^- $ & $ 2 k_7 V_{ub}V_{cs}^*$\\\hline
$\Omega_{bbb}^{-}\to \Xi_{bb}^{0}  D^-  \pi^0  $ & $ -\frac{\left(k_3+k_4-k_5+k_6\right) V_{ub}V_{cd}^*}{\sqrt{2}}$ & $\Omega_{bbb}^{-}\to \Xi_b^0  \overline B^0_s  D^- $ & $ \left(k_8+k_9\right) V_{ub}V_{cd}^*$\\\hline
$\Omega_{bbb}^{-}\to \Xi_{bb}^{0}  D^-  \overline K^0  $ & $ \left(k_4+k_6\right) V_{ub}V_{cs}^*$ & $\Omega_{bbb}^{-}\to \Xi_b^0  \overline B^0_s  D^-_s $ & $ \left(2 k_7+k_8+k_9\right) V_{ub}V_{cs}^*$\\\hline
$\Omega_{bbb}^{-}\to \Xi_{bb}^{0}  D^-  \eta  $ & $ \frac{\left(-k_3+k_4+k_5+k_6\right) V_{ub}V_{cd}^*}{\sqrt{6}}$ & $\Omega_{bbb}^{-}\to \Xi_b^-  \overline B^0  \overline D^0 $ & $ \left(k_8-k_9\right) V_{ub}V_{cs}^*$\\\hline
$\Omega_{bbb}^{-}\to \Xi_{bb}^{0}  D^-_s  \pi^0  $ & $ \frac{\left(k_5-k_3\right) V_{ub}V_{cs}^*}{\sqrt{2}}$ & $\Omega_{bbb}^{-}\to \Xi_b^-  \overline B^0_s  \overline D^0 $ & $ \left(k_9-k_8\right) V_{ub}V_{cd}^*$\\\hline
\Xcline{3-4}{1.2pt}
$\Omega_{bbb}^{-}\to \Xi_{bb}^{0}  D^-_s  K^0  $ & $ \left(k_4+k_6\right) V_{ub}V_{cd}^*$ & $\Omega_{bbb}^{-}\to \Sigma_{b}^{+}  B^-  D^- $ & $ \left(k_{10}+k_{12}\right) V_{ub}V_{cd}^*$\\\hline
$\Omega_{bbb}^{-}\to \Xi_{bb}^{0}  D^-_s  \eta  $ & $ -\frac{\left(k_3+2 k_4-k_5+2 k_6\right) V_{ub}V_{cs}^*}{\sqrt{6}}$ & $\Omega_{bbb}^{-}\to \Sigma_{b}^{+}  B^-  D^-_s $ & $ \left(k_{10}+k_{12}\right) V_{ub}V_{cs}^*$\\\hline
$\Omega_{bbb}^{-}\to \Xi_{bb}^{-}  \overline D^0  \pi^0  $ & $ -\frac{\left(k_3+k_4+k_5-k_6\right) V_{ub}V_{cd}^*}{\sqrt{2}}$ & $\Omega_{bbb}^{-}\to \Sigma_{b}^{0}  B^-  \overline D^0 $ & $ \frac{\left(-k_{10}+2 k_{11}+k_{12}\right) V_{ub}V_{cd}^*}{\sqrt{2}}$\\\hline
$\Omega_{bbb}^{-}\to \Xi_{bb}^{-}  \overline D^0  \overline K^0  $ & $ \left(k_3+k_5\right) V_{ub}V_{cs}^*$ & $\Omega_{bbb}^{-}\to \Sigma_{b}^{0}  \overline B^0  D^- $ & $ \frac{\left(k_{10}+2 k_{11}+k_{12}\right) V_{ub}V_{cd}^*}{\sqrt{2}}$\\\hline
$\Omega_{bbb}^{-}\to \Xi_{bb}^{-}  \overline D^0  \eta  $ & $ \frac{\left(k_3-k_4+k_5+k_6\right) V_{ub}V_{cd}^*}{\sqrt{6}}$ & $\Omega_{bbb}^{-}\to \Sigma_{b}^{0}  \overline B^0  D^-_s $ & $ \frac{\left(k_{10}+k_{12}\right) V_{ub}V_{cs}^*}{\sqrt{2}}$\\\hline
$\Omega_{bbb}^{-}\to \Xi_{bb}^{-}  D^-  \pi^+  $ & $ \left(-k_3-k_4+k_5+k_6\right) V_{ub}V_{cd}^*$ & $\Omega_{bbb}^{-}\to \Sigma_{b}^{0}  \overline B^0_s  D^-_s $ & $ \sqrt{2} k_{11} V_{ub}V_{cd}^*$\\\hline
$\Omega_{bbb}^{-}\to \Xi_{bb}^{-}  D^-_s  \pi^+  $ & $ \left(k_5-k_3\right) V_{ub}V_{cs}^*$ & $\Omega_{bbb}^{-}\to \Sigma_{b}^{-}  \overline B^0  \overline D^0 $ & $ \left(k_{12}-k_{10}\right) V_{ub}V_{cd}^*$\\\hline
$\Omega_{bbb}^{-}\to \Xi_{bb}^{-}  D^-_s  K^+  $ & $ \left(k_6-k_4\right) V_{ub}V_{cd}^*$ & $\Omega_{bbb}^{-}\to \Xi_{b}^{\prime0}  B^-  \overline D^0 $ & $ \frac{\left(-k_{10}+2 k_{11}+k_{12}\right) V_{ub}V_{cs}^*}{\sqrt{2}}$\\\hline
$\Omega_{bbb}^{-}\to \Omega_{bb}^{-}  \overline D^0  \pi^0  $ & $ \frac{\left(k_6-k_4\right) V_{ub}V_{cs}^*}{\sqrt{2}}$ & $\Omega_{bbb}^{-}\to \Xi_{b}^{\prime0}  \overline B^0  D^- $ & $ \sqrt{2} k_{11} V_{ub}V_{cs}^*$\\\hline
$\Omega_{bbb}^{-}\to \Omega_{bb}^{-}  \overline D^0  K^0  $ & $ \left(k_3+k_5\right) V_{ub}V_{cd}^*$ & $\Omega_{bbb}^{-}\to \Xi_{b}^{\prime0}  \overline B^0_s  D^- $ & $ \frac{\left(k_{10}+k_{12}\right) V_{ub}V_{cd}^*}{\sqrt{2}}$\\\hline
$\Omega_{bbb}^{-}\to \Omega_{bb}^{-}  \overline D^0  \eta  $ & $ -\frac{\left(2 k_3+k_4+2 k_5-k_6\right) V_{ub}V_{cs}^*}{\sqrt{6}}$ & $\Omega_{bbb}^{-}\to \Xi_{b}^{\prime0}  \overline B^0_s  D^-_s $ & $ \frac{\left(k_{10}+2 k_{11}+k_{12}\right) V_{ub}V_{cs}^*}{\sqrt{2}}$\\\hline
$\Omega_{bbb}^{-}\to \Omega_{bb}^{-}  D^-  \pi^+  $ & $ \left(k_6-k_4\right) V_{ub}V_{cs}^*$ & $\Omega_{bbb}^{-}\to \Xi_{b}^{\prime-}  \overline B^0  \overline D^0 $ & $ \frac{\left(k_{12}-k_{10}\right) V_{ub}V_{cs}^*}{\sqrt{2}}$\\\hline
$\Omega_{bbb}^{-}\to \Omega_{bb}^{-}  D^-  K^+  $ & $ \left(k_5-k_3\right) V_{ub}V_{cd}^*$ & $\Omega_{bbb}^{-}\to \Xi_{b}^{\prime-}  \overline B^0_s  \overline D^0 $ & $ \frac{\left(k_{12}-k_{10}\right) V_{ub}V_{cd}^*}{\sqrt{2}}$\\\hline
$\Omega_{bbb}^{-}\to \Omega_{bb}^{-}  D^-_s  K^+  $ & $ \left(-k_3-k_4+k_5+k_6\right) V_{ub}V_{cs}^*$ & $\Omega_{bbb}^{-}\to \Omega_{b}^{-}  \overline B^0_s  \overline D^0 $ & $ \left(k_{12}-k_{10}\right) V_{ub}V_{cs}^*$\\\hline
\hline
\end{tabular}
\end{table}

For the anti-charm production, the operator having the quark contents $(\bar ub)(\bar qc)$  is given by
\begin{eqnarray}
{\cal H}_{e.w.} &=& \frac{G_{F}}{\sqrt{2}}
     V_{ub} V_{cq}^{*} \big[
     C_{1}  O^{\bar uc}_{1}
  +  C_{2}  O^{\bar uc}_{2}\Big]+ {\rm h.c.}.
\end{eqnarray}
The two light anti-quarks form the ${\bf  \bar 3}$ and ${\bf  6}$ representations.
The anti-symmetric tensor $H_{\bar 3}''$ and the symmetric tensor
$H_{ 6}$ have nonzero components
\begin{eqnarray}
 (H_{\bar 3}'')^{13} =- (H_{\bar 3}'')^{31} =V_{ub}V_{cs}^*,\;\;\; (H_{  6})^{13}=(H_{  6})^{31} =V_{ub}V_{cs}^*, \label{eq:btoucbarq}
\end{eqnarray}
for the $b\to u\bar cs$ transition. For the transition $b\to
u\bar cd$ one requests the interchange of $2\leftrightarrow 3$ in the
subscripts, and $V_{cs}$ replaced by $V_{cd}$.

The effective Hamiltonian is derived as:
\begin{eqnarray}
  {\cal H}_{\rm{eff}} &= &  k_1  \Omega_{bbb}  (\overline T_{bb})_i D_j (H_{\bar 3})^{ij} +k_2  \Omega_{bbb}  (\overline T_{bb})_i D_j (H_{6})^{ij}\nonumber\\
  && + k_3  \Omega_{bbb} (\overline T_{bb})_i D_j (M_8)^i_k (H_{\bar 3})^{jk} + k_4\Omega_{bbb}(\overline T_{bb})_i D_j (M_8)^j_k  (H_{\bar 3})^{ik} \nonumber\\
  && + k_5 \Omega_{bbb} (\overline T_{bb})_i D_j (M_8)^i_k (H_{6})^{jk} + k_6\Omega_{bbb} (\overline T_{bb})_i D_j (M_8)^j_k (H_{6})^{ik}   \nonumber\\
  &&+ k_7 \Omega_{bbb}(\overline T_{b\bar 3})_{[ij]} \overline B^k  D_k  (H_{\bar 3})^{ij} + k_{8}\Omega_{bbb}(\overline T_{b\bar 3})_{[ik]} \overline B^k  D_j(H_{\bar 3})^{ij}  \nonumber\\
  && + k_{9}  \Omega_{bbb} (\overline T_{b\bar 3})_{[ik]} \overline B^k  D_j (H_{6})^{ij} +k_{10}  \Omega_{bbb} (\overline T_{b6})_{\{ik\}} \overline B^k  D_j (H_{\bar 3})^{ij}\nonumber\\
  && +k_{11}   \Omega_{bbb}(\overline T_{b6})_{\{ij\}} \overline B^k  D_k (H_{6})^{ij}  +  k_{12}  \Omega_{bbb} (\overline T_{b6})_{\{ik\}} \overline B^k  D_j(H_{6})^{ij}.
\end{eqnarray}
Decay amplitudes for different channels are obtained by expanding the above Hamiltonian and are collected in Tab.~\ref{tab:bbb_C}.

\subsection{Charmless $b\to q_1 \bar q_2 q_3$ Decays}
\subsubsection{Decays into a doubly bottom baryon $bbq$ and a light meson}

 \begin{table}
\caption{Amplitudes for $\Omega_{bbb}$ decays into a doubly bottom baryon and a light meson. The $b\to d$ transitions are given in the left columns, and the $b\to s$ ones are shown in the right columns. }\label{tab:bbb_Dtwobody}
\begin{tabular}{|cc|cc|c|c|c|c}\hline\hline
channel & amplitude &channel & amplitude \\\hline
$\Omega_{bbb}^{-}\to \Xi_{bb}^{0}  \pi^-  $ & $ l_1+l_2+3 l_3$ & $\Omega_{bbb}^{-}\to \Xi_{bb}^{0}  K^-  $ & $ l_1^{\prime}+l_2^{\prime}+3 l_3^{\prime}$\\\hline
$\Omega_{bbb}^{-}\to \Xi_{bb}^{-}  \pi^0  $ & $ -\frac{l_1+l_2-5 l_3}{\sqrt{2}}$ & $\Omega_{bbb}^{-}\to \Xi_{bb}^{-}  \overline K^0  $ & $ l_1^{\prime}-l_2^{\prime}-l_3^{\prime}$\\\hline
$\Omega_{bbb}^{-}\to \Xi_{bb}^{-}  \eta  $ & $ \frac{l_1-3 l_2+3 l_3}{\sqrt{6}}$ & $\Omega_{bbb}^{-}\to \Omega_{bb}^{-}  \pi^0  $ & $ -\sqrt{2} \left(l_2^{\prime}-2 l_3^{\prime}\right)$\\\hline
$\Omega_{bbb}^{-}\to \Omega_{bb}^{-}  K^0  $ & $ l_1-l_2-l_3$ & $\Omega_{bbb}^{-}\to \Omega_{bb}^{-}  \eta  $ & $ -\sqrt{\frac{2}{3}} \left(l_1^{\prime}-3 l_3^{\prime}\right)$\\\hline
\hline
\end{tabular}
\end{table}

The  charmless $b\to q$ ($q=d,s$) transition is controlled by the weak Hamiltonian ${\cal H}_{eff}$:
 \begin{eqnarray}
 {\cal H}_{e.w.} &=& \frac{G_{F}}{\sqrt{2}}
     \bigg\{ V_{ub} V_{uq}^{*} \big[
     C_{1}  O^{\bar uu}_{1}
  +  C_{2}  O^{\bar uu}_{2}\Big]- V_{tb} V_{tq}^{*} \big[{\sum\limits_{i=3}^{10}} C_{i}  O_{i} \Big]\bigg\}+ \mbox{h.c.} ,
 \label{eq:hamiltonian}
\end{eqnarray}
where  $O_{i}$ is a four-quark operator or a moment type operator. In the SU(3) group,   penguin operators  behave as the ${\bf  3}$ representation while  tree operators   can
be decomposed in terms of a vector $H_{\bf 3}$, a traceless
tensor antisymmetric in upper indices, $H_{\bf\overline6}$, and a
traceless tensor symmetric in  upper indices,
$H_{\bf{15}}$.

  \begin{table}
\caption{Amplitudes for $\Omega_{bbb}$ decays into a doubly bottom baryon and two light mesons. The $b\to d$ transitions are given in the left columns, and the $b\to s$ ones are shown in the right columns.  The amplitude $l_7$ is neglected since  it correlates  with $l_6$.   }\label{tab:bbb_Dthreebody}
\begin{tabular}{|cc|cc|c|c|c|c}\hline\hline
channel & amplitude &channel & amplitude \\\hline
$\Omega_{bbb}^{-}\to \Xi_{bb}^{0}  \pi^-   \pi^0  $ & $ 4 \sqrt{2} l_9$ & $\Omega_{bbb}^{-}\to \Xi_{bb}^{0}  \overline K^0   \pi^-  $ & $ l^{\prime}_4+l^{\prime}_6 +3 l^{\prime}_8-l^{\prime}_9$\\\hline
$\Omega_{bbb}^{-}\to \Xi_{bb}^{0}  K^-   K^0  $ & $ l_4+l_6 +3 l_8-l_9$ & $\Omega_{bbb}^{-}\to \Xi_{bb}^{0}  K^-   \pi^0  $ & $ \frac{l^{\prime}_4+l^{\prime}_6 +3 l^{\prime}_8+7 l^{\prime}_9}{\sqrt{2}}$\\\hline
$\Omega_{bbb}^{-}\to \Xi_{bb}^{0}  \eta   \pi^-  $ & $ \sqrt{\frac{2}{3}} \left(l_4+l_6 +3 l_8+3 l_9\right)$ & $\Omega_{bbb}^{-}\to \Xi_{bb}^{0}  K^-   \eta  $ & $ -\frac{l^{\prime}_4+l^{\prime}_6 +3 l^{\prime}_8-9 l^{\prime}_9}{\sqrt{6}}$\\\hline
$\Omega_{bbb}^{-}\to \Xi_{bb}^{-}  \pi^0   \pi^0  $ & $ l_4+2 l_5-l_6 +l_8-5 l_9$ & $\Omega_{bbb}^{-}\to \Xi_{bb}^{-}  \pi^+   K^-  $ & $ l^{\prime}_4-l^{\prime}_6 -l^{\prime}_8+3 l^{\prime}_9$\\\hline
$\Omega_{bbb}^{-}\to \Xi_{bb}^{-}  \pi^-   \pi^+  $ & $ l_4+2 l_5-l_6 +l_8+3 l_9$ & $\Omega_{bbb}^{-}\to \Xi_{bb}^{-}  \overline K^0   \pi^0  $ & $ \frac{-l^{\prime}_4+l^{\prime}_6 +l^{\prime}_8+5 l^{\prime}_9}{\sqrt{2}}$\\\hline
$\Omega_{bbb}^{-}\to \Xi_{bb}^{-}  \overline K^0   K^0  $ & $ l_4+2 l_5+l_6 -3 l_8-l_9$ & $\Omega_{bbb}^{-}\to \Xi_{bb}^{-}  \overline K^0   \eta  $ & $ \frac{-l^{\prime}_4+l^{\prime}_6 +l^{\prime}_8+5 l^{\prime}_9}{\sqrt{6}}$\\\hline
$\Omega_{bbb}^{-}\to \Xi_{bb}^{-}  K^-   K^+  $ & $ 2 \left(l_5+l_8\right)$ & $\Omega_{bbb}^{-}\to \Omega_{bb}^{-}  \pi^+   \pi^-  $ & $ 2 \left(l^{\prime}_5+l^{\prime}_8\right)$\\\hline
$\Omega_{bbb}^{-}\to \Xi_{bb}^{-}  \eta   \pi^0  $ & $ \frac{-l_4-l_6 +5 l_8+l_9}{\sqrt{3}}$ & $\Omega_{bbb}^{-}\to \Omega_{bb}^{-}  \pi^0   \pi^0  $ & $ 2l^{\prime}_5+2l^{\prime}_8$\\\hline
$\Omega_{bbb}^{-}\to \Xi_{bb}^{-}  \eta   \eta  $ & $ \frac{1}{3} \left(l_4+3 \left(2 l_5+l_6 -l_8+l_9\right)\right)$ & $\Omega_{bbb}^{-}\to \Omega_{bb}^{-}  \pi^0   \eta  $ & $ -\frac{2 \left(l^{\prime}_6 -2 l^{\prime}_8+2 l^{\prime}_9\right)}{\sqrt{3}}$\\\hline
$\Omega_{bbb}^{-}\to \Omega_{bb}^{-}  \pi^0   K^0  $ & $ \frac{-l_4+l_6 +l_8+5 l_9}{\sqrt{2}}$ & $\Omega_{bbb}^{-}\to \Omega_{bb}^{-}  K^+   K^-  $ & $ l^{\prime}_4+2 l^{\prime}_5-l^{\prime}_6 +l^{\prime}_8+3 l^{\prime}_9$\\\hline
$\Omega_{bbb}^{-}\to \Omega_{bb}^{-}  K^+   \pi^-  $ & $ l_4-l_6 -l_8+3 l_9$ & $\Omega_{bbb}^{-}\to \Omega_{bb}^{-}  \overline K^0   K^0  $ & $ l^{\prime}_4+2 l^{\prime}_5+l^{\prime}_6 -3 l^{\prime}_8-l^{\prime}_9$\\\hline
$\Omega_{bbb}^{-}\to \Omega_{bb}^{-}  \eta   K^0  $ & $ \frac{-l_4+l_6 +l_8+5 l_9}{\sqrt{6}}$ & $\Omega_{bbb}^{-}\to \Omega_{bb}^{-}  \eta   \eta  $ & $ 2\left(\frac{2 l^{\prime}_4}{3}+l^{\prime}_5-l^{\prime}_8-2 l^{\prime}_9\right)$\\\hline
\hline
\end{tabular}
\end{table}

For the $\Delta S=0 (b\to d)$ decays, the non-zero components of the effective Hamiltonian are:
\begin{eqnarray}
 (H_3)^2=1,\;\;\;(H_{\overline6})^{12}_1=-(H_{\overline6})^{21}_1=(H_{\overline6})^{23}_3=-(H_{\overline6})^{32}_3=1,\nonumber\\
 2(H_{15})^{12}_1= 2(H_{15})^{21}_1=-3(H_{15})^{22}_2=
 -6(H_{15})^{23}_3=-6(H_{15})^{32}_3=6,\label{eq:H3615_bb}
\end{eqnarray}
and  all other remaining entries are zero. For the $\Delta S=1(b\to s)$
decays the nonzero entries in the $H_{\bf{3}}$, $H_{\bf\overline6}$,
$H_{\bf{15}}$ are obtained from Eq.~\eqref{eq:H3615_bb}
with the exchange  $2\leftrightarrow 3$.

If the final state contains one light meson, the effective Hamiltonian is given as:
\begin{eqnarray}
  {\cal H}_{\rm{eff}} &= &  l_1 \Omega_{bbb} (\overline T_{bb})_j (M_8)^j_i (H_{3})^i
 + l_2  \Omega_{bbb} (\overline T_{bb})_i  (M_8)^k_j (H_{\bar 6})^{ij}_k +l_3 \Omega_{bbb}(\overline T_{bb})_i  (M_8)^k_j (H_{15})^{ij}_k.
\end{eqnarray}
Decay amplitudes for different channels are obtained by expanding the above Hamiltonian and are collected in Tab.~\ref{tab:bbb_Dtwobody}.

With one additional light meson, we have
\begin{eqnarray}
  {\cal H}_{\rm{eff}} &= &  l_4\Omega_{bbb} (\overline T_{bb})_i  (M_8)^i_j (M_8)^j_k (H_{3})^k +l_5\Omega_{bbb}   (\overline T_{bb})_i  (M_8)^j_k  (M_8)^k_j (H_{3})^i \nonumber\\
  && + l_6 \Omega_{bbb} (\overline T_{bb})_i  (M_8)^k_l (M_8)^l_j (H_{\bar 6})^{ij}_k +l_7 \Omega_{bbb} (\overline T_{bb})_i  (M_8)^i_j (M_8)^l_k (H_{\bar 6})^{jk}_l \nonumber\\
  &&  + l_8 \Omega_{bbb} (\overline T_{bb})_i  (M_8)^k_l (M_8)^l_j (H_{15})^{ij}_k +l_9\Omega_{bbb} (\overline T_{bb})_i  (M_8)^i_j (M_8)^l_k (H_{15})^{jk}_l . \label{eq:bbb2bb2M}
\end{eqnarray}
Decay amplitudes for different channels are obtained by expanding the above Hamiltonian and are collected in Tab.~\ref{tab:bbb_Dthreebody}.
A few remarks are given in order.
\begin{itemize}
\item In Tab.~\ref{tab:bbb_Dtwobody} and Tab.~\ref{tab:bbb_Dthreebody}, both $b\to d$ and $b\to s$ channels are included. Since the CKM matrix elements are different, the SU(3) irreducible amplitudes for the $b\to s$ transition are primed.

\item Expanding Eq.~\eqref{eq:bbb2bb2M}, one can find the amplitudes $l_6$ and $l_7$ are not independent and they always appear in the product $l_6-l_7$.  So in the two tables, we did not show the $l_7$.

\item Inspired from the $B$ meson decay data, we can infer that the typical branching fractions are at the order $10^{-6}$. Thus these channels are rare decays, and can be studied with a large amount of data. However, the direct CP asymmetries in these channels are typically sizable.

\item
For  the $b\to q_1\bar q_2q_3$ decays, there are two amplitudes with different CKM factors. One can  consider the  U-spin connected decays with the decay amplitudes
\begin{eqnarray}
  A(\Delta S=0)&=& r\left( V_{ub}V^*_{ud}A^T+ V_{tb}V_{td}^*A^P  \right), \nonumber\\
   A(\Delta S=1)&=&  V_{ub}V^*_{us}A^T+ V_{tb}V_{ts}^*A^P .
\end{eqnarray}
where $r$ is a constant factor, and $A^T$ and $A^P$ are the amplitudes without the CKM factors.
Such channel pairs  include $[\Omega_{bbb}^-\to \Xi_{bb}^0\pi^-, \Omega_{bbb}^-\to \Xi_{bb}^0K^-]$, $[\Omega_{bbb}^-\to \Omega_{bb}^-K^0, \Omega_{bbb}^-\to \Omega_{bb}^-\overline K^0]$ and etc.
As pointed out in Refs.~\cite{Deshpande:1994ii,He:1998rq,Gronau:2000zy}, there exists a relation for the CP violating quantity $\Delta=\Gamma-\bar \Gamma$:
\begin{eqnarray}\label{cprelation}
  \frac{A_{CP}(\Delta S=0)}{A_{CP}(\Delta S=1)}= -r^2 \frac{\Gamma(\Delta S=1)}{\Gamma(\Delta S=0)}.
\end{eqnarray}
The future  experiment data   will be valuable   to test flavor SU(3) symmetry and  the  CKM mechanism for CP violation.

\end{itemize}


\subsubsection{ Decays into a bottom meson and a bottom  baryon $bqq$}

\begin{table}
\caption{Amplitudes for $\Omega_{bbb}$ decays into a bottom baryon (anti-triplet)}\label{tab:bbb_D2triplet}
\begin{tabular}{|cc|cc|c|c|c|c}\hline\hline
channel & amplitude &channel & amplitude \\\hline
$\Omega_{bbb}^{-}\to \Lambda_b^0  B^- $ & $ m_1+2 m_2$ & $\Omega_{bbb}^{-}\to \Xi_b^0  B^- $ & $ m^{\prime}_1+2 m^{\prime}_2$\\\hline
$\Omega_{bbb}^{-}\to \Xi_b^-  \overline B^0_s $ & $ 2 m_2-m_1$ & $\Omega_{bbb}^{-}\to \Xi_b^-  \overline B^0 $ & $ m^{\prime}_1-2 m^{\prime}_2$\\\hline
\Xcline{1-4}{1.2pt}
$\Omega_{bbb}^{-}\to \Lambda_b^0  B^-  \pi^0  $ & $ \frac{m_3-m_4+2 m_5-m_6-5 m_8+6 m_9}{\sqrt{2}}$ & $\Omega_{bbb}^{-}\to \Lambda_b^0  B^-  \overline K^0  $ & $ m^{\prime}_4-m^{\prime}_6-m^{\prime}_7+m^{\prime}_8-3 m^{\prime}_9$\\\hline
$\Omega_{bbb}^{-}\to \Lambda_b^0  B^-  \eta  $ & $ \frac{m_3+m_4+2 m_5-3 m_6-2 m_7-3 m_8}{\sqrt{6}}$ & $\Omega_{bbb}^{-}\to \Lambda_b^0  \overline B^0  K^-  $ & $ -m^{\prime}_4-m^{\prime}_6-m^{\prime}_7+3 m^{\prime}_8-m^{\prime}_9$\\\hline
$\Omega_{bbb}^{-}\to \Lambda_b^0  \overline B^0  \pi^-  $ & $ m_3-m_4+2 m_5-m_6+3 m_8-2 m_9$ & $\Omega_{bbb}^{-}\to \Xi_b^0  B^-  \pi^0  $ & $ \frac{m^{\prime}_3+2 m^{\prime}_5-2 m^{\prime}_6-m^{\prime}_7-4 m^{\prime}_8+3 m^{\prime}_9}{\sqrt{2}}$\\\hline
$\Omega_{bbb}^{-}\to \Lambda_b^0  \overline B^0_s  K^-  $ & $ m_3+2 m_5+m_7-m_9$ & $\Omega_{bbb}^{-}\to \Xi_b^0  B^-  \eta  $ & $ \frac{m^{\prime}_3-2 m^{\prime}_4+2 m^{\prime}_5+m^{\prime}_7-6 m^{\prime}_8+9 m^{\prime}_9}{\sqrt{6}}$\\\hline
$\Omega_{bbb}^{-}\to \Xi_b^0  B^-  K^0  $ & $ m_4-m_6-m_7+m_8-3 m_9$ & $\Omega_{bbb}^{-}\to \Xi_b^0  \overline B^0  \pi^-  $ & $ m^{\prime}_3+2 m^{\prime}_5+m^{\prime}_7-m^{\prime}_9$\\\hline
$\Omega_{bbb}^{-}\to \Xi_b^0  \overline B^0_s  \pi^-  $ & $ -m_4-m_6-m_7+3 m_8-m_9$ & $\Omega_{bbb}^{-}\to \Xi_b^0  \overline B^0_s  K^-  $ & $ m^{\prime}_3-m^{\prime}_4+2 m^{\prime}_5-m^{\prime}_6+3 m^{\prime}_8-2 m^{\prime}_9$\\\hline
$\Omega_{bbb}^{-}\to \Xi_b^-  B^-  K^+  $ & $ -m_3+2 m_5+m_7-3 m_9$ & $\Omega_{bbb}^{-}\to \Xi_b^-  B^-  \pi^+  $ & $ m^{\prime}_3-2 m^{\prime}_5-m^{\prime}_7+3 m^{\prime}_9$\\\hline
$\Omega_{bbb}^{-}\to \Xi_b^-  \overline B^0  K^0  $ & $ -m_3+m_4+2 m_5-m_6+m_8+2 m_9$ & $\Omega_{bbb}^{-}\to \Xi_b^-  \overline B^0  \pi^0  $ & $ -\frac{m^{\prime}_3-2 m^{\prime}_5+2 m^{\prime}_6+m^{\prime}_7+4 m^{\prime}_8-m^{\prime}_9}{\sqrt{2}}$\\\hline
$\Omega_{bbb}^{-}\to \Xi_b^-  \overline B^0_s  \pi^0  $ & $ \frac{m_4+m_6+m_7+5 m_8+m_9}{\sqrt{2}}$ & $\Omega_{bbb}^{-}\to \Xi_b^-  \overline B^0  \eta  $ & $ \frac{m^{\prime}_3-2 m^{\prime}_4-2 m^{\prime}_5-m^{\prime}_7-6 m^{\prime}_8-3 m^{\prime}_9}{\sqrt{6}}$\\\hline
$\Omega_{bbb}^{-}\to \Xi_b^-  \overline B^0_s  \eta  $ & $ \frac{2 m_3-m_4-4 m_5+3 m_6+m_7+3 m_8-3 m_9}{\sqrt{6}}$ & $\Omega_{bbb}^{-}\to \Xi_b^-  \overline B^0_s  \overline K^0  $ & $ m^{\prime}_3-m^{\prime}_4-2 m^{\prime}_5+m^{\prime}_6-m^{\prime}_8-2 m^{\prime}_9$\\\hline
\hline
\end{tabular}
\end{table}

\begin{table}
\caption{Amplitudes for $\Omega_{bbb}$ decays into a bottom baryon (sextet)}\label{tab:bbb_D2sextet}
\begin{tabular}{|cc|cc|c|c|c|c}\hline\hline
channel & amplitude &channel & amplitude \\\hline
$\Omega_{bbb}^{-}\to \Sigma_{b}^{0}  B^- $ & $ \frac{n_1+6 n_2}{\sqrt{2}}$ & $\Omega_{bbb}^{-}\to \Xi_{b}^{\prime0}  B^- $ & $ \frac{n^{\prime}_1+6 n^{\prime}_2}{\sqrt{2}}$\\\hline
$\Omega_{bbb}^{-}\to \Sigma_{b}^{-}  \overline B^0 $ & $ n_1-2 n_2$ & $\Omega_{bbb}^{-}\to \Xi_{b}^{\prime-}  \overline B^0 $ & $ \frac{n^{\prime}_1-2 n^{\prime}_2}{\sqrt{2}}$\\\hline
$\Omega_{bbb}^{-}\to \Xi_{b}^{\prime-}  \overline B^0_s $ & $ \frac{n_1-2 n_2}{\sqrt{2}}$ & $\Omega_{bbb}^{-}\to \Omega_{b}^{-}  \overline B^0_s $ & $ n^{\prime}_1-2 n^{\prime}_2$\\\hline
\Xcline{1-4}{1.2pt}
$\Omega_{bbb}^{-}\to \Sigma_{b}^{+}  B^-  \pi^-  $ & $ n_4+3 n_6+3 n_7+n_8+n_9$ & $\Omega_{bbb}^{-}\to \Sigma_{b}^{+}  B^-  K^-  $ & $ n^{\prime}_4+3 n^{\prime}_6+3 n^{\prime}_7+n^{\prime}_8+n^{\prime}_9$\\\hline
$\Omega_{bbb}^{-}\to \Sigma_{b}^{0}  B^-  \pi^0  $ & $ \frac{1}{2} \left(n_3-n_4+6 n_5+5 n_6-n_8-2 n_9\right)$ & $\Omega_{bbb}^{-}\to \Sigma_{b}^{0}  B^-  \overline K^0  $ & $ \frac{n^{\prime}_4-n^{\prime}_6+3 n^{\prime}_7-n^{\prime}_8+n^{\prime}_9}{\sqrt{2}}$\\\hline
$\Omega_{bbb}^{-}\to \Sigma_{b}^{0}  B^-  \eta  $ & $ \frac{n_3+n_4+6 n_5+3 n_6+6 n_7-3 n_8}{2 \sqrt{3}}$ & $\Omega_{bbb}^{-}\to \Sigma_{b}^{0}  \overline B^0  K^-  $ & $ \frac{n^{\prime}_4+3 n^{\prime}_6-n^{\prime}_7+n^{\prime}_8-n^{\prime}_9}{\sqrt{2}}$\\\hline
$\Omega_{bbb}^{-}\to \Sigma_{b}^{0}  \overline B^0  \pi^-  $ & $ \frac{n_3+n_4+6 n_5+3 n_6-2 n_7+n_8}{\sqrt{2}}$ & $\Omega_{bbb}^{-}\to \Sigma_{b}^{-}  \overline B^0  \overline K^0  $ & $ n^{\prime}_4-n^{\prime}_6-n^{\prime}_7-n^{\prime}_8-n^{\prime}_9$\\\hline
$\Omega_{bbb}^{-}\to \Sigma_{b}^{0}  \overline B^0_s  K^-  $ & $ \frac{n_3+6 n_5-n_7+n_9}{\sqrt{2}}$ & $\Omega_{bbb}^{-}\to \Xi_{b}^{\prime0}  B^-  \pi^0  $ & $ \frac{1}{2} \left(n^{\prime}_3+6 n^{\prime}_5+4 n^{\prime}_6+3 n^{\prime}_7-2 n^{\prime}_8-n^{\prime}_9\right)$\\\hline
$\Omega_{bbb}^{-}\to \Sigma_{b}^{-}  B^-  \pi^+  $ & $ n_3-2 n_5+3 n_7-n_9$ & $\Omega_{bbb}^{-}\to \Xi_{b}^{\prime0}  B^-  \eta  $ & $ \frac{n^{\prime}_3-2 n^{\prime}_4+6 n^{\prime}_5+6 n^{\prime}_6-3 n^{\prime}_7-3 n^{\prime}_9}{2 \sqrt{3}}$\\\hline
$\Omega_{bbb}^{-}\to \Sigma_{b}^{-}  \overline B^0  \pi^0  $ & $ -\frac{n_3+n_4-2 n_5-5 n_6-2 n_7+n_8}{\sqrt{2}}$ & $\Omega_{bbb}^{-}\to \Xi_{b}^{\prime0}  \overline B^0  \pi^-  $ & $ \frac{n^{\prime}_3+6 n^{\prime}_5-n^{\prime}_7+n^{\prime}_9}{\sqrt{2}}$\\\hline
$\Omega_{bbb}^{-}\to \Sigma_{b}^{-}  \overline B^0  \eta  $ & $ \frac{n_3+n_4-2 n_5+3 n_6-2 n_7-3 n_8}{\sqrt{6}}$ & $\Omega_{bbb}^{-}\to \Xi_{b}^{\prime0}  \overline B^0_s  K^-  $ & $ \frac{n^{\prime}_3+n^{\prime}_4+6 n^{\prime}_5+3 n^{\prime}_6-2 n^{\prime}_7+n^{\prime}_8}{\sqrt{2}}$\\\hline
$\Omega_{bbb}^{-}\to \Sigma_{b}^{-}  \overline B^0_s  \overline K^0  $ & $ n_3-2 n_5-n_7+n_9$ & $\Omega_{bbb}^{-}\to \Xi_{b}^{\prime-}  B^-  \pi^+  $ & $ \frac{n^{\prime}_3-2 n^{\prime}_5+3 n^{\prime}_7-n^{\prime}_9}{\sqrt{2}}$\\\hline
$\Omega_{bbb}^{-}\to \Xi_{b}^{\prime0}  B^-  K^0  $ & $ \frac{n_4-n_6+3 n_7-n_8+n_9}{\sqrt{2}}$ & $\Omega_{bbb}^{-}\to \Xi_{b}^{\prime-}  \overline B^0  \pi^0  $ & $ \frac{1}{2} \left(-n^{\prime}_3+2 n^{\prime}_5+4 n^{\prime}_6+n^{\prime}_7-2 n^{\prime}_8-n^{\prime}_9\right)$\\\hline
$\Omega_{bbb}^{-}\to \Xi_{b}^{\prime0}  \overline B^0_s  \pi^-  $ & $ \frac{n_4+3 n_6-n_7+n_8-n_9}{\sqrt{2}}$ & $\Omega_{bbb}^{-}\to \Xi_{b}^{\prime-}  \overline B^0  \eta  $ & $ \frac{n^{\prime}_3-2 n^{\prime}_4-2 n^{\prime}_5+6 n^{\prime}_6+n^{\prime}_7+3 n^{\prime}_9}{2 \sqrt{3}}$\\\hline
$\Omega_{bbb}^{-}\to \Xi_{b}^{\prime-}  B^-  K^+  $ & $ \frac{n_3-2 n_5+3 n_7-n_9}{\sqrt{2}}$ & $\Omega_{bbb}^{-}\to \Xi_{b}^{\prime-}  \overline B^0_s  \overline K^0  $ & $ \frac{n^{\prime}_3+n^{\prime}_4-2 n^{\prime}_5-n^{\prime}_6-2 n^{\prime}_7-n^{\prime}_8}{\sqrt{2}}$\\\hline
$\Omega_{bbb}^{-}\to \Xi_{b}^{\prime-}  \overline B^0  K^0  $ & $ \frac{n_3+n_4-2 n_5-n_6-2 n_7-n_8}{\sqrt{2}}$ & $\Omega_{bbb}^{-}\to \Omega_{b}^{-}  B^-  K^+  $ & $ n^{\prime}_3-2 n^{\prime}_5+3 n^{\prime}_7-n^{\prime}_9$\\\hline
$\Omega_{bbb}^{-}\to \Xi_{b}^{\prime-}  \overline B^0_s  \pi^0  $ & $ \frac{1}{2} \left(-n_4+5 n_6+n_7-n_8+n_9\right)$ & $\Omega_{bbb}^{-}\to \Omega_{b}^{-}  \overline B^0  K^0  $ & $ n^{\prime}_3-2 n^{\prime}_5-n^{\prime}_7+n^{\prime}_9$\\\hline
$\Omega_{bbb}^{-}\to \Xi_{b}^{\prime-}  \overline B^0_s  \eta  $ & $ \frac{-2 n_3+n_4+4 n_5+3 n_6+n_7-3 n_8-3 n_9}{2 \sqrt{3}}$ & $\Omega_{bbb}^{-}\to \Omega_{b}^{-}  \overline B^0_s  \pi^0  $ & $ \sqrt{2} \left(2 n^{\prime}_6-n^{\prime}_8\right)$\\\hline
$\Omega_{bbb}^{-}\to \Omega_{b}^{-}  \overline B^0_s  K^0  $ & $ n_4-n_6-n_7-n_8-n_9$ & $\Omega_{bbb}^{-}\to \Omega_{b}^{-}  \overline B^0_s  \eta  $ & $ -\sqrt{\frac{2}{3}} \left(n^{\prime}_3+n^{\prime}_4-2 n^{\prime}_5-3 n^{\prime}_6-2 n^{\prime}_7\right)$\\\hline
\hline
\end{tabular}
\end{table}

If the bottom baryon is an anti-triplet, we have the effective Hamiltonian for two-body and three-body decays:
\begin{eqnarray}
{\cal H}_{\rm eff}&=& m_1 \Omega_{bbb} (\overline  T_{\bf b\bar 3})_{[ij]} \overline B^i (H_{3})^j + m_2 \Omega_{bbb} (\overline  T_{\bf b\bar 3})_{[ij]} \overline B^k (H_{\bar 6})^{ij}_k \nonumber\\
&&+m_3 \Omega_{bbb} (\overline  T_{\bf b\bar 3})_{[ij]} \overline B^k (M_8)_k^i  (H_{3})^j + m_4 \Omega_{bbb} (\overline  T_{\bf b\bar 3})_{[ij]} \overline B^i (M_8)_k^j  (H_{3})^k
\nonumber\\
&&+m_5\Omega_{bbb}  (\overline  T_{\bf b\bar 3})_{[ij]} \overline B^l (M_8)^k_l (H_{\bar 6})^{ij}_k +m_6\Omega_{bbb}  (\overline  T_{\bf b\bar 3})_{[ij]} \overline B^i (M_8)^k_l (H_{\bar 6})^{jl}_k   \nonumber\\
&& +m_7\Omega_{bbb}  (\overline  T_{\bf b\bar 3})_{[ij]} \overline B^k (M_8)^i_l (H_{\bar 6})^{jl}_k +m_{8}\Omega_{bbb}  (\overline  T_{\bf b\bar 3})_{[ij]} \overline B^j (M_8)^l_k (H_{15})^{ik}_l \nonumber\\
&& +m_{9}\Omega_{bbb}  (\overline  T_{\bf b\bar 3})_{[ij]} \overline B^l (M_8)^i_k (H_{15})^{jk}_l.
\end{eqnarray}
Decay amplitudes for different channels are obtained by expanding the above Hamiltonian and are collected in Tab.~\ref{tab:bbb_D2triplet}.

In the case of a sextet, we have
\begin{eqnarray}
{\cal H}_{\rm eff}&=& n_1 \Omega_{bbb} (\overline  T_{\bf b6})_{\{ij\}} \overline B^i (H_{3})^j + n_2 \Omega_{bbb} (\overline  T_{\bf b6})_{\{ij\}} \overline B^k (H_{15})^{ij}_k \nonumber\\
&&+n_3 \Omega_{bbb} (\overline  T_{\bf b6})_{\{ij\}} \overline B^k (M_8)_k^i  (H_{3})^j+ n_4 \Omega_{bbb} (\overline  T_{\bf b6})_{\{ij\}} \overline B^i (M_8)_k^j  (H_{3})^k\nonumber\\
&&
+n_5\Omega_{bbb}  (\overline  T_{\bf b6})_{\{ij\}} \overline B^l (M_8)^k_l (H_{15})^{ij}_k +n_6\Omega_{bbb}  (\overline  T_{\bf b6})_{\{ij\}} \overline B^i (M_8)^k_l (H_{15})^{jl}_k   \nonumber\\
&& +n_7\Omega_{bbb}  (\overline  T_{\bf b6})_{\{ij\}} \overline B^k (M_8)^i_l (H_{15})^{jl}_k +n_{8}\Omega_{bbb}  (\overline  T_{\bf b6})_{\{ij\}} \overline B^j (M_8)^l_k (H_{\bar 6})^{ik}_l \nonumber\\
&& +n_{9}\Omega_{bbb}  (\overline  T_{\bf b6})_{\{ij\}} \overline B^l (M_8)^i_k (H_{\bar 6})^{jk}_l .
\end{eqnarray}
Decay amplitudes for different channels are obtained by expanding the above Hamiltonian and are collected in Tab.~\ref{tab:bbb_D2sextet}.

\section{Non-Leptonic   $\Omega_{ccb}$ and $\Omega_{cbb}$ decays}
\label{sec:ccb_cbb_nonleptonic}

\begin{table}
\caption{Amplitudes for $\Omega_{cbb}$ decays into a singly bottom baryon (triplet)}\label{tab:cbb_1triplet}
\begin{tabular}{|cc|cc|c|c|c|c}\hline\hline
channel & amplitude &channel & amplitude \\\hline
$\Omega_{cbb}^{0}\to \Lambda_b^0  \pi^0  $ & $ -\sqrt{2} a_5 V_{cd}^*$ & $\Omega_{cbb}^{0}\to \Xi_b^0  K^0  $ & $ \left(a_1+a_5\right) V_{cd}^*$\\\hline
$\Omega_{cbb}^{0}\to \Lambda_b^0  \overline K^0  $ & $ \left(a_1+a_5\right) V_{cs}^*$ & $\Omega_{cbb}^{0}\to \Xi_b^0  \eta  $ & $ -\frac{\left(a_1+3 a_5\right) V_{cs}^*}{\sqrt{6}}$\\\hline
$\Omega_{cbb}^{0}\to \Lambda_b^0  \eta  $ & $ \sqrt{\frac{2}{3}} a_1 V_{cd}^*$ & $\Omega_{cbb}^{0}\to \Xi_b^-  \pi^+  $ & $ \left(a_1-a_5\right) V_{cs}^*$\\\hline
$\Omega_{cbb}^{0}\to \Xi_b^0  \pi^0  $ & $ \frac{\left(a_1-a_5\right) V_{cs}^*}{\sqrt{2}}$ & $\Omega_{cbb}^{0}\to \Xi_b^-  K^+  $ & $ \left(a_5-a_1\right) V_{cd}^*$\\\hline
\hline
$\Omega_{cbb}^{0}\to \Lambda_b^0  \pi^0   \pi^0  $ & $ 2\left(2 a_2+a_3-a_4\right) V_{cd}^*$ & $\Omega_{cbb}^{0}\to \Xi_b^0  K^+   \pi^-  $ & $ \left(a_3-2 a_4+a_6\right) V_{cd}^*$\\\hline
$\Omega_{cbb}^{0}\to \Lambda_b^0  \pi^0   \overline K^0  $ & $ -\frac{\left(a_3-2 a_4+a_6\right) V_{cs}^*}{\sqrt{2}}$ & $\Omega_{cbb}^{0}\to \Xi_b^0  K^0   \pi^0  $ & $ -\frac{\left(a_3-2 a_4+a_6\right) V_{cd}^*}{\sqrt{2}}$\\\hline
$\Omega_{cbb}^{0}\to \Lambda_b^0  \pi^-   \pi^+  $ & $ 2 \left(2 a_2+a_3-a_4\right) V_{cd}^*$ & $\Omega_{cbb}^{0}\to \Xi_b^0  \overline K^0   K^0  $ & $ \left(4 a_2+a_3+a_6\right) V_{cs}^*$\\\hline
$\Omega_{cbb}^{0}\to \Lambda_b^0  K^+   K^-  $ & $ \left(4 a_2+a_3-a_6\right) V_{cd}^*$ & $\Omega_{cbb}^{0}\to \Xi_b^0  K^-   K^+  $ & $ 2 \left(2 a_2+a_3-a_4\right) V_{cs}^*$\\\hline
$\Omega_{cbb}^{0}\to \Lambda_b^0  K^0   \overline K^0  $ & $ \left(4 a_2+a_3+a_6\right) V_{cd}^*$ & $\Omega_{cbb}^{0}\to \Xi_b^0  \eta   K^0  $ & $ -\frac{\left(a_3-2 a_4+a_6\right) V_{cd}^*}{\sqrt{6}}$\\\hline
$\Omega_{cbb}^{0}\to \Lambda_b^0  \overline K^0   \eta  $ & $ -\frac{\left(a_3-2 a_4+a_6\right) V_{cs}^*}{\sqrt{6}}$ & $\Omega_{cbb}^{0}\to \Xi_b^0  \eta   \eta  $ & $ \frac{1}{3} \left(12 a_2+5 a_3-4 a_4+3 a_6\right) V_{cs}^*$\\\hline
$\Omega_{cbb}^{0}\to \Lambda_b^0  K^-   \pi^+  $ & $ \left(a_3-2 a_4+a_6\right) V_{cs}^*$ & $\Omega_{cbb}^{0}\to \Xi_b^-  \pi^+   K^0  $ & $ \left(-a_3+2 a_4+a_6\right) V_{cd}^*$\\\hline
$\Omega_{cbb}^{0}\to \Lambda_b^0  \eta   \pi^0  $ & $ -\frac{2 a_6 V_{cd}^*}{\sqrt{3}}$ & $\Omega_{cbb}^{0}\to \Xi_b^-  \pi^+   \eta  $ & $ \sqrt{\frac{2}{3}} \left(a_3-2 a_4-a_6\right) V_{cs}^*$\\\hline
$\Omega_{cbb}^{0}\to \Lambda_b^0  \eta   \eta  $ & $ \frac{2}{3} \left(6 a_2+a_3+a_4\right) V_{cd}^*$ & $\Omega_{cbb}^{0}\to \Xi_b^-  K^+   \pi^0  $ & $ \frac{\left(-a_3+2 a_4+a_6\right) V_{cd}^*}{\sqrt{2}}$\\\hline
$\Omega_{cbb}^{0}\to \Xi_b^0  \pi^+   \pi^-  $ & $ \left(4 a_2+a_3-a_6\right) V_{cs}^*$ & $\Omega_{cbb}^{0}\to \Xi_b^-  K^+   \eta  $ & $ \frac{\left(a_3-2 a_4-a_6\right) V_{cd}^*}{\sqrt{6}}$\\\hline
$\Omega_{cbb}^{0}\to \Xi_b^0  \pi^0   \pi^0  $ & $ \left(4 a_2+a_3-a_6\right) V_{cs}^*$ & $\Omega_{cbb}^{0}\to \Xi_b^-  \overline K^0   K^+  $ & $ \left(a_3-2 a_4-a_6\right) V_{cs}^*$\\\hline
$\Omega_{cbb}^{0}\to \Xi_b^0  \pi^0   \eta  $ & $ \frac{\left(a_3-2 a_4-a_6\right) V_{cs}^*}{\sqrt{3}}$ &&\\\hline
\hline
\end{tabular}
\end{table}

\begin{table}
\caption{Amplitudes for $\Omega_{cbb}$ decays into a singly bottom baryon (sextet)}\label{tab:cbb_1sextet}
\begin{tabular}{|cc|cc|c|c|c|c}\hline\hline
channel & amplitude &channel & amplitude \\\hline
$\Omega_{cbb}^{0}\to \Sigma_{b}^{+}  \pi^-  $ & $ \left(b_1+b_4\right) V_{cd}^*$ & $\Omega_{cbb}^{0}\to \Xi_{b}^{\prime0}  \pi^0  $ & $ \frac{1}{2} \left(b_4-b_1\right) V_{cs}^*$\\\hline
$\Omega_{cbb}^{0}\to \Sigma_{b}^{+}  K^-  $ & $ \left(b_1+b_4\right) V_{cs}^*$ & $\Omega_{cbb}^{0}\to \Xi_{b}^{\prime0}  K^0  $ & $ \frac{\left(b_1+b_4\right) V_{cd}^*}{\sqrt{2}}$\\\hline
$\Omega_{cbb}^{0}\to \Sigma_{b}^{0}  \pi^0  $ & $ -b_1 V_{cd}^*$ & $\Omega_{cbb}^{0}\to \Xi_{b}^{\prime0}  \eta  $ & $ -\frac{\left(3 b_1+b_4\right) V_{cs}^*}{2 \sqrt{3}}$\\\hline
$\Omega_{cbb}^{0}\to \Sigma_{b}^{0}  \overline K^0  $ & $ \frac{\left(b_1+b_4\right) V_{cs}^*}{\sqrt{2}}$ & $\Omega_{cbb}^{0}\to \Xi_{b}^{\prime-}  \pi^+  $ & $ \frac{\left(b_4-b_1\right) V_{cs}^*}{\sqrt{2}}$\\\hline
$\Omega_{cbb}^{0}\to \Sigma_{b}^{0}  \eta  $ & $ \frac{b_4 V_{cd}^*}{\sqrt{3}}$ & $\Omega_{cbb}^{0}\to \Xi_{b}^{\prime-}  K^+  $ & $ \frac{\left(b_4-b_1\right) V_{cd}^*}{\sqrt{2}}$\\\hline
$\Omega_{cbb}^{0}\to \Sigma_{b}^{-}  \pi^+  $ & $ \left(b_4-b_1\right) V_{cd}^*$ & $\Omega_{cbb}^{0}\to \Omega_{b}^{-}  K^+  $ & $ \left(b_4-b_1\right) V_{cs}^*$\\\hline
\hline
$\Omega_{cbb}^{0}\to \Sigma_{b}^{-}  \pi^+   \overline K^0  $ & $ 2 b_7 V_{cs}^*$   & $\Omega_{cbb}^{0}\to \Sigma_{b}^{-}  \overline K^0   K^+  $ & $ \left(b_6-b_2\right) V_{cd}^*$\\\hline
$\Omega_{cbb}^{0}\to \Sigma_{b}^{+}  \pi^0   K^-  $ & $ \frac{\left(b_2+b_6+2 b_7\right) V_{cs}^*}{\sqrt{2}}$ & $\Omega_{cbb}^{0}\to \Sigma_{b}^{-}  \eta   \pi^+  $ & $ \sqrt{\frac{2}{3}} \left(-b_2+b_6+b_7\right) V_{cd}^*$\\\hline
$\Omega_{cbb}^{0}\to \Sigma_{b}^{+}  \pi^-   \pi^0  $ & $ \sqrt{2} b_7 V_{cd}^*$ & $\Omega_{cbb}^{0}\to \Xi_{b}^{\prime0}  \pi^+   \pi^-  $ & $ \frac{\left(-b_2+4 b_5+b_6\right) V_{cs}^*}{\sqrt{2}}$\\\hline
$\Omega_{cbb}^{0}\to \Sigma_{b}^{+}  \pi^-   \overline K^0  $ & $ \left(b_2+b_6\right) V_{cs}^*$ & $\Omega_{cbb}^{0}\to \Xi_{b}^{\prime0}  \pi^0   \pi^0  $ & $ \frac{\left(-b_2+4 b_5+b_6\right) V_{cs}^*}{ \sqrt{2}}$\\\hline
$\Omega_{cbb}^{0}\to \Sigma_{b}^{+}  \pi^-   \eta  $ & $ \sqrt{\frac{2}{3}} \left(b_2+b_6+b_7\right) V_{cd}^*$ & $\Omega_{cbb}^{0}\to \Xi_{b}^{\prime0}  \pi^0   K^0  $ & $ -\frac{1}{2} \left(b_2+b_6-2 b_7\right) V_{cd}^*$\\\hline
$\Omega_{cbb}^{0}\to \Sigma_{b}^{+}  K^0   K^-  $ & $ \left(b_2+b_6\right) V_{cd}^*$ & $\Omega_{cbb}^{0}\to \Xi_{b}^{\prime0}  \pi^0   \eta  $ & $ \frac{\left(-b_2+b_6-2 b_7\right) V_{cs}^*}{\sqrt{6}}$\\\hline
$\Omega_{cbb}^{0}\to \Sigma_{b}^{0}  \eta   \eta  $ & $ \frac{2\left(6 b_5+b_6+b_7\right) V_{cd}^*}{3 \sqrt{2}}$ & $\Omega_{cbb}^{0}\to \Xi_{b}^{\prime0}  K^+   \pi^-  $ & $ \frac{\left(b_2+b_6+2 b_7\right) V_{cd}^*}{\sqrt{2}}$\\\hline
$\Omega_{cbb}^{0}\to \Sigma_{b}^{+}  K^-   \eta  $ & $ -\frac{\left(b_2+b_6-2 b_7\right) V_{cs}^*}{\sqrt{6}}$ & $\Omega_{cbb}^{0}\to \Xi_{b}^{\prime0}  K^+   K^-  $ & $ \sqrt{2} \left(2 b_5+b_6+b_7\right) V_{cs}^*$\\\hline
$\Omega_{cbb}^{0}\to \Omega_{b}^{-}  K^+   \eta  $ & $ \frac{\left(b_2-b_6-4 b_7\right) V_{cs}^*}{\sqrt{6}}$ & $\Omega_{cbb}^{0}\to \Xi_{b}^{\prime0}  K^0   \overline K^0  $ & $ \frac{\left(b_2+4 b_5+b_6\right) V_{cs}^*}{\sqrt{2}}$\\\hline
$\Omega_{cbb}^{0}\to \Sigma_{b}^{0}  \pi^+   \pi^-  $ & $ \sqrt{2} \left(2 b_5+b_6+b_7\right) V_{cd}^*$ & $\Omega_{cbb}^{0}\to \Xi_{b}^{\prime0}  K^0   \eta  $ & $ -\frac{\left(b_2+b_6-2 b_7\right) V_{cd}^*}{2 \sqrt{3}}$\\\hline
$\Omega_{cbb}^{0}\to \Sigma_{b}^{0}  \pi^+   K^-  $ & $ \frac{\left(b_2+b_6+2 b_7\right) V_{cs}^*}{\sqrt{2}}$ & $\Omega_{cbb}^{0}\to \Omega_{b}^{-}  K^+   \pi^0  $ & $ \frac{\left(b_6-b_2\right) V_{cs}^*}{\sqrt{2}}$ \\\hline
$\Omega_{cbb}^{0}\to \Sigma_{b}^{0}  \pi^0   \pi^0  $ & $ \frac{2\left(2 b_5+b_6-b_7\right) V_{cd}^*}{\sqrt{2}}$ & $\Omega_{cbb}^{0}\to \Omega_{b}^{-}  \pi^+   K^0  $ & $ \left(b_6-b_2\right) V_{cs}^*$ \\\hline
$\Omega_{cbb}^{0}\to \Sigma_{b}^{0}  \pi^0   \overline K^0  $ & $ -\frac{1}{2} \left(b_2+b_6-2 b_7\right) V_{cs}^*$ & $\Omega_{cbb}^{0}\to \Xi_{b}^{\prime-}  K^+   \eta  $ & $ \frac{\left(b_2-b_6+2 b_7\right) V_{cd}^*}{2 \sqrt{3}}$ \\\hline
$\Omega_{cbb}^{0}\to \Sigma_{b}^{0}  \pi^0   \eta  $ & $ -\sqrt{\frac{2}{3}} b_2 V_{cd}^*$ & $\Omega_{cbb}^{0}\to \Xi_{b}^{\prime-}  K^+   \overline K^0  $ & $ \frac{\left(-b_2+b_6+2 b_7\right) V_{cs}^*}{\sqrt{2}}$ \\\hline
$\Omega_{cbb}^{0}\to \Omega_{b}^{-}  K^+   K^0  $ & $ 2 b_7 V_{cd}^*$ & $\Omega_{cbb}^{0}\to \Xi_{b}^{\prime0}  \eta   \eta  $ & $ \frac{\left(3 b_2+12 b_5+5 b_6-4 b_7\right) V_{cs}^*}{3 \sqrt{2}}$\\\hline
$\Omega_{cbb}^{0}\to \Sigma_{b}^{0}  K^+   K^-  $ & $ \frac{\left(-b_2+4 b_5+b_6\right) V_{cd}^*}{\sqrt{2}}$ & $\Omega_{cbb}^{0}\to \Xi_{b}^{\prime-}  \pi^+   K^0  $ & $ \frac{\left(-b_2+b_6+2 b_7\right) V_{cd}^*}{\sqrt{2}}$\\\hline
$\Omega_{cbb}^{0}\to \Sigma_{b}^{0}  K^0   \overline K^0  $ & $ \frac{\left(b_2+4 b_5+b_6\right) V_{cd}^*}{\sqrt{2}}$ & $\Omega_{cbb}^{0}\to \Xi_{b}^{\prime-}  \pi^+   \eta  $ & $ \frac{\left(-b_2+b_6-2 b_7\right) V_{cs}^*}{\sqrt{3}}$\\\hline
$\Omega_{cbb}^{0}\to \Sigma_{b}^{-}  \pi^+   \pi^0  $ & $ -\sqrt{2} b_7 V_{cd}^*$ & $\Omega_{cbb}^{0}\to \Xi_{b}^{\prime-}  K^+   \pi^0  $ & $ -\frac{1}{2} \left(b_2-b_6+2 b_7\right) V_{cd}^*$\\\hline
$\Omega_{cbb}^{0}\to \Sigma_{b}^{0}  \overline K^0   \eta  $ & $ -\frac{\left(b_2+b_6-2 b_7\right) V_{cs}^*}{2 \sqrt{3}}$ & & \\\hline
\hline
\end{tabular}
\end{table}

For the mixed triply heavy baryons, $\Omega_{ccb}$ and $\Omega_{cbb}$, most of their weak decays can be obtained from the $\Omega_{ccc}$ and $\Omega_{bbb}$ decay channels with some replacements. For instance, decays of $\Omega_{ccb}$ induced by the charm quark can be obtained from the ones of $\Omega_{ccc}$ by replacing one charmed meson by the corresponding bottom meson,  or a charmed baryon by the corresponding   bottom baryon, or a doubly charmed baryon $H_{cc}$  by its  counterpart $H_{bc}$.

\begin{figure}
\includegraphics[width=0.5\columnwidth]{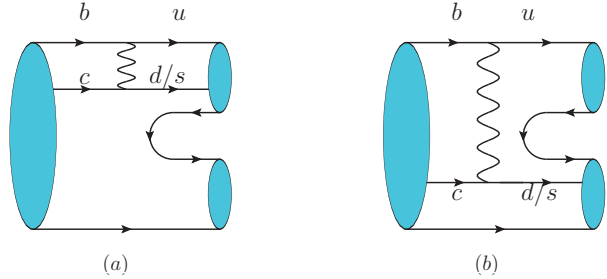}
\caption{ Feynman diagrams for W-exchange. The spectator is a bottom or a charm quark.  If the final $u$ quark is replaced by a charm quark, the W-exchange contribution is an SU(3) triplet, and this triplet contribution has been incorporated in the $b\to q_1\bar q_2 q_3$.  }
\label{fig:W-exchange}
\end{figure}

In addition, there   are  new decay modes, which are   induced by W-exchange transition shown in Fig.~\ref{fig:W-exchange}~\footnote{If the final $u$ quark is replaced by a charm quark, the W-exchange contribution is an SU(3) triplet, and this triplet contribution has been incorporated in the $b\to q_1\bar q_2 q_3$. In this case, the CKM matrix element is $V_{cb}V_{cd/cs}^*$.  },  $bc\to u d$ or $bc\to us$, with two heavy quarks annihilating into two light quarks. The spectator quark is a bottom or charm quark.  These diagrams  are dynamically suppressed by factors of  $1/m_{b,c}$.   The electroweak  Hamiltonian   is similar with Eq.~\eqref{eq:btoucbarq}. Taking  $\Omega_{cbb}$ as the example, one should notice that the final state contains only one heavy bottom quark. Thus at hadron level, the final state can be a bottom baryon (antitriplet) which has:
\begin{eqnarray}
 {\cal H}_{\rm eff} &=& a_1 \Omega_{cbb} (\overline T_{b\bar 3})_{[ik]} (M_8)^{k}_{j} (H_{\bar 3}'')^{[ij]} + a_2 \Omega_{cbb} (\overline T_{b\bar 3})_{[ij]} (M_8)^{k}_{l} (M_8)^{l}_{k} (H_{\bar 3}'')^{[ij]} \nonumber\\
 &&  +  a_3 \Omega_{cbb}  (\overline T_{b\bar 3})_{[ik]} (M_8)^{k}_{l} (M_8)^{l}_{j} (H_{\bar 3}'')^{[ij]}+a_4 \Omega_{cbb}  (\overline T_{b\bar 3})_{[kl]} (M_8)^{k}_{i} (M_8)^{l}_{j}(H_{\bar 3}'')^{[ij]} \nonumber\\
 && +a_5 \Omega_{cbb} (\overline T_{b\bar 3})_{[ik]} (M_8)^{k}_{j} (H_{6}'')^{\{ij\}}+  a_{6} \Omega_{cbb}  (\overline T_{b\bar 3})_{[ik]} (M_8)^{k}_{l} (M_8)^{l}_{j} (H_{6}'')^{\{ij\}}.
\end{eqnarray}
Decay amplitudes for different channels are obtained by expanding the above Hamiltonian and are collected in Tab.~\ref{tab:cbb_1triplet}.

The final state can be a bottom baryon (sextet) which has:
\begin{eqnarray}
 {\cal H}_{\rm eff} &=& b_1 \Omega_{cbb} (\overline T_{b 6})_{\{ik\}} (M_8)^{k}_{j} (H_{\bar 3}'')^{[ij]} +b_2 \Omega_{cbb}  (\overline T_{b 6})_{\{ik\}} (M_8)^{k}_{l} (M_8)^{l}_{j} (H_{\bar 3}'')^{[ij]}  \nonumber\\
 &&  +b_3 \Omega_{cbb} (\overline T_{b 6})_{\{ik\}} (M_8)^{k}_{j} (H_{6}'')^{\{ij\}}  \nonumber\\
 &&  + b_{5} \Omega_{cbb} (\overline T_{b 6})_{\{ij\}} (M_8)^{k}_{l} (M_8)^{l}_{k} (H_{6}'')^{\{ij\}} +b_{6} \Omega_{cbb}  (\overline T_{b 6})_{\{ik\}} (M_8)^{k}_{l} (M_8)^{l}_{j} (H_{6}'')^{\{ij\}}  \nonumber\\
 && +b_{7} \Omega_{cbb}  (\overline T_{b 6})_{\{kl\}} (M_8)^{k}_{i} M^{l}_{j}(H_{6}'')^{\{ij\}}.
\end{eqnarray}
Decay amplitudes for different channels are obtained by expanding the above Hamiltonian and are collected in Tab.~\ref{tab:cbb_1sextet}.

 \begin{table}
\tiny
\caption{Amplitudes for $\Omega_{cbb}$ decays into a bottom meson and a light baryon (octet)}\label{tab:cbb_2octet}
\begin{tabular}{|cc|cc|c|c|c|c}\hline\hline
channel & amplitude &channel & amplitude \\\hline
$\Omega_{cbb}^{0}\to B^-  \Sigma^+ $ & $ \left(-2 c_1-c_8\right) V_{ub}V_{cs}^*$ & $\Omega_{cbb}^{0}\to \overline B^0_s  \Lambda^0 $ & $ \sqrt{\frac{2}{3}} \left(-2 c_1\right) V_{ub}V_{cd}^*$\\\hline
$\Omega_{cbb}^{0}\to \overline B^0  \Lambda^0 $ & $ \frac{\left(-2 c_1+3 c_8\right) V_{ub}V_{cs}^*}{\sqrt{6}}$ & $\Omega_{cbb}^{0}\to \overline B^0_s  \Sigma^0 $ & $ -\sqrt{2} c_8 V_{ub}V_{cd}^*$\\\hline
$\Omega_{cbb}^{0}\to \overline B^0  \Sigma^0 $ & $ \frac{\left(2 c_1+c_8\right) V_{ub}V_{cs}^*}{\sqrt{2}}$ & $\Omega_{cbb}^{0}\to \overline B^0_s  \Xi^0 $ & $ \left(-2 c_1+c_8\right) V_{ub}V_{cs}^*$\\\hline
$\Omega_{cbb}^{0}\to \overline B^0  {n} $ & $ \left(2 c_1-c_8\right) V_{ub}V_{cd}^*$ && \\\hline
\hline
$\Omega_{cbb}^{0}\to B^-  \Lambda^0  \pi^+  $ & $ -\frac{\left(2 c_3+c_4-c_5+c_6-2 c_7-c_9+c_{10}+3 c_{11}+2 c_{12}\right) V_{ub}V_{cs}^*}{\sqrt{6}}$ & $\Omega_{cbb}^{0}\to \overline B^0  {n}  \overline K^0  $ & $ \left(c_4+c_5+c_9+c_{10}\right) V_{ub}V_{cs}^*$\\\hline
$\Omega_{cbb}^{0}\to B^-  \Lambda^0  K^+  $ & $ -\frac{\left(4 c_3+2 c_4+c_5+2 c_6-c_7-2 c_9-c_{10}+c_{12}\right) V_{ub}V_{cd}^*}{\sqrt{6}}$ & $\Omega_{cbb}^{0}\to \overline B^0  {n}  \eta  $ & $ \frac{\left(2 c_3+c_4+2 c_5+c_6+c_7+c_9+c_{11}-c_{12}\right) V_{ub}V_{cd}^*}{\sqrt{6}}$\\\hline
$\Omega_{cbb}^{0}\to B^-  \Sigma^+  \pi^0  $ & $ \frac{\left(-2 c_3-c_4-c_5-c_6+c_9+c_{10}+c_{11}\right) V_{ub}V_{cs}^*}{\sqrt{2}}$ & $\Omega_{cbb}^{0}\to \overline B^0  \Xi^-  K^+  $ & $ \left(c_4-c_7-c_9+c_{12}\right) V_{ub}V_{cs}^*$\\\hline
$\Omega_{cbb}^{0}\to B^-  \Sigma^+  K^0  $ & $ -\left(c_5+c_7+c_{10}+c_{12}\right) V_{ub}V_{cd}^*$ & $\Omega_{cbb}^{0}\to \overline B^0  \Xi^0  K^0  $ & $ -\left(2 c_3+c_6+c_{11}\right) V_{ub}V_{cs}^*$\\\hline
$\Omega_{cbb}^{0}\to B^-  \Sigma^+  \eta  $ & $ \frac{\left(-2 c_3-c_4+c_5-c_6+2 c_7+c_9+3 c_{10}+c_{11}+2 c_{12}\right) V_{ub}V_{cs}^*}{\sqrt{6}}$ & $\Omega_{cbb}^{0}\to \overline B^0_s  \Lambda^0  \pi^0  $ & $ \frac{\left(c_9+2 c_{10}+c_{12}\right) V_{ub}V_{cd}^*}{\sqrt{3}}$\\\hline
$\Omega_{cbb}^{0}\to B^-  \Sigma^0  \pi^+  $ & $ \frac{\left(2 c_3+c_4+c_5+c_6-c_9-c_{10}-c_{11}\right) V_{ub}V_{cs}^*}{\sqrt{2}}$ & $\Omega_{cbb}^{0}\to \overline B^0_s  \Lambda^0  \overline K^0  $ & $ -\frac{\left(2 c_3+c_4+2 c_5+c_6+c_7+c_9+2 c_{10}+3 c_{11}+c_{12}\right) V_{ub}V_{cs}^*}{\sqrt{6}}$\\\hline
$\Omega_{cbb}^{0}\to B^-  \Sigma^0  K^+  $ & $ \frac{\left(-c_5-c_7+c_{10}+2 c_{11}+c_{12}\right) V_{ub}V_{cd}^*}{\sqrt{2}}$ & $\Omega_{cbb}^{0}\to \overline B^0_s  \Lambda^0  \eta  $ & $ \frac{1}{3} \left(4 c_3-c_4-2 c_5+2 c_6-c_7\right) V_{ub}V_{cd}^*$\\\hline
$\Omega_{cbb}^{0}\to B^-  {n}  \pi^+  $ & $ \left(2 c_3+c_4+c_6-c_7-c_9+c_{11}+c_{12}\right) V_{ub}V_{cd}^*$ & $\Omega_{cbb}^{0}\to \overline B^0_s  \Sigma^+  \pi^-  $ & $ \left(-c_4+c_7-c_9+c_{12}\right) V_{ub}V_{cd}^*$\\\hline
$\Omega_{cbb}^{0}\to B^-  \Xi^0  K^+  $ & $ -\left(2 c_3+c_4+c_6-c_7-c_9+c_{11}+c_{12}\right) V_{ub}V_{cs}^*$ & $\Omega_{cbb}^{0}\to \overline B^0_s  \Sigma^+  K^-  $ & $ \left(-2 c_3-c_4-c_6+c_7-c_9+c_{11}+c_{12}\right) V_{ub}V_{cs}^*$\\\hline
$\Omega_{cbb}^{0}\to \overline B^0  \Lambda^0  \pi^0  $ & $ \frac{\left(2 c_3+c_4-c_5+c_6-2 c_7-c_9+c_{10}+3 c_{11}+2 c_{12}\right) V_{ub}V_{cs}^*}{2 \sqrt{3}}$ & $\Omega_{cbb}^{0}\to \overline B^0_s  \Sigma^0  \pi^0  $ & $ \left(c_7-c_4\right) V_{ub}V_{cd}^*$\\\hline
$\Omega_{cbb}^{0}\to \overline B^0  \Lambda^0  K^0  $ & $ -\frac{\left(4 c_3+2 c_4+c_5+2 c_6-c_7+2 c_9+c_{10}-c_{12}\right) V_{ub}V_{cd}^*}{\sqrt{6}}$ & $\Omega_{cbb}^{0}\to \overline B^0_s  \Sigma^0  \overline K^0  $ & $ \frac{\left(2 c_3+c_4+c_6-c_7+c_9-c_{11}-c_{12}\right) V_{ub}V_{cs}^*}{\sqrt{2}}$\\\hline
$\Omega_{cbb}^{0}\to \overline B^0  \Lambda^0  \eta  $ & $ \frac{1}{6} \left(-2 c_3+5 c_4+c_5-c_6-4 c_7+3 c_9+3 c_{10}-3 c_{11}\right) V_{ub}V_{cs}^*$ & $\Omega_{cbb}^{0}\to \overline B^0_s  \Sigma^0  \eta  $ & $ \frac{\left(c_9-2 c_{11}-c_{12}\right) V_{ub}V_{cd}^*}{\sqrt{3}}$\\\hline
$\Omega_{cbb}^{0}\to \overline B^0  \Sigma^+  \pi^-  $ & $ \left(-2 c_3-c_6+c_{11}\right) V_{ub}V_{cs}^*$ &¡¡$\Omega_{cbb}^{0}\to \overline B^0_s  \Sigma^-  \pi^+  $ & $ \left(-c_4+c_7+c_9-c_{12}\right) V_{ub}V_{cd}^*$\\\hline
$\Omega_{cbb}^{0}\to \overline B^0  \Sigma^0  \pi^0  $ & $ -\frac{1}{2} \left(2 c_3-c_4-c_5+c_6+c_9+c_{10}-c_{11}\right) V_{ub}V_{cs}^*$¡¡& $\Omega_{cbb}^{0}\to \overline B^0_s  {n}  \overline K^0  $ & $ \left(2 c_3+c_6+c_{11}\right) V_{ub}V_{cd}^*$\\\hline
$\Omega_{cbb}^{0}\to \overline B^0  \Sigma^0  K^0  $ & $ \frac{\left(c_5+c_7+c_{10}+2 c_{11}+c_{12}\right) V_{ub}V_{cd}^*}{\sqrt{2}}$ & $\Omega_{cbb}^{0}\to \overline B^0_s  \Xi^-  \pi^+  $ & $ \left(c_5+c_7-c_{10}-c_{12}\right) V_{ub}V_{cs}^*$\\\hline
$\Omega_{cbb}^{0}\to \overline B^0  \Sigma^0  \eta  $ & $ \frac{\left(2 c_3+c_4-c_5+c_6-2 c_7-c_9-3 c_{10}-c_{11}-2 c_{12}\right) V_{ub}V_{cs}^*}{2 \sqrt{3}}$ & $\Omega_{cbb}^{0}\to \overline B^0_s  \Xi^-  K^+  $ & $ \left(-c_4-c_5+c_9+c_{10}\right) V_{ub}V_{cd}^*$\\\hline
$\Omega_{cbb}^{0}\to \overline B^0  \Sigma^-  \pi^+  $ & $ \left(c_4+c_5-c_9-c_{10}\right) V_{ub}V_{cs}^*$ & $\Omega_{cbb}^{0}\to \overline B^0_s  \Xi^0  \pi^0  $ & $ \frac{\left(-c_5-c_7+c_{10}+c_{12}\right) V_{ub}V_{cs}^*}{\sqrt{2}}$\\\hline
$\Omega_{cbb}^{0}\to \overline B^0  \Sigma^-  K^+  $ & $ \left(-c_5-c_7+c_{10}+c_{12}\right) V_{ub}V_{cd}^*$ & $\Omega_{cbb}^{0}\to \overline B^0_s  \Xi^0  K^0  $ & $ -\left(c_4+c_5+c_9+c_{10}\right) V_{ub}V_{cd}^*$\\\hline
$\Omega_{cbb}^{0}\to \overline B^0  {n}  \pi^0  $ & $ -\frac{\left(2 c_3+c_4+c_6-c_7+c_9+2 c_{10}+c_{11}+c_{12}\right) V_{ub}V_{cd}^*}{\sqrt{2}}$ & $\Omega_{cbb}^{0}\to \overline B^0_s  \Xi^0  \eta  $ & $ \frac{\left(4 c_3+2 c_4+c_5+2 c_6-c_7+2 c_9+3 c_{10}+2 c_{11}+c_{12}\right) V_{ub}V_{cs}^*}{\sqrt{6}}$\\\hline
\hline
\end{tabular}
\end{table}

If the final state contains a bottom meson and a light baryon (octet), we have
\begin{eqnarray}
 {\cal H}_{\rm eff} &=& c_1 \Omega_{cbb} \overline B^{k}\epsilon_{ijl} (\overline T_8)^{l}_{k}  (H_{\bar 3}'')^{[ij]} +c_2 \Omega_{cbb} \overline B^{i}\epsilon_{ijl} (\overline T_8)^{l}_{k}  (H_{\bar 3}'')^{[jk]} \nonumber\\
 &&+c_3 \Omega_{cbb} \overline B^{m}\epsilon_{ijl} (\overline T_8)^{l}_{k} M^k_m  (H_{\bar 3}'')^{[ij]} +c_4 \Omega_{cbb} \overline B^{j}\epsilon_{ijl} (\overline T_8)^{l}_{k} M^k_m  (H_{\bar 3}'')^{[im]} \nonumber\\
 &&+c_5 \Omega_{cbb} \overline B^{k}\epsilon_{ijl} (\overline T_8)^{l}_{k} M^j_m  (H_{\bar 3}'')^{[im]} +c_6 \Omega_{cbb} \overline B^{m}\epsilon_{ijl} (\overline T_8)^{l}_{k} M^j_m  (H_{\bar 3}'')^{[ik]} \nonumber\\
 &&+c_7 \Omega_{cbb} \overline B^{i}\epsilon_{ijl} (\overline T_8)^{l}_{k} M^j_m  (H_{\bar 3}'')^{[km]} +c_{8} \Omega_{cbb} \overline B^{i}\epsilon_{ijl} (\overline T_8)^{l}_{k}  (H_{6}'')^{\{jk\}} \nonumber\\
 &&+c_{9} \Omega_{cbb} \overline B^{j}\epsilon_{ijl} (\overline T_8)^{l}_{k} M^k_m  (H_{6}'')^{\{im\}} +c_{10} \Omega_{cbb} \overline B^{k}\epsilon_{ijl} (\overline T_8)^{l}_{k} M^j_m  (H_{6}'')^{\{im\}}\nonumber\\
 && +c_{11} \Omega_{cbb} \overline B^{m}\epsilon_{ijl} (\overline T_8)^{l}_{k} M^j_m  (H_{6}'')^{\{ik\}}+c_{12} \Omega_{cbb} \overline B^{i}\epsilon_{ijl} (\overline T_8)^{l}_{k} M^j_m  (H_{6}'')^{\{km\}}.
\end{eqnarray}
Decay amplitudes for different channels are obtained by expanding the above Hamiltonian and are collected in Tab.~\ref{tab:cbb_2octet}. One can find the amplitudes $c_1$ and $c_2$ are not independent, they always appear in the product $2c_1-c_2$, So in Tab.~\ref{tab:cbb_2octet}, we did not show $c_2$.

\begin{table}
\tiny
\caption{Amplitudes for $\Omega_{cbb}$ decays into a bottom meson and a light baryon (decuplet)}\label{tab:cbb_2decuplet}
\begin{tabular}{|cc|cc|c|c|c|c}\hline\hline
channel & amplitude &channel & amplitude \\\hline
$\Omega_{cbb}^{0}\to B^-  \Delta^{+} $ & $ \frac{2 d_4 V_{ub}V_{cd}^*}{\sqrt{3}}$ & $\Omega_{cbb}^{0}\to \overline B^0  \Sigma^{\prime0} $ & $ \sqrt{\frac{2}{3}} d_4 V_{ub}V_{cs}^*$\\\hline
$\Omega_{cbb}^{0}\to B^-  \Sigma^{\prime+} $ & $ \frac{2 d_4 V_{ub}V_{cs}^*}{\sqrt{3}}$ & $\Omega_{cbb}^{0}\to \overline B^0_s  \Sigma^{\prime0} $ & $ \sqrt{\frac{2}{3}} d_4 V_{ub}V_{cd}^*$\\\hline
$\Omega_{cbb}^{0}\to \overline B^0  \Delta^{0} $ & $ \frac{2 d_4 V_{ub}V_{cd}^*}{\sqrt{3}}$ & $\Omega_{cbb}^{0}\to \overline B^0_s  \Xi^{\prime0} $ & $ \frac{2 d_4 V_{ub}V_{cs}^*}{\sqrt{3}}$\\\hline
\hline
$\Omega_{cbb}^{0}\to B^-  \Delta^{++}  \pi^-  $ & $ \left(d_1+d_7\right) V_{ub}V_{cd}^*$ & $\Omega_{cbb}^{0}\to \overline B^0  \Sigma^{\prime0}  K^0  $ & $ \frac{\left(d_1+2 d_6+d_7\right) V_{ub}V_{cd}^*}{\sqrt{6}}$\\\hline
$\Omega_{cbb}^{0}\to B^-  \Delta^{++}  K^-  $ & $ \left(d_1+d_7\right) V_{ub}V_{cs}^*$ & $\Omega_{cbb}^{0}\to \overline B^0  \Sigma^{\prime0}  \eta  $ & $ -\frac{1}{6} \left(3 d_1-2 d_6+d_7\right) V_{ub}V_{cs}^*$\\\hline
$\Omega_{cbb}^{0}\to B^-  \Delta^{+}  \pi^0  $ & $ -\sqrt{\frac{2}{3}} \left(d_1-d_6\right) V_{ub}V_{cd}^*$ &¡¡$\Omega_{cbb}^{0}\to \overline B^0  \Sigma^{\prime-}  \pi^+  $ & $ \frac{\left(-d_1+d_7\right) V_{ub}V_{cs}^*}{\sqrt{3}}$\\\hline
$\Omega_{cbb}^{0}\to B^-  \Delta^{+}  \overline K^0  $ & $ \frac{\left(d_1+d_7\right) V_{ub}V_{cs}^*}{\sqrt{3}}$ & $\Omega_{cbb}^{0}\to \overline B^0  \Sigma^{\prime-}  K^+  $ & $ \frac{\left(-d_1+d_7\right) V_{ub}V_{cd}^*}{\sqrt{3}}$\\\hline
$\Omega_{cbb}^{0}\to B^-  \Delta^{+}  \eta  $ & $ \frac{1}{3} \sqrt{2} \left(d_6+d_7\right) V_{ub}V_{cd}^*$ & $\Omega_{cbb}^{0}\to \overline B^0  \Xi^{\prime0}  K^0  $ & $ \frac{2 d_6 V_{ub}V_{cs}^*}{\sqrt{3}}$\\\hline
$\Omega_{cbb}^{0}\to B^-  \Delta^{0}  \pi^+  $ & $ \frac{\left(-d_1+2 d_6+d_7\right) V_{ub}V_{cd}^*}{\sqrt{3}}$ & $\Omega_{cbb}^{0}\to \overline B^0  \Xi^{\prime-}  K^+  $ & $ \frac{\left(-d_1+d_7\right) V_{ub}V_{cs}^*}{\sqrt{3}}$\\\hline
$\Omega_{cbb}^{0}\to B^-  \Sigma^{\prime+}  \pi^0  $ & $ \frac{\left(-d_1+2 d_6+d_7\right) V_{ub}V_{cs}^*}{\sqrt{6}}$ & $\Omega_{cbb}^{0}\to \overline B^0_s  \Delta^{+}  K^-  $ & $ \frac{2 d_6 V_{ub}V_{cd}^*}{\sqrt{3}}$\\\hline
$\Omega_{cbb}^{0}\to B^-  \Sigma^{\prime+}  K^0  $ & $ \frac{\left(d_1+d_7\right) V_{ub}V_{cd}^*}{\sqrt{3}}$ & $\Omega_{cbb}^{0}\to \overline B^0_s  \Delta^{0}  \overline K^0  $ & $ \frac{2 d_6 V_{ub}V_{cd}^*}{\sqrt{3}}$\\\hline
$\Omega_{cbb}^{0}\to B^-  \Sigma^{\prime+}  \eta  $ & $ -\frac{\left(3 d_1-2 d_6+d_7\right) V_{ub}V_{cs}^*}{3 \sqrt{2}}$ & $\Omega_{cbb}^{0}\to \overline B^0_s  \Sigma^{\prime+}  \pi^-  $ & $ \frac{\left(d_1+d_7\right) V_{ub}V_{cd}^*}{\sqrt{3}}$\\\hline
$\Omega_{cbb}^{0}\to B^-  \Sigma^{\prime0}  \pi^+  $ & $ \frac{\left(-d_1+2 d_6+d_7\right) V_{ub}V_{cs}^*}{\sqrt{6}}$ & $\Omega_{cbb}^{0}\to \overline B^0_s  \Sigma^{\prime+}  K^-  $ & $ \frac{\left(d_1+2 d_6+d_7\right) V_{ub}V_{cs}^*}{\sqrt{3}}$\\\hline
$\Omega_{cbb}^{0}\to B^-  \Sigma^{\prime0}  K^+  $ & $ \frac{\left(-d_1+2 d_6+d_7\right) V_{ub}V_{cd}^*}{\sqrt{6}}$ & $\Omega_{cbb}^{0}\to \overline B^0_s  \Sigma^{\prime0}  \pi^0  $ & $ -\frac{d_1 V_{ub}V_{cd}^*}{\sqrt{3}}$\\\hline
$\Omega_{cbb}^{0}\to B^-  \Xi^{\prime0}  K^+  $ & $ \frac{\left(-d_1+2 d_6+d_7\right) V_{ub}V_{cs}^*}{\sqrt{3}}$ & $\Omega_{cbb}^{0}\to \overline B^0_s  \Sigma^{\prime0}  \overline K^0  $ & $ \frac{\left(d_1+2 d_6+d_7\right) V_{ub}V_{cs}^*}{\sqrt{6}}$\\\hline
$\Omega_{cbb}^{0}\to \overline B^0  \Delta^{+}  \pi^-  $ & $ \frac{\left(d_1+2 d_6+d_7\right) V_{ub}V_{cd}^*}{\sqrt{3}}$ & $\Omega_{cbb}^{0}\to \overline B^0_s  \Sigma^{\prime0}  \eta  $ & $ \frac{1}{3} \left(-2 d_6+d_7\right) V_{ub}V_{cd}^*$\\\hline
$\Omega_{cbb}^{0}\to \overline B^0  \Delta^{+}  K^-  $ & $ \frac{\left(d_1+d_7\right) V_{ub}V_{cs}^*}{\sqrt{3}}$ & $\Omega_{cbb}^{0}\to \overline B^0_s  \Sigma^{\prime-}  \pi^+  $ & $ \frac{\left(-d_1+d_7\right) V_{ub}V_{cd}^*}{\sqrt{3}}$\\\hline
$\Omega_{cbb}^{0}\to \overline B^0  \Delta^{0}  \pi^0  $ & $ -\sqrt{\frac{2}{3}} \left(d_1+d_6\right) V_{ub}V_{cd}^*$ & $\Omega_{cbb}^{0}\to \overline B^0_s  \Xi^{\prime0}  \pi^0  $ & $ \frac{\left(-d_1+d_7\right) V_{ub}V_{cs}^*}{\sqrt{6}}$\\\hline
$\Omega_{cbb}^{0}\to \overline B^0  \Delta^{0}  \overline K^0  $ & $ \frac{\left(d_1+d_7\right) V_{ub}V_{cs}^*}{\sqrt{3}}$ & $\Omega_{cbb}^{0}\to \overline B^0_s  \Xi^{\prime0}  K^0  $ & $ \frac{\left(d_1+d_7\right) V_{ub}V_{cd}^*}{\sqrt{3}}$\\\hline
$\Omega_{cbb}^{0}\to \overline B^0  \Delta^{0}  \eta  $ & $ \frac{1}{3} \sqrt{2} \left(d_6+d_7\right) V_{ub}V_{cd}^*$ & $\Omega_{cbb}^{0}\to \overline B^0_s  \Xi^{\prime0}  \eta  $ & $ -\frac{\left(3 d_1+4 d_6+d_7\right) V_{ub}V_{cs}^*}{3 \sqrt{2}}$\\\hline
$\Omega_{cbb}^{0}\to \overline B^0  \Delta^{-}  \pi^+  $ & $ \left(-d_1+d_7\right) V_{ub}V_{cd}^*$ & $\Omega_{cbb}^{0}\to \overline B^0_s  \Xi^{\prime-}  \pi^+  $ & $ \frac{\left(-d_1+d_7\right) V_{ub}V_{cs}^*}{\sqrt{3}}$\\\hline
$\Omega_{cbb}^{0}\to \overline B^0  \Sigma^{\prime+}  \pi^-  $ & $ \frac{2 d_6 V_{ub}V_{cs}^*}{\sqrt{3}}$ & $\Omega_{cbb}^{0}\to \overline B^0_s  \Xi^{\prime-}  K^+  $ & $ \frac{\left(-d_1+d_7\right) V_{ub}V_{cd}^*}{\sqrt{3}}$\\\hline
$\Omega_{cbb}^{0}\to \overline B^0  \Sigma^{\prime0}  \pi^0  $ & $ -\frac{\left(d_1+2 d_6-d_7\right) V_{ub}V_{cs}^*}{2 \sqrt{3}}$ & $\Omega_{cbb}^{0}\to \overline B^0_s  \Omega^-  K^+  $ & $ \left(-d_1+d_7\right) V_{ub}V_{cs}^*$\\\hline
\hline
\end{tabular}
\end{table}

If the final state contains a bottom meson and a light baryon (decuplet), we have
\begin{eqnarray}
 {\cal H}_{\rm eff} &=& d_1 \Omega_{cbb} \overline B^{j}(\overline T_{10})_{ijk} M^k_m  (H_{\bar 3}'')^{[im]}+d_2 \Omega_{cbb} \overline B^{k}(\overline T_{10})_{ijk} M^j_m  (H_{\bar 3}'')^{[im]}  \nonumber\\
 &&+d_3 \Omega_{cbb} \overline B^{i}(\overline T_{10})_{ijk} M^j_m  (H_{\bar 3}'')^{[km]} +d_{4} \Omega_{cbb} \overline B^{k}(\overline T_{10})_{ijk}  (H_{6}'')^{\{ij\}} \nonumber\\
 && +d_{5} \Omega_{cbb} \overline B^{i}(\overline T_{10})_{ijk}  (H_{6}'')^{\{jk\}} +d_{6} \Omega_{cbb} \overline B^{m}(\overline T_{10})_{ijk} M^k_m  (H_{6}'')^{\{ij\}} \nonumber\\
 &&+d_{7} \Omega_{cbb} \overline B^{j}(\overline T_{10})_{ijk} M^k_m  (H_{6}'')^{\{im\}} +d_{8} \Omega_{cbb} \overline B^{k}(\overline T_{10})_{ijk} M^j_m  (H_{6}'')^{\{im\}} \nonumber\\
 && +d_{9} \Omega_{cbb} \overline B^{m}(\overline T_{10})_{ijk} M^j_m  (H_{6}'')^{\{ik\}} +d_{10} \Omega_{cbb} \overline B^{i}(\overline T_{10})_{ijk} M^j_m  (H_{6}'')^{\{km\}}.
\end{eqnarray}
Decay amplitudes for different channels are obtained by expanding the above Hamiltonian and are collected in Tab.~\ref{tab:cbb_2decuplet}.
It is interesting to notice that the above amplitudes $d_i$s are not all independent.  The products $d_1+d_2+d_3$, $d_4+d_5$, $d_6+d_9$, $d_7+d_8+d_{10}$, appear in the expansion, thus we have removed the $d_2, d_3$ and $d_5$ and $d_9$ and $d_8$, $d_{10}$ in Tab.~\ref{tab:cbb_2decuplet}.

\section{Golden Channels}
\label{eq:goldenmodes}

\begin{table}
\caption{Cabibbo allowed decays of   $\Omega_{ccc}$  with typical branching fractions at a few percent level.  A light meson $\overline  K^0$ can be replaced by a $\overline K^{*0}$. }\label{tab:ccc_golden}
\begin{tabular}{c|c|c|c}\hline\hline
channel & channel & channel & channel \\\hline
$\Omega_{ccc}^{++}\to \Omega_{cc}^{+}\ell^+\nu_{\ell} $ & $\Omega_{ccc}^{++}\to \Xi_{c}^{\prime+}  D^0 \ell^+\nu_{\ell} $ & $\Omega_{ccc}^{++}\to \Xi_c^0  D^+ \ell^+\nu_{\ell} $ & $\Omega_{ccc}^{++}\to \Xi_{cc}^{+}  \overline K^0  \ell^+\nu_{\ell} $ \\\hline
$\Omega_{ccc}^{++}\to \Xi_c^+  D^0 \ell^+\nu_{\ell} $ & $\Omega_{ccc}^{++}\to \Xi_{c}^{\prime0}  D^+ \ell^+\nu_{\ell} $  & $\Omega_{ccc}^{++}\to \Xi_{cc}^{++}  K^-  \ell^+\nu_{\ell} $ & $\Omega_{ccc}^{++}\to \Omega_{c}^{0}  D^+_s \ell^+\nu_{\ell} $ \\\hline
\Xcline{1-4}{0.8pt}
$\Omega_{ccc}^{++}\to \Xi_{cc}^{++}  \overline K^0  $   & $\Omega_{ccc}^{++}\to \Omega_{cc}^{+}  \pi^+  $ & &    \\\hline
\Xcline{1-4}{0.8pt}
$\Omega_{ccc}^{++}\to \Xi_{cc}^{++}  \overline K^0   \pi^0  $ & $\Omega_{ccc}^{++}\to \Xi_{cc}^{+}  \pi^+   \overline K^0  $ &  $\Omega_{ccc}^{++}\to \Xi_{cc}^{++}  K^-   \pi^+  $ & $\Omega_{ccc}^{++}\to \Omega_{cc}^{+}  K^+   \overline K^0  $   \\\hline
\Xcline{1-4}{0.8pt}
$\Omega_{ccc}^{++}\to \Xi_c^+  D^+ $  & $\Omega_{ccc}^{++}\to \Xi_{c}^{\prime+}  D^+ $  & & \\\hline
\Xcline{1-4}{0.8pt}
$\Omega_{ccc}^{++}\to \Lambda_c^+  D^+  \overline K^0  $ & $\Omega_{ccc}^{++}\to \Xi_c^+  D^+_s  \overline K^0  $ & $\Omega_{ccc}^{++}\to \Xi_c^+  D^+  \pi^0  $ & $\Omega_{ccc}^{++}\to \Xi_c^0  D^+  \pi^+  $ \\\hline
$\Omega_{ccc}^{++}\to \Xi_c^+  D^0  \pi^+  $ &  & & \\\hline
\Xcline{1-4}{0.8pt}
$\Omega_{ccc}^{++}\to \Sigma_{c}^{++}  D^0  \overline K^0  $ & $\Omega_{ccc}^{++}\to \Xi_{c}^{\prime0}  D^+  \pi^+  $ & $\Omega_{ccc}^{++}\to \Xi_{c}^{\prime+}  D^0  \pi^+  $ & $\Omega_{ccc}^{++}\to \Omega_{c}^{0}  D^+  K^+  $ \\\hline
$\Omega_{ccc}^{++}\to \Sigma_{c}^{++}  D^+  K^-  $ & $\Omega_{ccc}^{++}\to \Xi_{c}^{\prime+}  D^+_s  \overline K^0  $ & $\Omega_{ccc}^{++}\to \Xi_{c}^{\prime+}  D^+  \pi^0  $ & $\Omega_{ccc}^{++}\to \Omega_{c}^{0}  D^+_s  \pi^+  $ \\\hline
$\Omega_{ccc}^{++}\to \Sigma_{c}^{+}  D^+  \overline K^0  $ &  & & \\\hline
\end{tabular}
\end{table}

\begin{table}
\caption{CKM allowed decay channels of  $\Omega_{bbb}^{-}$. }\label{tab:bbb_golden}
\begin{tabular}{c|c|c|c}\hline\hline
channel & channel & channel & channel \\\hline
$\Omega_{bbb}^{-}\to \Xi_{bb}^{0}  D^0 \ell^-\bar\nu_{\ell} $ & $\Omega_{bbb}^{-}\to \Xi_{bc}^{+}  B^- \ell^-\bar\nu_{\ell} $  & $\Omega_{bbb}^{-}\to \Omega_{bb}^{-}  D^+_s \ell^-\bar\nu_{\ell} $ & $\Omega_{bbb}^{-}\to \Omega_{bc}^{0}  \overline B^0_s \ell^-\bar\nu_{\ell} $ \\\hline
$\Omega_{bbb}^{-}\to \Xi_{bb}^{-}  D^+ \ell^-\bar\nu_{\ell} $ & $\Omega_{bbb}^{-}\to \Xi_{bc}^{0}  \overline B^0 \ell^-\bar\nu_{\ell} $ & & \\\hline
\Xcline{1-4}{0.8pt}
$\Omega_{bbb}^{-}\to \Omega_{bb}^{-}J/\psi $ & $\Omega_{bbb}^{-}\to \Xi_{b}^{\prime0}  B^- J/\psi $ & $\Omega_{bbb}^{-}\to \Xi_{bb}^{0}  K^-  J/\psi $ & $\Omega_{bbb}^{-}\to \Omega_{b}^{-}  \overline B^0_s J/\psi $ \\\hline
$\Omega_{bbb}^{-}\to \Xi_b^0  B^- J/\psi $ & $\Omega_{bbb}^{-}\to \Xi_{b}^{\prime-}  \overline B^0 J/\psi $ & $\Omega_{bbb}^{-}\to \Xi_{bb}^{-}  \overline K^0  J/\psi $ & $\Omega_{bbb}^{-}\to \Xi_b^-  \overline B^0 J/\psi $  \\\hline
$\Omega_{bbb}^{-}\to \Omega_{bbc}^{0}  D^-_s $ & $\Omega_{bbb}^{-}\to \Omega_{bbc}^{0}  D^-  \overline K^0  $ & $\Omega_{bbb}^{-}\to \Omega_{bbc}^{0}  \overline D^0  K^-  $ & \\\hline
$\Omega_{bbb}^{-}\to \Omega_{bbc}^{0}  \pi^-  $ & $\Omega_{bbb}^{-}\to \Omega_{bbc}^{0}  K^0   K^-  $ &  & \\\hline
$\Omega_{bbb}^{-}\to \Xi_{bb}^{-}  D^0 $ & $\Omega_{bbb}^{-}\to \Xi_{bb}^{-}  D^+  \pi^-  $  & $\Omega_{bbb}^{-}\to \Xi_{bb}^{-}  D^0  \pi^0  $ & $\Omega_{bbb}^{-}\to \Omega_{bb}^{-}  D^0  K^0  $ \\\hline
$\Omega_{bbb}^{-}\to \Lambda_b^0  B^-  D^0 $ & $\Omega_{bbb}^{-}\to \Sigma_{b}^{-}  B^-  D^+ $ & $\Omega_{bbb}^{-}\to \Xi_b^-  \overline B^0_s  D^0 $ &  $\Omega_{bbb}^{-}\to \Xi_{b}^{\prime-}  B^-  D^+_s $ \\\hline
$\Omega_{bbb}^{-}\to \Xi_{bb}^{0}  D^0  \pi^-  $ & $\Omega_{bbb}^{-}\to \Sigma_{b}^{-}  \overline B^0  D^0 $ & $\Omega_{bbb}^{-}\to \Omega_{bb}^{-}  D^+_s  \pi^-  $ & $\Omega_{bbb}^{-}\to \Xi_{b}^{\prime-}  \overline B^0_s  D^0 $ \\\hline
$\Omega_{bbb}^{-}\to \Xi_b^-  B^-  D^+_s $ & $\Omega_{bbb}^{-}\to \Xi_{bb}^{-}  D^+_s  K^-  $ & $\Omega_{bbb}^{-}\to \Sigma_{b}^{0}  B^-  D^0 $ &  \\\hline
\end{tabular}
\end{table}


Based on the above analysis, we give a collection of the CKM allowed decay channels for the $\Omega_{ccc}^{++}$ in Tab.~\ref{tab:ccc_golden} and for the $\Omega_{bbb}$ in Tab.~\ref{tab:bbb_golden}.  The ones for $\Omega_{ccb}$ and $\Omega_{cbb}$ can be obtained by the replacements as discussed in the above section.

\begin{itemize}

\item The light pseudoscalar meson in these two tables can be replaced by its vector counterpart. For instance a $\overline K^0$ can be replaced by  a $\overline K^{*0}$ decaying into $K^-\pi^+$.

\item Branching fractions for semileptonic $\Omega_{ccc}$ decay channels in Tab.~\ref{tab:ccc_golden}  can reach a few percents, but there is a neutrino in the final state, reducing somewhat experimental efficiency.

\item Nonleptonic $\Omega_{ccc}$ such as $\Omega_{ccc}^{+++}\to \Xi_{cc}^{++}K^-\pi^+$ might be used  to search for $\Omega_{ccc}$ especially at LHC, since their branching fractions are sizable, and the final state can be easily to identify.   This will make use of the doubly heavy baryon $\Xi_{cc}^{++}$ which has been just discovered by LHCb.

\item For nonleptonic decays of  $\Omega_{bbb}^{-}$,  the largest branching fraction might reach   $10^{-3}$.  Taking into account its daughter decays, we expect the branching fraction for  $\Omega_{bbb}^-$ decaying into charmless final state is at most $10^{-9}$. Thus the triply bottom baryon can be only observed with a large amount of data in future, such as the high luminosity LHC.

\end{itemize}

\section{Conclusions}
\label{sec:conclusions}

Up to date, quark model is a most successful theoretical tools to describe the hadron spectrum especially  the lowest lying hadrons. Since the charm and bottom quarks are much heavier than the lighter ones,   hadrons with a different number of heavy quarks will have distinct dynamics. On experimental side, light hadrons with no heavy quark, singly heavy baryons, and doubly heavy baryons have been established, but   triply heavy baryons are still missing.  Thus it deserves more theoretical and experimental efforts to study various properties of triply heavy baryons from both theoretical and experimental sides.

In this work, we have systematically analyzed weak decays of triply heavy baryons for the first time in the literature.  Decay amplitudes for various transitions have been parametrized in terms of the SU(3) independent amplitudes. Using these results, we find a number of relations for the partial decay widths.
We also give a list of decay channels with sizable branching fractions.   We suggest our experimental colleagues to perform a search at hadron colliders and  the electron and positron collisions in future.

\section*{Acknowledgements}

The authors are   grateful to Prof. Ji-Bo He  for useful discussions and valuable comments. WW thanks the hospitality from Prof.~Xiao-Gang He at Tsinghua University, Hsinchu where this work was finalized.
This work is supported  in part   by National  Natural
Science Foundation of China under Grant
 No.11575110, 11655002, 11735010,  Natural  Science Foundation of Shanghai under Grant  No.~15DZ2272100.

\end{document}